\documentclass[aps,prd,twocolumn]{revtex4-2}
\usepackage{graphicx}
\usepackage{amssymb}
\usepackage{amsmath}
\usepackage{float}

\begin{document}

\title{Gravitational Lensing in the Schwarzschild Spacetime: Photon Rings in Vacuum and in the Presence of a Plasma}

\author{Torben C. Frost}

\affiliation{Kavli Institute for Astronomy and Astrophysics, Peking University, 100871 Beijing, China.\\
e-mail: torben.frost@pku.edu.cn}

\date{August 1st, 2025}

\begin{abstract}
Astrophysical black holes are usually surrounded by an accretion disk. At least parts of these accretion disks consist of a plasma in which light rays with different energies are dispersed. However, we usually do not know the exact configurations of these plasmas. In this paper we will now use the example of a Schwarzschild black hole embedded in an inhomogeneous pressureless and nonmagnetised plasma to investigate how the structural changes of the photon rings can help us to determine the properties of a plasma surrounding a black hole using multifrequency observations. For this purpose we will use a simple analytic model which describes a plasma whose electron density increases towards the equatorial plane when we approach the event horizon. For the chosen model we will first derive and then analytically solve the equations of motion. Then we will place an observer in the domain of outer communication and introduce an orthonormal tetrad to relate the constants of motion to latitude-longitude coordinates on the observer's celestial sphere. In the next step we will use the analytic solutions to investigate the geometric structures of the direct image as well as the photon rings of first and second order on the observer's celestial sphere. We will write down a lens equation and calculate the redshift and the travel time. We will compare the obtained results to results for photon rings in vacuum and in the presence of a homogeneous plasma. Finally, we will discuss which of these quantities can be used to extract information about the properties of the plasma.
\end{abstract}

\maketitle
\section{Introduction}
The recently published images of the shadows of the supermassive compact objects at the centres of the galaxy M87 \cite{EHTCollaboration2019a} and the Milky Way \cite{EHTCollaboration2022} show two different features. A dark area at their centre, which is commonly interpreted as the shadow of a black hole, and a bright ring, which is usually interpreted as emission from an accretion disk. While the shadow is strongly blurred and thus alone does not allow to determine whether we have a spherically symmetric and static black hole or a rotating black hole, the shadow and the observed emission from the accretion disk in comparison with results from numerical accretion disk models allowed to determine that both compact objects can be described by the Kerr spacetime \cite{Kerr1963} from general relativity.

Although the observation of the shadow alone is already a big scientific achievement in itself one problem for the analysis of the observations is that we do not know the exact configuration of the accretion disk. While we can be sure that it at least partially consists of a plasma we do neither know the plasma's exact distribution around the black hole nor its plasma frequency. Here, in the case of the supermassive black hole at the centre of the galaxy M87 the Event Horizon Telescope Collaboration managed to infer information about the properties of the plasma \cite{EHTCollaboration2019b,EHTCollaboration2021}, however, these are based on assuming specific emission models. Because we do not know whether these emission models are correct or not, for future analyses it is desirable to find methods which allow to directly infer information about the plasma surrounding a black hole using means not relying on specific model assumptions. 

When we want to determine the nature of a supermassive black hole candidate and its surroundings we need to perform observations in the strong field regime. Thus when we want to construct models which accurately describe such a scenario, we need to use the full formalism of general relativity. However, since real astrophysical scenarios are rather complex this automatically requires that we use numerical methods to investigate different scenarios. Luckily we can already get a good idea about which features and effects we may be able to observe by using simplified plasma models so that we can treat the whole problem analytically. Thus, although still most of the shadow calculations in the presence of an accretion disk are done numerically, in the last years also analytical approaches to the problem attracted more and more attentention.

In the context of general relativity the propagation of light in a pressureless and magnetised plasma was investigated by Breuer and Ehlers \cite{Breuer1980,Breuer1981} and the propagation of light in a pressureless and nonmagentised plasma was treated by Perlick \cite{Perlick2000}. In the context of his work Perlick also derived a formula for the deflection angle of light in a plasma whose density only depends on the radius coordinate for the Schwarzschild spacetime and the equatorial plane of the Kerr spacetime.

Tsupko and Bisnovatyi-Kogan \cite{Tsupko2013} investigated gravitational lensing of light rays in the Schwarzschild spacetime in the presence of a homogeneous plasma. In their work they only considered light rays which make more than one full turn around the black hole and found that the angular size of relativistic rings, the angular distance of point images from the centre of the shadow, and the magnification associated with relativistic images increase. Similarly, in the weak-field limit it was shown by, e.g., Er and Mao \cite{Er2014}, that when light rays travel through a plasma the positions of the observed images as well as the travel time of the light rays change in comparison to light rays propagating in vacuum. In the strong deflection limit gravitational lensing of light in an inhomogeneous plasma was investigated by Feleppa, Bozza, and Tsupko \cite{Feleppa2024}. In their work they derived analytical formulas for the strong deflection limit and also calculated the positions and the magnifications of higher-order images. The effects of a plasma on the shadow of a Schwarzschild black hole was treated in the work of Perlick, Tsupko, and Bisnovatyi-Kogan \cite{Perlick2015}. Building on their earlier results Perlick and Tsupko \cite{Perlick2017} then extended this work to investigate the effects of a plasma on the shadow of a Kerr black hole. In addition, they demonstrated under which conditions the equations of motion can be separated in the presence of a pressureless and nonmagnetised plasma. This work was then extended to general stationary and axisymmetric spacetimes by Bezd\v{e}kov\'{a}, Perlick, and Bi\v{c}\'{a}k \cite{Bezdekova2022}. Perlick and Tsupko \cite{Perlick2024} showed that in the Kerr spacetime in the presence of an inhomogeneous pressureless and nonmagnetised plasma non-equatorial closed circular photon orbits can exist outside the equatorial plane, and for light rays in the equatorial plane they derived a formula which can be used to calculate the deflection angle for light rays inside and outside the ergoregion. While these works generally assume that the light rays propagate in a pressureless and nonmagnetised plasma, Tsupko \cite{Tsupko2021} investigated the deflection of light in general spherically symmetric and static spacetimes in the presence of a dispersive medium.

In addition to the conventional lensing observables in the strong field regime around black holes we can also find one very characteristic feature the so-called \emph{photon rings}. That these ring-like structures exist is already known since the 70s, see, e.g., the work of Luminet \cite{Luminet1979} but only attracted increased attention relatively recently. The photon rings are very thin rings formed by light emitted by the accretion disk, and their shape and width are expected to encode characteristic information about the black hole spacetime, see, e.g., the work of Johannsen and Psaltis \cite{Johannsen2010}. In addition, when the photon rings do not overlap also the gaps between them are expected to contain information about the characteristic parameters of a compact object, see, e.g., the work of Aratore, Tsupko, and Perlick \cite{Aratore2024}. While it was already claimed that for the supermassive black hole in the centre of the galaxy M87 hints for the existence of the first-order photon ring were detected, see the work of Broderick \emph{et al.} \cite{Broderick2022}, this claim is highly disputed, see, e.g., the work of Lockhart and Gralla \cite{Lockhart2022}. However, detecting them will be one of the main objectives of the Black Hole Explorer \cite{Johnson2024} and thus we can expect that more concrete evidence will become available once the mission is successfully launched. 

In the presence of a plasma we can now expect that the photons forming the rings are dispersed and thus for each observation frequency we observe a series of rings whose positions, shapes, and widths depend on the exact plasma distribution. Thus observing the photon rings at different frequencies will provide us with valuable information about the configuration of the plasma in the accretion disk. That the visibility and the structure of the shadow depends on the observation frequency was already shown by Falcke, Melia, and Agol \cite{Falcke2000}. In their work they investigated the structural changes of the shadow for different plasma models with the Schwarzschild spacetime and the Kerr spacetime as backgrounds. First investigations on the frequency dependency of the photon rings themselves on the other hand were also already performed by, e.g., Desire, C\'{a}rdenas-Avenda\~{n}o, and Chael \cite{Desire2025} and Kobialko, Gal'tsov, and Mochanov \cite{Kobialko2025}. In the first work the authors coupled an analytic ray tracing code with an accretion disk model to calculate the photon ring structures at three different frequencies in the radio. In the second work the authors calculated shadow images as well as intensity maps for a plasma density profile which follows an inverse power-law. In both works the authors found that the resulting images, and in particular the visibility of the secondary image, strongly depend on the frequency of the photons. 

In the present paper we now use the formalism of Perlick and Tsupko \cite{Perlick2017} to demonstrate that in the presence of a plasma the photon rings do not only contain characteristic information about the spacetime but also about the plasma density. For this purpose we will assume that we have an idealised inhomogeneous pressureless and nonmagnetised plasma in the Schwarzschild spacetime described by a generalised version of the inhomogeneous plasma distribution of Perlick \cite{Perlick2023}. The plasma distribution describes an inhomogeneous pressureless and nonmagnetised plasma which has its highest plasma frequency in the equatorial plane close to the event horizon. For $\vartheta=0$ and $\vartheta=\pi$ and at spatial infinity on the other hand the plasma frequency of this plasma distribution approaches the plasma freqency of a homogeneous plasma. 

As first step we will derive the equations of motion. We will see that, when we rewrite them in terms of the Mino parameter \cite{Mino2003}, the equations of motion can be analytically solved using elementary and elliptic functions, and elliptic integrals. Here, for this purpose we will use Jacobi's elliptic functions and Legendre's elliptic integrals following the approaches in Gralla and Lupsasca \cite{Gralla2020} and Frost \cite{Frost2023}. In the next step we will then distribute static light sources in a luminous disk between the horizon and an outer boundary marked by the radius coordinate $r_{\text{out}}$. Then we will place a static observer in the domain of outer communication and introduce an orthonormal tetrad following the approach described in Perlick and Tsupko \cite{Perlick2017} to relate the constants of motion to latitude-longitude coordinates on the observer's celestial sphere and, for light rays in a plasma, the energy of the light rays measured by the observer. Then we use the analytic solutions to the equations of motion to calculate the direct image and the photon rings of first and second order for different plasma distributions and compare them to the direct image and the photon rings for light rays travelling in vacuum. Here, besides the general structure of the direct image and the photon rings, for each of them we will also define a lens equation, and calculate the redshift and the travel time. Finally, we will discuss which of these quantities can be used to determine the properties of the plasma.

The remainder of this paper is now structured as follows. In Sec.~\ref{Sec:SST} we will write down and briefly discuss the properties of the Schwarzschild spacetime. Then we will derive and solve the equations of motion. In Sec.~\ref{Sec:CCS} we will relate the constants of motion to the latitude-longitude coordinates on the celestial sphere of a static observer and derive the angular radius of the shadow. In Sec.~\ref{Sec:PR} we will then discuss the geometric structure, the lens equation, the redshift, and the travel time of the direct images and the photon rings. We will also discuss the implications of our results for the observation of the photon rings and determining the properties of plasmas in accretion disks around supermassive black holes in space. Finally, in Sec.~\ref{Sec:SC} we will summarise our results. Note that throughout the whole paper we will use geometric units such that $c=G=1$ and the metric signature $(-,+,+,+)$.

\section{Spacetime and Equations of Motion}\label{Sec:SST}
In this paper we want to investigate how the presence of a pressureless and nonmagnetised plasma affects the photon ring structure using the Schwarzschild metric as an example. For this purpose in this section we will first write down its line element, and then derive and solve the equations of motion for light rays travelling in vacuum and through homogeneous and inhomogeneous plasmas.

\subsection{The Schwarzschild Spacetime}
The Schwarzschild spacetime is the first exact solution to Einstein's vacuum field equations which was discovered after Einstein formulated them. It is spherically symmetric and static, and in geometric units with $c=G=1$ its line element reads
\begin{eqnarray}
&g_{\mu\nu}\text{d}x^{\mu}\text{d}x^{\nu}=-P(r)\text{d}t^2+\frac{\text{d}r^2}{P(r)}\\
&+r^2\left(\text{d}\vartheta^2+\sin^2\vartheta\text{d}\varphi^2\right),\nonumber
\end{eqnarray}
where, $P(r)=1-2m/r$, $m$ is the mass parameter and we have $t\in\mathbb{R}$, $0<r$, $\vartheta\in[0,\pi]$, and $\varphi\in[0,2\pi)$. In addition, we recall that the Schwarzschild spacetime has a spacelike curvature singularity at $r=0$ and a coordinate singularity at $r_{\text{H}}=2m$. Here, in the domain $0<r<r_{\text{H}}$ the spacetime is nonstatic, and in the domain $r_{\text{H}}<r$ the spacetime is static. Note that the static region outside the horizon is commonly also referred to as \emph{domain of outer communication} and we will follow this convention throughout the rest of this paper.

\subsection{The Equations of Motion in the Presence of A Plasma}
Now we want to derive the equations of motion for light rays travelling in the Schwarzschild spacetime in vacuum and through pressureless and nonmagnetised homogeneous and inhomogeneous plasmas. For this purpose we will first write down the Hamiltonian. It reads
\begin{eqnarray}
\mathcal{H}\left(x,p\right)=\frac{1}{2}\left(g^{\mu\nu}p_{\mu}p_{\nu}+E_{\text{pl}}(x)^2\right),
\end{eqnarray}
where the $g^{\mu\nu}$ are the coefficients of the inverse metric tensor, the $p_{\mu}$ are the components of the four-momentum, and $E_{\text{pl}}(x)$ is what we refer to as \emph{plasma energy}. Note that in many other works, see, e.g., Perlick, Tsupko, and Bisnovatyi-Kogan \cite{Perlick2015} or Perlick and Tsupko \cite{Perlick2017,Perlick2024}, the authors commonly write the Hamiltonian in units such that it contains a plasma frequency $\omega_{\text{pl}}(x)$. Here, the plasma energy $E_{\text{pl}}(x)$ and the plasma frequency $\omega_{\text{pl}}(x)$ are related by
\begin{eqnarray}
E_{\text{pl}}(x)=\hbar\omega_{\text{pl}}(x).
\end{eqnarray}
The plasma frequency $\omega_{\text{pl}}(x)$ is related to the charge of the electrons in the plasma $e$, the mass of the electrons $m_{e}$, and the electron number density $n_{e}(x)$ by
\begin{eqnarray}
\omega_{\text{pl}}^2(x)=\frac{4\pi e^2}{m_{e}}n_{e}(x).
\end{eqnarray}
Now for a light ray with local energy $E_{\text{loc}}(x)$ propagating through a plasma with the plasma energy $E_{\text{pl}}(x)$ at an event $x$ the refractive index reads
\begin{eqnarray}
n_{\text{rf}}(x)^2=1-\frac{E_{\text{pl}}(x)^2}{E_{\text{loc}}(x)^2}.
\end{eqnarray}
Thus, light rays can only propagate through the plasma as long as the refractive index is positive.

In this paper we want to analytically investigate the effects of an inhomogeneous plasma on the photon rings and compare the obtained results to results obtained for photon rings in vacuum and in a homogeneous plasma. Thus in the following we will derive and solve the equations of motion for light rays travelling in vacuum and through pressureless and nonmagnetised homogeneous and inhomogeneous plasmas. While for general inhomogeneous plasma distributions the equations of motion for light rays are not analytically solvable the plasma profile presented by Perlick \cite{Perlick2023} is one of the few exceptions. In this paper we use a generalised version of this profile. It reads
\begin{eqnarray}\label{eq:PlasmaEn}
E_{\text{pl}}(r,\vartheta)^2=E_{\text{c}}^2\frac{r^2+\omega_{\text{p}}^2\sin^2\vartheta}{r^2},
\end{eqnarray}
where $E_{\text{C}}$ is a scaling parameter and $\omega_{\text{p}}$ is a parameter which characterises the inhomogeneity of the plasma close to the black hole. This plasma distribution describes a plasma whose plasma energy takes the highest value in the equatorial plane close to the event horizon of the black hole and approaches the plasma energy of a homogeneous plasma for $\sin\vartheta=0$ and at spatial infinity. Note that in the case $\omega_{\text{p}}=0$ (\ref{eq:PlasmaEn}) reduces to the plasma energy for a homogeneous plasma. For $\omega_{\text{p}}=4m$ on the other hand we obtain the original plasma profile of Perlick \cite{Perlick2023}. 

Now we use Hamilton's equations 
\begin{eqnarray}
\dot{x}^{\mu}=\frac{\partial\mathcal{H}}{\partial p_{\mu}}~~~\text{and}~~~\dot{p}_{\mu}=-\frac{\partial\mathcal{H}}{\partial x^{\mu}}
\end{eqnarray}
and the constraint $\mathcal{H}(x,p)=0$ for light rays in vacuum and in plasma to derive the equations of motion. They read
\begin{eqnarray}\label{eq:EoMt}
\frac{\text{d}t}{\text{d}\lambda}=\frac{r^2E}{P(r)},
\end{eqnarray}
\begin{eqnarray}\label{eq:EoMr}
\left(\frac{\text{d}r}{\text{d}\lambda}\right)^2=E^2r^4-E_{\text{C}}^2 r^4P(r)-r^2 P(r)K,
\end{eqnarray}
\begin{eqnarray}\label{eq:EoMtheta}
\left(\frac{\text{d}\vartheta}{\text{d}\lambda}\right)^2=K-\omega_{\text{p}}^2 E_{\text{c}}^2\sin^2\vartheta-\frac{L_{z}^2}{\sin^2\vartheta},
\end{eqnarray}
\begin{eqnarray}\label{eq:EoMphi}
\frac{\text{d}\varphi}{\text{d}\lambda}=\frac{L_{z}}{\sin^2\vartheta},
\end{eqnarray}
where we introduced three constants of motion, namely the energy $E$, the angular momentum about the $z$-axis $L_{z}$, and the Carter constant $K$. In addition, we reparameterised the equations of motion in terms of the so-called Mino parameter $\lambda$ \cite{Mino2003}, which is related to the affine parameter $s$ by 
\begin{eqnarray}
\frac{\text{d}\lambda}{\text{d}s}=\frac{1}{r^2}.
\end{eqnarray}
Note that here we choose the Mino parameter such that it is positive for future-directed trajectories and negative for past-directed trajectories.

\subsection{Solving the Equations of Motion}\label{Sec:EoM}
We already derived the equations of motion and reparameterised them in terms of the Mino parameter. Now all of them have structures so that we can analytically solve them using elementary and Jacobi's elliptic functions as well as Legendre's elliptic integrals. While in general depending on the type of motion we consider the equations of motion have various different solutions, in this paper we are only interested in the photon rings and thus light rays which can leave the close vicinity of the black hole. While in vacuum this is the case for all light rays in plasma this is effectively only the case for unbound light rays. Note that we can also have marginally bound light rays but they barely reach spatial infinity and thus are not of interest for us.

In addition, in this paper for investigating the photon rings we will assume that we have a luminous disk in the equatorial plane. Since the photon rings are only defined for light rays that can leave the equatorial plane we do not have to consider light rays associated with the other types of motion. Therefore, in the following we will only derive the solutions for the types of motion which can contribute to the formation of the photon rings. Here, for now, we assume that we have general initial conditions $x_{i}=x\left(\lambda_{i}\right)=\left(t_{i},r_{i},\vartheta_{i},\varphi_{i}\right)$, where $\lambda_{i}$ is the Mino parameter at the initial coordinates $x_{i}$. 

\subsubsection{The $r$ Motion}\label{Sec:EoMr}
We start with the $r$ motion. Since in Sec.~\ref{Sec:PR} we will only consider a luminous disk in the equatorial plane we can exclude light rays travelling along radial trajectories from our discussion since outside the horizon these light rays can either not come from the equatorial plane or for them, in the case of light rays travelling in the equatorial plane, the photon rings are not well-defined. For light rays in vacuum and a homogeneous plasma radial motion always occurs for $K=0$, while for light rays travelling in the inhomogeneous plasma described by the distribution given by (\ref{eq:PlasmaEn}) this type of motion only occurs for light rays with $L_{z}=0$ and $K=0$ travelling along $\vartheta=0$ and $\vartheta=\pi$ and light rays with $L_{z}=0$ and $K=\omega_{\text{p}}^2E_{\text{C}}^2$ in the equatorial plane.

Before we start to discuss the remaining types of motion we will now first calculate the radius coordinates of the unstable photon orbits in the domain of outer communication. For these orbits we require that we have $\text{d}r/\text{d}\lambda=\text{d}^2 r/\text{d}\lambda^2=0$. From these two conditions we obtain the two conditional equations
\begin{eqnarray}\label{eq:EoMrf1}
\left(E^2-E_{\text{C}}^2\right)r^4+2mE_{\text{C}}^2 r^3-Kr^2+2mKr=0
\end{eqnarray}
and
\begin{eqnarray}\label{eq:EoMrf2}
2\left(E^2-E_{\text{C}}^2\right)r^3+3mE_{\text{C}}^2 r^2-Kr+mK=0
\end{eqnarray}
Now we multiply (\ref{eq:EoMrf1}) with 2 and divide it by $r$. Then we subtract (\ref{eq:EoMrf2}) from the resulting equation. We solve for $K$, insert the obtained result in (\ref{eq:EoMrf1}), and rewrite the equation as a polynomial of $r$. We divide by the coefficient of the highest order term and obtain
\begin{eqnarray}
r^2+m\frac{4E_{\text{C}}^2-3E^2}{E^2-E_{\text{C}}^2}r-\frac{4m^2 E_{\text{C}}^2}{E^2-E_{\text{C}}^2}=0.
\end{eqnarray}
This equation now has two different radius coordinates as solutions. The first radius coordinate is positive and lies in the domain of outer communication. The other radius coordinate is negative and thus cannot represent the radius coordinate of a photon orbit. Therefore, only the first radius coordinate marks the position of a photon orbit. It reads
\begin{eqnarray}
r_{\text{ph}}=\frac{m\left(3E^2-4E_{\text{C}}^2+E\sqrt{9E^2-8E_{\text{C}}^2}\right)}{2\left(E^2-E_{\text{C}}^2\right)}.
\end{eqnarray}
Using the radius coordinate of the unstable photon orbit $r_{\text{ph}}$ and the Carter constant $K$ we can now distinguish three more different types of motion. When we define $K_{\text{ph}}$ as the Carter constant for the light rays on or asymptotically coming from (or going to) the unstable photon orbit, we can use them to distinguish the following three different types of motion.
 
In the first case the Carter constant $K$ fulfills $0<K<K_{\text{ph}}$ and the right-hand side of (\ref{eq:EoMr}) has two distinct real roots and a pair of complex conjugate roots. We now label and sort the real roots such that $r_{2}<r_{1}=0$. The complex conjugate roots on the other hand are labelled such that $r_{3}=\bar{r}_{4}=R_{3}+iR_{4}$, where we choose the imaginary part such that $0<R_{4}$. Please note that here and for the other two cases we only write down the general characterisation of the roots. The analytical forms of the roots can be found in Appendix~\ref{Sec:rRoots}. Light rays travelling along these trajectories cannot pass through a turning point in the domain of outer communication. As first step we now rewrite (\ref{eq:EoMr}) in terms of the roots. In the next step we separate variables. Then we integrate from $r(\lambda_{i})=r_{i}$ to $r(\lambda)=r$. We get
\begin{eqnarray}
&\hspace{-0.5cm}\lambda-\lambda_{i}=i_{r_{i}}\int_{r_{i}}^{r}\frac{\text{d}r'}{\sqrt{\left(E^2-E_{\text{C}}^2\right)r'(r'-r_{2})\left(\left(R_{3}-r'\right)^2+R_{4}^2\right)}},
\end{eqnarray}
where $i_{r_{i}}=\text{sgn}\left(\left.\text{d}r/\text{d}\lambda\right|_{r=r_{i}}\right)$. In the next step we define two new constants of motion. They read
\begin{eqnarray}\label{eq:R}
R=\sqrt{R_{3}^2+R_{4}^2}
\end{eqnarray}
and
\begin{eqnarray}\label{eq:Rbar}
\bar{R}=\sqrt{\left(R_{3}-r_{2}\right)^2+R_{4}^2}.
\end{eqnarray}
In the next step we substitute \cite{Hancock1917,Gralla2020}
\begin{eqnarray}\label{eq:sub1}
r=\frac{r_{2}R\left(\cos\chi-1\right)}{\bar{R}-R+\left(\bar{R}+R\right)\cos\chi},
\end{eqnarray}
to put the integral into the Legendre form defined by (\ref{eq:LFE}). Then we follow the steps outlined in Appendix~\ref{eq:JEF} to derive the solution to the equation of motion. It is given in terms of Jacobi's elliptic cn function and reads
\begin{widetext}
\begin{eqnarray}\label{eq:solr1}
r(\lambda)=\frac{r_{2}R\left(\text{cn}\left(i_{r_{i}}\sqrt{\left(E^2-E_{\text{C}}^2\right)R\bar{R}}\left(\lambda-\lambda_{i}\right)+\lambda_{r_{i},k_{1}},k_{1}\right)-1\right)}{\bar{R}-R+\left(\bar{R}+R\right)\text{cn}\left(i_{r_{i}}\sqrt{\left(E^2-E_{\text{C}}^2\right)R\bar{R}}\left(\lambda-\lambda_{i}\right)+\lambda_{r_{i},k_{1}},k_{1}\right)},
\end{eqnarray}
\end{widetext}
where the square of the elliptic modulus $k_{1}$, $\lambda_{r_{i},k_{1}}$, and the initial condition $\chi_{i}$ are given by
\begin{eqnarray}\label{eq:k1}
k_{1}=\frac{\left(R+\bar{R}\right)^2-r_{2}^2}{4R\bar{R}},
\end{eqnarray}
\begin{eqnarray}
\lambda_{r_{i},k_{1}}=F_{L}\left(\chi_{i},k_{1}\right),
\end{eqnarray}
and 
\begin{eqnarray}\label{eq:chi1}
\chi_{i}=\arccos\left(\frac{(r_{i}-r_{2})R-r_{i}\bar{R}}{(r_{i}-r_{2})R+r_{i}\bar{R}}\right).
\end{eqnarray}
Note that here for $E_{\text{C}}=0$ (\ref{eq:solr1}) reduces to the solution for light rays in vacuum.

In the second case we have $K=K_{\text{ph}}$ and the right-hand side of (\ref{eq:EoMr}) has four real roots. We sort and label the roots such that we have $r_{4}<r_{3}=0<r_{\text{H}}=2m<r_{2}=r_{\text{ph}}=r_{1}$. These are light rays which asymptotically come from the unstable photon orbit. Again we first rewrite (\ref{eq:EoMr}) in terms of the roots. Then we separate variables and integrate from $r(\lambda_{i})=r_{i}$ to $r(\lambda)=r$. We get
\begin{eqnarray}
&\lambda-\lambda_{i}=i_{r_{i}}\int_{r_{i}}^{r}\frac{\text{d}r'}{\sqrt{\left(E^2-E_{\text{C}}^2\right)(r'-r_{\text{ph}})^2 r'(r'-r_{4})}}.
\end{eqnarray}
In the next step we substitute 
\begin{eqnarray}\label{eq:suby}
r=\frac{6mK}{12y+K}
\end{eqnarray}
and get
\begin{eqnarray}
\lambda-\lambda_{i}=-\frac{i_{r_{i}}}{2}\int_{y_{i}}^{y}\frac{\text{d}y'}{\sqrt{\left(y'-y_{\text{ph}}\right)^2\left(y'-y_{1}\right)}},
\end{eqnarray}
where $y_{i}$, $y$, $y_{\text{ph}}$, and $y_{1}$ are related to $r_{i}$, $r$, $r_{\text{ph}}$, and $r_{4}$ by (\ref{eq:suby}), respectively. In this case we can easily check that we have $y'<y_{\text{ph}}$. Now we pull the term $y_{\text{ph}}-y'$ out of the root. Now it is easy to see that we can rewrite the right-hand side in the form of the elementary integral $I_{1}$ given by (\ref{eq:EI1}) in Appendix~\ref{Sec:EI}. We integrate and obtain $I_{1_{3}}$ given by (\ref{eq:EII13}). We insert the result and solve for $r$ to obtain the solution to the equation of motion. It reads
\begin{widetext}
\begin{eqnarray}\label{eq:solr2}
r(\lambda)=\frac{r_{\text{ph}}r_{4}}{r_{\text{ph}}-(r_{\text{ph}}-r_{4})\tanh^2\left(\text{artanh}\left(\sqrt{\frac{r_{\text{ph}}(r_{i}-r_{4})}{(r_{\text{ph}}-r_{4})r_{i}}}\right)-i_{r_{i}}\sqrt{-\frac{mK\left(r_{\text{ph}}-r_{4}\right)}{2r_{\text{ph}}r_{4}}}\right)\left(\lambda-\lambda_{i}\right)}.
\end{eqnarray}
\end{widetext}
Note that here the solution $r(\lambda)$ is structurally the same for light rays travelling in vacuum, through a homogeneous plasma, and through the inhomogeneous plasma distribution described by (\ref{eq:PlasmaEn}), however, when we set $E_{\text{C}}=0$ the roots of the right-hand side of the equation of motion reduce to the roots for light rays travelling in vacuum.

In the third and last case we have $K_{\text{ph}}<K$ and the right-hand side of (\ref{eq:EoMr}) has four distinct real roots. We sort and label them such that we have $r_{4}<r_{3}=0<r_{\text{H}}=2m<r_{2}<r_{\text{ph}}<r_{1}$. The light rays travelling along these trajectories can pass through a turning point in the domain of outer communication. For light rays travelling outside the photon sphere this turning point will always be a minimum and it is given by $r_{\text{min}}=r_{1}$. Now we follow the same procedure as before. We first separate variables and integrate from $r(\lambda_{i})=r_{i}$ to $r(\lambda)=r$ and get
\begin{eqnarray}\label{eq:SolrInt3}
&\lambda-\lambda_{i}=i_{r_{i}}\int_{r_{i}}^{r}\frac{\text{d}r'}{\sqrt{\left(E^2-E_{\text{C}}^2\right)(r'-r_{1})(r'-r_{2})r'(r'-r_{4})}}.
\end{eqnarray}
Note that here this integral is formally only valid up to the first turning point. Now we substitute \cite{Hancock1917,Gralla2020}
\begin{eqnarray}\label{eq:sub2}
r=r_{2}+\frac{(r_{1}-r_{2})(r_{2}-r_{4})}{r_{2}-r_{4}-(r_{1}-r_{4})\sin^2\chi},
\end{eqnarray}
to put the integral into the Legendre form given by (\ref{eq:LFE}). Then we follow the steps outlined in Appendix~\ref{eq:JEF} to derive the solution to the equation of motion. In this case it is given in terms of Jacobi's elliptic sn function and reads
\begin{widetext}
\begin{eqnarray}\label{eq:solr3}
r(\lambda)=r_{2}+\frac{(r_{1}-r_{2})(r_{2}-r_{4})}{r_{2}-r_{4}-(r_{1}-r_{4})\text{sn}^2\left(\frac{i_{r_{i}}}{2}\sqrt{\left(E^2-E_{\text{C}}^2\right)r_{1}(r_{2}-r_{4})}\left(\lambda-\lambda_{i}\right)+\lambda_{r_{i},k_{2}},k_{2}\right)},
\end{eqnarray}
\end{widetext}
where the square of the elliptic modulus $k_{2}$, $\lambda_{r_{i},k_{2}}$, and the initial condition $\chi_{i}$ are given by 
\begin{eqnarray}\label{eq:k2}
k_{2}=\frac{r_{2}(r_{1}-r_{4})}{r_{1}(r_{2}-r_{4})},
\end{eqnarray}
\begin{eqnarray}
\lambda_{r_{i},k_{2}}=F_{L}\left(\chi_{i},k_{2}\right),
\end{eqnarray}
and 
\begin{eqnarray}\label{eq:chi2}
\chi_{i}=\arcsin\left(\sqrt{\frac{(r_{i}-r_{1})(r_{2}-r_{4})}{(r_{i}-r_{2})(r_{1}-r_{4})}}\right).
\end{eqnarray}
Note that as stated above formally the integral on the right-hand side of (\ref{eq:SolrInt3}) is only valid up to the turning point. However, due to the periodicity of Jacobi's elliptic sn function the derived solution for $r$ is also valid beyond the turning point and thus it can be used to describe the $r$ motion along the whole trajectory. Note that also here when we set $E_{\text{C}}=0$ (\ref{eq:solr3}) reduces to the solution for light rays in vacuum.

\subsubsection{The $\vartheta$ Motion}\label{Sec:EoMtheta}
We continue with deriving the solutions to the equation of motion for $\vartheta$. Note that although in this paper we only need the integral form for the calculations in Sec.~\ref{Sec:PR}, for light rays travelling in vacuum and through the homogeneous plasma we also need the solution for the $\vartheta$ motion to derive the solution to the equation of motion for $\varphi$ and thus we will also derive it here. Note that for completeness we will also derive the corresponding solution for light rays propagating through the inhomogeneous plasma described by the distribution given by Eq.~(\ref{eq:PlasmaEn}). 

While in general we can have three different types of motion for light rays travelling in vacuum or through a homogeneous plasma, and four different types of motion for light rays travelling through the inhomogeneous plasma described by (\ref{eq:PlasmaEn}), in our case the only types of motion we need to consider are those for light rays which can leave the equatorial plane. Please note that while an observer in the equatorial plane can also observe light rays travelling in the equatorial plane as already mentioned above for these light rays the photon rings are not well-defined and thus we exclude them from our discussion. 

We start with light rays travelling in vacuum or through a homogeneous plasma. We can obtain the associated equation of motion from (\ref{eq:EoMtheta}) by setting $\omega_{\text{p}}=0$. It reads
\begin{eqnarray}\label{eq:EoMthetaVHP}
\left(\frac{\text{d}\vartheta}{\text{d}\lambda}\right)^2=K-\frac{L_{z}^2}{\sin^2\vartheta}.
\end{eqnarray}
Now when we exclude light rays with $L_{z}=0$ and $K=0$ travelling along radial trajectories, and light rays with $\vartheta_{i}=\pi/2$ and $K=L_{z}^2$ travelling in the equatorial plane, we are only left with one type of motion. This type of motion describes the trajectories of light rays which generally oscillate between the two turning points 
\begin{eqnarray}
\vartheta_{\text{min}}=\arccos\left(\sqrt{1-\frac{L_{z}^2}{K}}\right)
\end{eqnarray}
and 
\begin{eqnarray}
\vartheta_{\text{max}}=\arccos\left(-\sqrt{1-\frac{L_{z}^2}{K}}\right).
\end{eqnarray}
For solving the equation of motion we now first multiply (\ref{eq:EoMthetaVHP}) with $\sin^2\vartheta$. Then we use that $-\sin\vartheta\text{d}\vartheta=\text{d}\cos\vartheta$ and rewrite the right-hand side in terms of $x=\cos\vartheta$. The result reads
\begin{eqnarray}
\left(\frac{\text{d}x}{\text{d}\lambda}\right)^2=-Kx^2+K-L_{z}^2.
\end{eqnarray}
As we can see the right-hand side is a polynomial of second order in $x$ and thus we can solve this differential equation using elementary functions. For this purpose we now first separate variables and integrate from $x\left(\lambda_{i}\right)=\cos\vartheta\left(\lambda_{i}\right)=\cos\vartheta_{i}$ to $x\left(\lambda\right)=\cos\vartheta$. Now we integrate and solve for $\vartheta$. The result reads \cite{Frost2023}
\begin{widetext}
\begin{eqnarray}\label{eq:solSVHP}
\vartheta(\lambda)=\arccos\left(\sqrt{1-\frac{L_{z}^2}{K}}\sin\left(\arcsin\left(\frac{\cos\vartheta_{i}}{\sqrt{1-\frac{L_{z}^2}{K}}}\right)-i_{\vartheta_{i}}\sqrt{K}\left(\lambda-\lambda_{i}\right)\right)\right).
\end{eqnarray}
\end{widetext}
Please note that also here the integral we obtained after separating variables  is only valid up to the first turning point. The same is also the case for the evaluated integral. However, due to the periodicity of the sine the solution given by (\ref{eq:solSVHP}) is also valid beyond the first turning point and can be used to describe the motion along the whole trajectory.

For light rays travelling in the inhomogeneous plasma we proceed analogously. Please note that in this case we can have two different types of motion for which light rays can come from the equatorial plane. The first type of motion is a special type of vortical motion. However, for this type of motion light rays can only come from the equatorial plane for $\lambda \rightarrow \pm\infty$ and thus we exclude this type of motion from our discussion. This has as consequence that also here we are only left with one type of motion. Namely, light rays generally oscillating between the two turning points $\vartheta_{\text{min}}$ and $\vartheta_ {\text{max}}$ above and below the equatorial plane. For calculating the roots and solving the equations of motion, we now first multiply (\ref{eq:EoMtheta}) by $\sin^2\vartheta$. Then we again use that $-\sin\vartheta\text{d}\vartheta=\text{d}\cos\vartheta$ and rewrite the right-hand side in terms of $x=\cos\vartheta$. The result reads
\begin{widetext}
\begin{eqnarray}\label{eq:EoMx}
\left(\frac{\text{d}x}{\text{d}\lambda}\right)^2=-\omega_{\text{p}}^2 E_{\text{C}}^2 x^4+\left(2\omega_{\text{p}}^2 E_{\text{C}}^2-K\right)x^2+K-\omega_{\text{p}}^2 E_{\text{C}}^2-L_{z}^2.
\end{eqnarray}
\end{widetext}
As we can see in this form the polynomial on the right-hand side is biquadratic. Now motion oscillating between two turning points on opposite sides of the equatorial plane can only exist when we have $0<K-\omega_{\text{p}}^2 E_{\text{C}}^2-L_{z}^2$. In this case the right-hand side of (\ref{eq:EoMx}) has two distinct real roots and a pair of complex conjugate roots. The analytic expressions for these roots can be found in Appendix~\ref{Sec:thetaRoots}. We now sort and label the real roots such that we have $x_{2}=\cos\vartheta_{\text{max}}<0<x_{1}=\cos\vartheta_{\text{min}}$. On the other hand we sort and label the complex conjugate roots such that we have $x_{3}=\bar{x}_{4}=X_{3}+iX_{4}$, where in our case we have $X_{3}=0$ and we choose $0<X_{4}$.

Now we separate variables and integrate from $x\left(\lambda_{i}\right)=\cos\vartheta\left(\lambda_{i}\right)=\cos\vartheta_{i}$ to $x\left(\lambda\right)=\cos\vartheta$. We get
\begin{widetext}
\begin{eqnarray}\label{eq:Intx}
\lambda-\lambda_{i}=-i_{\vartheta_{i}}\int_{\cos\vartheta_{i}}^{\cos\vartheta}\frac{\text{d}x'}{\sqrt{-\omega_{\text{p}}^2 E_{\text{C}}^2 x'^4+\left(2\omega_{\text{p}}^2 E_{\text{C}}^2-K\right)x'^2+K-\omega_{\text{p}}^2 E_{\text{C}}^2-L_{z}^2}},
\end{eqnarray}
\end{widetext}
where $i_{\vartheta_{i}}=\text{sgn}\left(\left.\text{d}\vartheta/\text{d}\lambda\right|_{\vartheta=\vartheta_{i}}\right)$. Note that again formally this integral is only valid up to the first turning point. Now we substitute 
\begin{eqnarray}\label{eq:subtheta}
x=\frac{x_{1}X_{4}\sin\chi}{\sqrt{x_{1}^2\cos^2\chi+X_{4}^2}}
\end{eqnarray}
to put (\ref{eq:Intx}) into the Legendre form given by (\ref{eq:LFE}). Then we follow the steps outlined in Appendix~\ref{eq:JEF} to derive the solution to the equation of motion. This time it is given by Jacobi's elliptic sd function. In the next step we rewrite the equation in terms of $\cos\vartheta$ and solve for $\vartheta$. The solution to (\ref{eq:EoMtheta}) then reads
\begin{widetext}
\begin{eqnarray}\label{eq:solthetaIP}
\vartheta(\lambda)=\arccos\left(\frac{x_{1}X_{4}}{\sqrt{x_{1}^2+X_{4}^2}}\text{sd}\left(i_{\vartheta_{i}}\omega_{\text{p}}E_{\text{C}}\sqrt{x_{1}^2+X_{4}^2}\left(\lambda_{i}-\lambda\right)+\lambda_{\vartheta_{i},k_{3}},k_{3}\right)\right),
\end{eqnarray}
\end{widetext}
where the square of the elliptic modulus $k_{3}$, $\lambda_{\vartheta_{i},k_{3}}$, and the initial condition $\chi_{i}$ are given by 
\begin{eqnarray}\label{eq:k3}
k_{3}=\frac{x_{1}^2}{x_{1}^2+X_{4}^2},
\end{eqnarray}
\begin{eqnarray}
\lambda_{\vartheta_{i},k_{3}}=F_{L}\left(\chi_{i},k_{3}\right),
\end{eqnarray}
and 
\begin{eqnarray}\label{eq:chi3}
\chi_{i}=\arcsin\left(\frac{\sqrt{x_{1}^2+X_{4}^2}\cos\vartheta_{i}}{x_{1}\sqrt{X_{4}^2+\cos^2\vartheta_{i}}}\right).
\end{eqnarray}
Note that while the original integral given by (\ref{eq:Intx}) was only valid up to the first turning point because of the periodicity of Jacobi's elliptic sd function the solution given by (\ref{eq:solthetaIP}) is also valid beyond the first turning point.

\subsubsection{The $\varphi$ Motion}
Now we turn to the $\varphi$ motion. As we can see in our case (\ref{eq:EoMphi}) is the same for light rays travelling in vacuum, through a homogeneous plasma, and through the inhomogeneous plasma density described by Eq.~(\ref{eq:PlasmaEn}). However, since the equation of motion for $\vartheta$ for light rays travelling in vacuum or through a homogeneous plasma, and the equation of motion for light rays travelling through the inhomogeneous plasma are different, we have to use different approaches for solving them.

We again start with deriving the solution for light rays travelling in vacuum or through a homogeneous plasma. Since radial light rays and light rays travelling in the equatorial plane are not relevant for calculating the photon rings, we again exclude them from the discussion. This leaves us with two different types of motion. The first type of motion corresponds to light rays with $L_{z}=0$ and $K\neq 0$. These light rays can pass through $\vartheta=0$ and $\vartheta=\pi$ and the right-hand side of (\ref{eq:EoMphi}) vanishes. Thus in this case the $\varphi$ coordinate has a discontinuity whenever the light rays passes through $\vartheta=0$ or $\vartheta=\pi$. The equation of motion is formally not solvable but we can still define a solution. It reads

\begin{eqnarray}\label{eq:phi1}
\varphi\left(\lambda\right)=\varphi_{i}+n(\lambda)\pi,
\end{eqnarray}

where $n(\lambda)$ counts how often the light ray passes through $\vartheta=0$ and $\vartheta=\pi$.

The second type of motion describes light rays generally oscillating between the turning points at $\vartheta_{\text{min}}$ and $\vartheta_{\text{max}}$. In this case we first separate variables. Then we integrate from $\varphi\left(\lambda_{i}\right)=\varphi_{i}$ to $\varphi(\lambda)=\varphi$ and get
\begin{eqnarray}
\varphi(\lambda)=\varphi_{i}+\int_{\lambda_{i}}^{\lambda}\frac{\text{d}\lambda'}{1-\cos^2\vartheta(\lambda')}.
\end{eqnarray}
In the next step we insert the solution for $\vartheta(\lambda)$ given by (\ref{eq:solSVHP}) and evaluate the integral. The result reads \cite{Frost2023}
\begin{widetext}
\begin{eqnarray}
&\varphi\left(\lambda\right)=\varphi_{i}+i_{\vartheta_{i}}\left(\arctan\left(\frac{L_{z}\cos\vartheta_{i}}{\sqrt{K-L_{z}^2-K\cos^2\vartheta_{i}}}\right)-\arctan\left(\frac{L_{z}}{\sqrt{K}}\tan\left(\arcsin\left(\frac{\cos\vartheta_{i}}{\sqrt{1-\frac{L_{z}^2}{K}}}\right)-i_{\vartheta_{i}}\sqrt{K}\left(\lambda-\lambda_{i}\right)\right)\right)\right).
\end{eqnarray}
\end{widetext}
Note that here for the explicit evaluation of $\varphi\left(\lambda\right)$ the multivaluedness of the arctan has to be taken into account.

For light rays travelling through the inhomogeneous plasma described by the distribution given by Eq.~(\ref{eq:PlasmaEn}) we have to consider the same two cases as for light rays travelling in vacuum or through a homogeneous plasma. Again the other three types of motion, namely light rays travelling along radial trajectories, light rays travelling in the equatorial plane, and light rays travelling along vortical trajectories are not relevant for the calculation of the photon rings and thus they are not included in our discussion. 
Now for light rays with  $L_{z}=0$ and $K\neq 0$ the right-hand side of (\ref{eq:EoMphi}) is again zero and the $\varphi$ coordinate has a discontinuity when the light rays pass through $\vartheta=0$ or $\vartheta=\pi$. Thus we can now again define a solution describing this type of motion and this solution is again given by (\ref{eq:phi1}). 

Again the second type of motion describes light rays generally oscillating between the two turning points $\vartheta_{\text{min}}$ and $\vartheta_{\text{max}}$. In this case we first rewrite the right-hand side of (\ref{eq:EoMphi}) in terms of $x=\cos\vartheta$. Then we substitute using (\ref{eq:subtheta}) and rewrite the right-hand side as a constant term and a term containing $\sin^2\chi$ in the denominator. Then we integrate from $\varphi\left(\lambda_{i}\right)=\varphi_{i}$ to $\varphi\left(\lambda\right)=\varphi$. In the next step we split the integral into two separate integrals and evaluate the integral over the constant term. Then we use the root of (\ref{eq:LegDiffTrans}), adapted to the type of motion at hand, to rewrite the second integral, which again has to be split into two or more integrals when the light ray passes through one or more turning points, in terms of Legendre's elliptic integral of the third kind. We get
\begin{widetext}
\begin{eqnarray}
\varphi\left(\lambda\right)=\varphi_{i}+\frac{L_{z}}{1+X_{4}^2}\left(\lambda-\lambda_{i}+\sum_{n=1}^{N}\frac{i_{\vartheta_{i}n}X_{4}^2}{\omega_{\text{p}}E_{\text{C}}\sqrt{x_{1}^2+X_{4}^2}}\left(\Pi_{L}\left(\chi_{n-1},k_{3},n_{3}\right)-\Pi_{L}\left(\chi_{n},k_{3},n_{3}\right)\right)\right),
\end{eqnarray}
\end{widetext}
where $N$ is the number of integrals into which we had to split the second integral. In addition, we have $\chi_{0}=\chi_{i}$, $\chi_{N}=\chi\left(\lambda\right)$, and for $1<n<N$ $\chi_{n}$ and $\chi_{n-1}$ are given by $\chi_{\text{min}}=\pi/2$ or $\chi_{\text{max}}=-\pi/2$. For $i_{\vartheta_{i}n}$ on the other hand we have $i_{\vartheta_{i}n}=-i_{\vartheta_{i}n-1}$ and $i_{\vartheta_{i}1}=i_{\vartheta_{i}}$. $\chi_{i}$, $\chi_{\text{min}}$, $\chi_{\text{max}}$, and $\chi\left(\lambda\right)$ are related to $\cos\vartheta_{i}$, $\cos\vartheta_{\text{min}}$, $\cos\vartheta_{\text{max}}$, and $\cos\vartheta\left(\lambda\right)$ by (\ref{eq:chi3}), the square of the elliptic modulus $k_{3}$ is given by (\ref{eq:k3}), and the parameter $n_{3}$ is given by
\begin{eqnarray}
n_{3}=x_{1}^2\frac{1+X_{4}^2}{x_{1}^2+X_{4}^2}.
\end{eqnarray}

\subsubsection{The Time Coordinate $t$}\label{Sec:EoMtint}
The last equation of motion we have to solve is the equation describing the evolution of the time coordinate $t$. As first step we separate variables in (\ref{eq:EoMt}) and integrate from $t\left(\lambda_{i}\right)=t_{i}$ to $t\left(\lambda\right)=t$. Then we use the root of (\ref{eq:EoMr}) to rewrite the integral over the Mino parameter as an integral over $r$. We get
\begin{widetext}
\begin{eqnarray}\label{eq:coordt}
t(\lambda)=t_{i}+\int_{r_{i}...}^{...r(\lambda)}\frac{Er'^3\text{d}r'}{\left(r'-2m\right)\sqrt{\left(E^2-E_{\text{C}}^2\right)r'^4+2mE_{\text{C}}^2 r'^3-Kr'^2+2mKr'}},
\end{eqnarray}
\end{widetext}
where the dots in the limits of the integral shall indicate that we have to split it at turning points into two integrals, which then have to be evaluated separately. The sign of the root in the denominator has to be chosen such that it agrees with the direction of the $r$ motion.

For the evaluation of the integral we now have to distinguish the same types of motion as for $r$. 

We again start with light rays for which the $r$ motion is characterised by $0<K<K_{\text{ph}}$. We recall that in this case the right-hand side of (\ref{eq:EoMr}) has two distinct real roots $r_{2}<r_{1}=0$ and a pair of complex conjugate roots $r_{3}=\bar{r}_{4}=R_{3}+iR_{4}$, where we chose $0<R_{4}$. As first step we rewrite 
\begin{eqnarray}\label{eq:TCoeff}
\frac{r'^3}{r'-2m}=r'^2+2mr'+4m^2+\frac{8m^3}{r'-2m}.
\end{eqnarray}
In the next step we substitute using (\ref{eq:sub1}). We rewrite the resulting terms in terms of Legendre's elliptic integral of the first kind, and the two nonstandard elliptic integrals $I_{L_{1}}(\chi_{i},\chi,k,n)$ and $I_{L_{2}}(\chi_{i},\chi,k,n)$ given by (\ref{eq:IL1}) and (\ref{eq:IL2}) in Appendix~\ref{Sec:ELF}. We evaluate the integrals as described in Appendix~\ref{Sec:ELF} and obtain
\begin{widetext}
\begin{eqnarray}\label{eq:tnoturn}
&t\left(\lambda\right)=t_{i}+\frac{i_{r_{i}}E}{\sqrt{\left(E^2-E_{\text{C}}^2\right)R\bar{R}}}\left(\frac{r_{t}^2\left(F_{L}\left(\chi\left(\lambda\right),k_{1}\right)-F_{L}\left(\chi_{i},k_{1}\right)\right)}{P\left(r_{t}\right)}-\frac{4R\bar{R}\left(r_{t}+m\right)r_{2}I_{L_{1}}\left(\chi_{i},\chi\left(\lambda\right),k_{1},n_{1}\right)}{\bar{R}^2-R^2}\right.\\
&\left.+\frac{16m^3 R\bar{R} r_{2}I_{L_{1}}\left(\chi_{i},\chi\left(\lambda\right),k_{1},n_{2}\right)}{4m^2\bar{R}^2-\left(r_{2}-2m\right)^2 R^2}+\frac{4R^2\bar{R}^2 r_{2}^2 I_{L_{2}}\left(\chi_{i},\chi\left(\lambda\right),k_{1},n_{1}\right)}{\left(\bar{R}^2-R^2\right)^2}\right),\nonumber
\end{eqnarray}
\end{widetext}
where the integrals $I_{L_{1}}\left(\chi_{i},\chi\left(\lambda\right),k_{1},n_{1}\right)$ and $I_{L_{1}}\left(\chi_{i},\chi\left(\lambda\right),k_{1},n_{2}\right)$, and $I_{L_{2}}\left(\chi_{i},\chi\left(\lambda\right),k_{1},n_{1}\right)$ are given by (\ref{eq:EI1EV}) and (\ref{eq:EI2EV}), respectively. $R$ and $\bar{R}$ are given by (\ref{eq:R}) and (\ref{eq:Rbar}), and $\chi_{i}$ and $\chi\left(\lambda\right)$ are related to $r_{i}$ and $r\left(\lambda\right)$ by (\ref{eq:chi1}), respectively. The square of the elliptic modulus $k_{1}$ is given by (\ref{eq:k1}). The parameters $n_{1}$ and $n_{2}$ are given by (\ref{eq:n1}) and (\ref{eq:n2}), respectively, and $r_{t}$ is given by 
\begin{eqnarray}
r_{t}=\frac{r_{2}R}{\bar{R}+R}.
\end{eqnarray}
Note that for $E_{\text{C}}=0$ (\ref{eq:tnoturn}) reduces to the solution for light rays in vacuum.

In the second case we have light rays travelling along trajectories characterised by $K=K_{\text{ph}}$. We recall that in this case the right-hand side of (\ref{eq:EoMr}) has four real roots and that we sorted  and labelled them such that we have $r_{4}<r_{3}=0<r_{\text{H}}<r_{2}=r_{\text{ph}}=r_{1}$. Now we again first use (\ref{eq:TCoeff}) to rewrite the term outside the root. Then we substitute using (\ref{eq:suby}). We pull the factor $y_{\text{ph}}-y'$ out of the root and perform a partial fraction decomposition with respect to $y'$. We rewrite the resulting integrals in terms of the integrals $I_{1}$ and $I_{2}$ given by (\ref{eq:EI1}) and (\ref{eq:EI2}) in Appendix~\ref{Sec:EI}. In the next step we use $I_{1_{1}}$, $I_{1_{2}}$, $I_{1_{3}}$, and $I_{2}$ given by (\ref{eq:EII11}), (\ref{eq:EII12}), (\ref{eq:EII13}), and (\ref{eq:EII2}) to rewrite (\ref{eq:coordt}) as
\begin{widetext}
\begin{eqnarray}
&t\left(\lambda\right)=t_{i}+i_{r_{i}}r_{\text{ph}}E\left(\left(2m+r_{\text{ph}}+\frac{r_{4}}{2}\right)\sqrt{-\frac{2r_{4}}{mK}}\left(\text{arcoth}\left(\sqrt{\frac{r\left(\lambda\right)-r_{4}}{r\left(\lambda\right)}}\right)-\text{arcoth}\left(\sqrt{\frac{r_{i}-r_{4}}{r_{i}}}\right)\right)\right.\\
&\left.+\frac{r_{4}}{\sqrt{2mK}}\left(\sqrt{\frac{r_{i}\left(r_{4}-r_{i}\right)}{r_{4}}}-\sqrt{\frac{r\left(\lambda\right)\left(r_{4}-r\left(\lambda\right)\right)}{r_{4}}}\right)+\frac{8m^2}{r_{\text{ph}}-2m}\sqrt{\frac{r_{4}}{K\left(r_{4}-2m\right)}}\left(\text{artanh}\left(\sqrt{\frac{2m\left(r\left(\lambda\right)-r_{4}\right)}{\left(2m-r_{4}\right)r\left(\lambda\right)}}\right)-\text{artanh}\left(\sqrt{\frac{2m\left(r_{i}-r_{4}\right)}{\left(2m-r_{4}\right)r_{i}}}\right)\right)\right.\nonumber\\
&\left.+\frac{r_{\text{ph}}^2}{r_{\text{ph}}-2m}\sqrt{\frac{2r_{\text{ph}}r_{4}}{mK\left(r_{4}-r_{\text{ph}}\right)}}\left(\text{artanh}\left(\sqrt{\frac{r_{\text{ph}}\left(r_{i}-r_{4}\right)}{\left(r_{\text{ph}}-r_{4}\right)r_{i}}}\right)-\text{artanh}\left(\sqrt{\frac{r_{\text{ph}}\left(r\left(\lambda\right)-r_{4}\right)}{\left(r_{\text{ph}}-r_{4}\right)r\left(\lambda\right)}}\right)\right)\right).\nonumber
\end{eqnarray}
\end{widetext} 

In the last case we have light rays travelling along trajectories characterised by $K_{\text{ph}}<K$. We recall that in this case the right-hand side of (\ref{eq:EoMr}) has four distinct real roots and that we sorted and labelled the roots such that we have $r_{4}<r_{3}=0<r_{\text{H}}<r_{2}<r_{\text{ph}}<r_{1}$. Light rays travelling along these trajectories can pass through the turning point at $r_{1}=r_{\text{min}}$. We again first use (\ref{eq:TCoeff}) to rewrite the integrand. Then we substitute using (\ref{eq:sub2}) and rewrite the result in terms of Legendre's elliptic integrals of the first and third kind, and the nonstandard elliptic integral $I_{L_{3}}\left(\chi_{i},\chi,\tilde{k},n\right)$ given by (\ref{eq:IL3}) in Appendix~\ref{Sec:ELF}. Now we use (\ref{eq:EI3EV}) to rewrite it in terms of elementary functions and Legendre's elliptic integrals of the first, second, and third kind. In the case that we have light rays not passing through the turning point we get as solution for $t(\lambda)$
\begin{widetext}
\begin{eqnarray}\label{eq:tmin1}
&t\left(\lambda\right)=t_{i}+\frac{i_{r_{i}}E}{\sqrt{\left(E^2-E_{\text{C}}^2\right)r_{1}\left(r_{2}-r_{4}\right)}}\bigg(\frac{2r_{2}^2\left(F_{L}\left(\chi\left(\lambda\right),k_{2}\right)-F_{L}\left(\chi_{i},k_{2}\right)\right)}{P\left(r_{2}\right)}+4\left(r_{1}-r_{2}\right)\bigg(\left(r_{2}+m\right)\left(\Pi_{L}\left(\chi\left(\lambda\right),k_{2},n_{4}\right)\right.\\
&\left.-\Pi_{L}\left(\chi_{i},k_{2},n_{4}\right)\right)+\frac{4m^3\left(\Pi_{L}\left(\chi_{i},k_{2},n_{5}\right)-\Pi_{L}\left(\chi\left(\lambda\right),k_{2},n_{5}\right)\right)}{\left(r_{1}-2m\right)\left(r_{2}-2m\right)}\bigg)+\left(r_{1}-r_{2}\right)^2\bigg(\frac{n_{4}^2}{2\left(n_{4}-k_{2}\right)\left(n_{4}-1\right)}\left(\frac{\sin\left(2\chi\left(\lambda\right)\right)\sqrt{1-k_{2}\sin^2\chi\left(\lambda\right)}}{1-n_{4}\sin^2\chi\left(\lambda\right)}\right.\nonumber\\
&\left.-\frac{\sin\left(2\chi_{i}\right)\sqrt{1-k_{2}\sin^2\chi_{i}}}{1-n_{4}\sin^2\chi_{i}}\right)+\frac{F_{L}\left(\chi\left(\lambda\right),k_{2}\right)-F_{L}\left(\chi_{i},k_{2}\right)}{n_{4}-1}+\frac{n_{4}\left(E_{L}\left(\chi_{i},k_{2}\right)-E_{L}\left(\chi\left(\lambda\right),k_{2}\right)\right)}{\left(n_{4}-k_{2}\right)\left(n_{4}-1\right)}\nonumber\\
&+\frac{\left(n_{4}\left(n_{4}-2\right)-\left(2n_{4}-3\right)k_{2}\right)\left(\Pi_{L}\left(\chi\left(\lambda\right),k_{2},n_{4}\right)-\Pi_{L}\left(\chi_{i},k_{2},n_{4}\right)\right)}{\left(n_{4}-k_{2}\right)\left(n_{4}-1\right)}\bigg)\bigg).\nonumber
\end{eqnarray}
\end{widetext}
For light rays passing through the turning point at $r_{1}=r_{\text{min}}$ on the other hand we have to split (\ref{eq:coordt}) at the turning point into two integrals and then we have to evaluate each integral separately. This time we get as solution for $t(\lambda)$
\begin{widetext}
\begin{eqnarray}\label{eq:tmin2}
&t\left(\lambda\right)=t_{i}-\frac{i_{r_{i}}E}{\sqrt{\left(E^2-E_{\text{C}}^2\right)r_{1}\left(r_{2}-r_{4}\right)}}\bigg(\frac{2r_{2}^2\left(F_{L}\left(\chi_{i},k_{2}\right)+F_{L}\left(\chi\left(\lambda\right),k_{2}\right)\right)}{P\left(r_{2}\right)}+4\left(r_{1}-r_{2}\right)\bigg(\left(r_{2}+m\right)\left(\Pi_{L}\left(\chi_{i},k_{2},n_{4}\right)\right.\\
&\left.+\Pi_{L}\left(\chi\left(\lambda\right),k_{2},n_{4}\right)\right)-\frac{4m^3\left(\Pi_{L}\left(\chi_{i},k_{2},n_{5}\right)+\Pi_{L}\left(\chi\left(\lambda\right),k_{2},n_{5}\right)\right)}{\left(r_{1}-2m\right)\left(r_{2}-2m\right)}\bigg)+\left(r_{1}-r_{2}\right)^2\bigg(\frac{n_{4}^2}{2\left(n_{4}-k_{2}\right)\left(n_{4}-1\right)}\left(\frac{\sin\left(2\chi_{i}\right)\sqrt{1-k_{2}\sin^2\chi_{i}}}{1-n_{4}\sin^2\chi_{i}}\right.\nonumber\\
&\left.+\frac{\sin\left(2\chi\left(\lambda\right)\right)\sqrt{1-k_{2}\sin^2\chi\left(\lambda\right)}}{1-n_{4}\sin^2\chi\left(\lambda\right)}\right)+\frac{F_{L}\left(\chi_{i},k_{2}\right)+F_{L}\left(\chi\left(\lambda\right),k_{2}\right)}{n_{4}-1}-\frac{n_{4}\left(E_{L}\left(\chi_{i},k_{2}\right)+E_{L}\left(\chi\left(\lambda\right),k_{2}\right)\right)}{\left(n_{4}-k_{2}\right)\left(n_{4}-1\right)}\nonumber\\
&+\frac{\left(n_{4}\left(n_{4}-2\right)-\left(2n_{4}-3\right)k_{2}\right)\left(\Pi_{L}\left(\chi_{i},k_{2},n_{4}\right)+\Pi_{L}\left(\chi\left(\lambda\right),k_{2},n_{4}\right)\right)}{\left(n_{4}-k_{2}\right)\left(n_{4}-1\right)}\bigg)\bigg).\nonumber
\end{eqnarray}
\end{widetext} 
Here in both cases $\chi_{i}$ and $\chi\left(\lambda\right)$ are related to $r_{i}$ and $r\left(\lambda\right)$ by (\ref{eq:chi2}), respectively, the square of the elliptic modulus $k_{2}$ is given by (\ref{eq:k2}), and the two parameters $n_{4}$ and $n_{5}$ are given by
\begin{eqnarray}
n_{4}=\frac{r_{1}-r_{4}}{r_{2}-r_{4}}
\end{eqnarray}
and 
\begin{eqnarray}
n_{5}=\frac{\left(r_{2}-2m\right)\left(r_{1}-r_{4}\right)}{\left(r_{1}-2m\right)\left(r_{2}-r_{4}\right)}.
\end{eqnarray}
Note that again for $E_{\text{C}}=0$ (\ref{eq:tmin1}) and (\ref{eq:tmin2}) reduce to the solutions for light rays in vacuum.

\section{The Celestial Coordinates and The Shadow}\label{Sec:CCS}
In Section~\ref{Sec:PR} we want to discuss the photon ring structures on the celestial sphere of an observer in the domain of outer communication. However, before we can do this we need to fix the coordinates on the observer's celestial sphere. In addition, for the calculation of the photon rings it is important to know the angular radius of the shadow. Thus in the first part of this section we will fix the coordinate system on the celestial sphere of the observer and in the second part of this section we will derive the angular radius of the shadow.

\subsection{The Celestial Coordinates}
For investigating the photon rings we assume that we have a static observer in the domain of outer communication of the Schwarzschild spacetime. At the position of the observer we now introduce an orthonormal tetrad, see, e.g., Grenzebach, Perlick, and L\"{a}mmerzahl \cite{Grenzebach2015}. The four tetrad vectors read
\begin{eqnarray}
&e_{0}=\left.\frac{\partial_{t}}{\sqrt{P(r)}}\right|_{x_{O}},~~~e_{1}=\left.\frac{\partial_{\vartheta}}{r}\right|_{x_{O}},\\
&e_{2}=-\left.\frac{\partial_{\varphi}}{r\sin\vartheta}\right|_{x_{O}},~\text{and}~~~e_{3}=-\sqrt{P(r)}\partial_{r}\bigg|_{x_{O}},
\end{eqnarray}
where $e_{0}$ is also the four-velocity of the observer. Now we have to relate the constants of motion $E$, $L_{z}$, and $K$ to the energy of the light rays in the reference frame of the observer and the latitude-longitude coordinates on the observer's celestial sphere. Here, we introduce the celestial latitude $\Sigma$ such that it is measured from the tetrad vector $e_{3}$. The celestial longitude $\Psi$ on the other hand is measured from the tetrad vector $e_{1}$ in the direction of the tetrad vector $e_{2}$. As first step we derive the relation between the energy $E$ of a light ray travelling along a trajectory and its energy measured in the reference frame of the static observer $E_{O}$. Here we have $E_{O}=-p_{\mu}u^{\mu}_{O}$, see, e.g., p.~69 in the book of Wald \cite{Wald1984}, and when we insert the components of the four-velocity of the observer and $p_{t}=-E$ we get
\begin{eqnarray}
E=\sqrt{P(r_{O})}E_{O}.
\end{eqnarray}
In the second step we relate the angular momentum about the $z$-axis $L_{z}$ and the Carter constant $K$ to $E_{O}$ and the latitude-longitude coordinates $\Sigma$ and $\Psi$ on the observer's celestial sphere. For this purpose let us first write down the tangent vector of the light ray in Mino parameterisation. It reads
\begin{eqnarray}\label{eq:tanvec}
\frac{\text{d}\eta}{\text{d}\lambda}=\frac{\text{d}t}{\text{d}\lambda}\partial_{t}+\frac{\text{d}r}{\text{d}\lambda}\partial_{r}+\frac{\text{d}\vartheta}{\text{d}\lambda}\partial_{\vartheta}+\frac{\text{d}\varphi}{\text{d}\lambda}\partial_{\varphi}.
\end{eqnarray}
At the position of the observer we can now also write the tangent vector in terms of the four tetrad vectors $e_{0}$, $e_{1}$, $e_{2}$, and $e_{3}$, the latitude-longitude coordinates on the observer's celestial sphere $\Sigma$ and $\Psi$, and two normalisation constants $\alpha$ and $\beta$. It reads \cite{Perlick2017}
\begin{eqnarray}\label{eq:tanvect}
&\left.\frac{\text{d}\eta}{\text{d}\lambda}\right|_{x_{O}}=-\alpha e_{0}+\beta\left(\sin\Sigma\cos\Psi e_{1}+\sin\Sigma\sin\Psi e_{2}\right.\\
&\left.+\cos\Sigma e_{3}\right).\nonumber
\end{eqnarray}
In the next step we now relate the normalisation constant $\beta$ to the normalisation constant $\alpha$. For this purpose we first calculate 
\begin{eqnarray}
g\left(\left.\frac{\text{d}\eta}{\text{d}\lambda}\right|_{x_{O}},\left.\frac{\text{d}\eta}{\text{d}\lambda}\right|_{x_{O}}\right)=-r_{O}^4 E_{\text{pl}}\left(r_{O},\vartheta_{O}\right)^2,
\end{eqnarray}
and get 
\begin{eqnarray}
\beta=-\sqrt{\alpha^2-r_{O}^4 E_{\text{pl}}\left(r_{O},\vartheta_{O}\right)^2}.
\end{eqnarray}
Note that here we originally had a sign ambiguity, however, we already fixed it such that for $E_{\text{pl}}\left(r_{O},\vartheta_{O}\right)\rightarrow 0$ the sign of the angular momentum about the $z$-axis is consistent with that for light rays in vacuum. 
Now we insert the plasma energy $E_{\text{pl}}\left(r_{O},\vartheta_{O}\right)$ given by (\ref{eq:PlasmaEn}) and get
\begin{eqnarray}\label{eq:NCB}
&\beta=-\sqrt{\alpha^2-E_{\text{C}}^2 r_{O}^2 \left(r_{O}^2+\omega_{\text{p}}^2\sin^2\vartheta_{O}\right)}.
\end{eqnarray}
Now we have to determine the normalisation constant $\alpha$. It is given by
\begin{eqnarray}
\alpha=g\left(\left.\frac{\text{d}\eta}{\text{d}\lambda}\right|_{x_{O}},e_{0}\right).
\end{eqnarray}
We insert the tangent vector given by (\ref{eq:tanvec}) and $e_{0}$ and evaluate the right-hand side. We get
\begin{eqnarray}
\alpha=-\frac{r_{O}^2E}{\sqrt{P(r_{O})}}.
\end{eqnarray}
We insert the obtained relation in (\ref{eq:NCB}) and get as result
\begin{eqnarray}
&\beta=-\sqrt{\frac{r_{O}^2\left(r_{O}^2\left(E^2-E_{\text{C}}^2 P(r_{O})\right)-\omega_{\text{p}}^2 P(r_{O})E_{\text{C}}^2\sin^2\vartheta_{O}\right)}{P(r_{O})}}.
\end{eqnarray}
Now we rewrite $E$ in terms of $E_{O}$ and get
\begin{eqnarray}
\alpha=-r_{O}^2E_{O},
\end{eqnarray}
and
\begin{eqnarray}
&\beta=-\sqrt{r_{O}^2\left(r_{O}^2\left(E_{O}^2-E_{\text{C}}^2\right)-\omega_{\text{p}}^2 E_{\text{C}}^2\sin^2\vartheta_{O}\right)}.
\end{eqnarray}
Now we insert the two normalisation constants in (\ref{eq:tanvect}) and compare coefficients with (\ref{eq:tanvec}). We insert the obtained relations in the equations of motion and solve for the constants of motion. Finally, we obtain for the relations between the constants of motion $E$, $L_{z}$, and $K$ and the energy of the light ray measured by the observer $E_{O}$, the celestial latitude $\Sigma$, and the celestial longitude $\Psi$:
\begin{eqnarray}\label{eq:CoME}
E=\sqrt{P(r_{O})}E_{O},
\end{eqnarray}
\begin{eqnarray}\label{eq:CoMLz}
&L_{z}=\left(r_{O}^2\left(E_{O}^2-E_{\text{C}}^2\right)
\right.\\
&\left.-\omega_{\text{p}}^2E_{\text{C}}^2\sin^2\vartheta_{O}\right)^{\frac{1}{2}}\sin\vartheta_{O}\sin\Sigma\sin\Psi,\nonumber
\end{eqnarray}
and
\begin{eqnarray}\label{eq:CoMK}
K=r_{O}^2\left(E_{O}^2-E_{\text{C}}^2\right)\sin^2\Sigma+\omega_{\text{p}}^2E_{\text{C}}^2\sin^2\vartheta_{O}\cos^2\Sigma.
\end{eqnarray}
Note that for light rays in vacuum ($E_{\text{C}}=0$) we can get the relations between the constants of motion and the angles on the observer's celestial sphere completely analogously. In this case one can easily show that we have $\alpha=\beta$. In addition, without loss of generality we can choose $\alpha=-r_{O}^2$ so that $E_{O}=1$. 
\newline

\subsection{The Shadow}
For calculating the photon rings the boundary of what we conventionally refer to as the shadow plays an important role. While for black holes surrounded by an accretion disk, in this paper we will simplify it as a luminous disk in the equatorial plane, we are not necessarily able to observe this shadow because we may also have light emitting matter inside the unstable photon orbit, knowing the position of its boundary on the celestial sphere of the observer will significantly ease our calculations. Thus we will calculate it in the following.

Here, we use the fact that light rays on the unstable photon orbit have the same constants of motion as light rays asymptotically coming from the photon orbit. In addition, we use the fact that for light rays asymptotically coming from the photon orbit the equation of motion for $r$ has a double root at the radius coordinate of the photon orbit. As a consequence we have $\left.\text{d}r/\text{d}\lambda\right|_{r=r_{\text{ph}}}=0$. Now the only thing we have to do is to rewrite the right-hand side of (\ref{eq:EoMr}) in terms of $r_{\text{ph}}$ and insert the obtained relations for the constants of motion $E$ and $K$ given by (\ref{eq:CoME}) and (\ref{eq:CoMK}). In the next step we solve for the angular radius of the shadow and obtain
\begin{widetext}
\begin{eqnarray}
\Sigma_{\text{ph}}=\arcsin\left(\sqrt{\frac{r_{\text{ph}}^2\left(P(r_{O})E_{O}^2-P(r_{\text{ph}})E_{\text{C}}^2\right)-\omega_{\text{p}}^2E_{\text{C}}^2P(r_{\text{ph}})\sin^2\vartheta_{O}}{P(r_{\text{ph}})\left(r_{O}^2\left(E_{O}^2-E_{\text{C}}^2\right)-\omega_{\text{p}}^2E_{\text{C}}^2\sin^2\vartheta_{O}\right)}}\right).
\end{eqnarray}
\end{widetext}
For $\omega_{\text{p}}=0$ the angular radius of the shadow for the inhomogeneous plasma distribution reduces to the angular radius of the shadow for a homogeneous plasma and for $E_{O}\rightarrow\infty$ it reduces to the angular radius of the shadow for light rays travelling in vacuum
\begin{eqnarray}
\Sigma_{\text{ph}}=\arcsin\left(\frac{r_{\text{ph}}}{r_{O}}\sqrt{\frac{P(r_{O})}{P(r_{\text{ph}})}}\right).
\end{eqnarray}
When we compare the angular radii with each other it is easy to see that for light rays propagating through a plasma the boundary of the shadow generally dependends on the energy of the light ray measured by the observer. In addition, for the inhomogeneous plasma the angular radius of the shadow also depends on the plasma parameter $\omega_{\text{p}}$ and the spacetime latitude of the observer $\vartheta_{O}$. Although it is not surprising this implies that for our inhomogeneous plasma density the size of the shadow varies depending on the spacetime latitude of the observer. Here, when the observer is located at $\vartheta_{O}=0$ or $\vartheta_{O}=\pi$ the second terms in the nominator and the denominator vanish and the angular radius of the shadow is the same as for light rays propagating through a homogeneous plasma. 

\section{Photon Rings in a Plasma}\label{Sec:PR}
In this section we will discuss and compare the photon rings in the Schwarzschild spacetime in vacuum with the photon rings in the presence of a homogeneous plasma and the photon rings in the presence of the inhomogeneous plasma described by the distribution given by Eq.~(\ref{eq:PlasmaEn}). Since in this paper we are mainly interested in how the plasma distributions affect the paths of the light rays we will only consider a rather simple setup. We will assume that we have a luminous disk of static sources between the radius coordinates $r_{\text{in}}$ and $r_{\text{out}}$ in the equatorial plane. In this paper we will choose the inner boundary of the luminous disk $r_{\text{in}}$ such that it corresponds to the radius coordinate of the horizon. On the other hand we will choose the outer boundary of the luminous disk such that we have $r_{\text{ph}}<r_{\text{out}}$. Note that in this case the photon rings will always be superposed with the direct image. Then we will assume that the observer detects the light rays emitted by the disk at the coordinates $t_{O}$, $r_{O}$, $\vartheta_{O}$, and $\varphi_{O}$ and thus we have $x_{i}=x_{O}$. 

Here, we will calculate three different observables. The first is a lens equation. It connects the sources in the luminous disk to their images on the observer's celestial sphere. The second is the redshift. It describes the energy shift a light ray experiences between the reference frame of the source and the reference frame of the observer. The last is the travel time. It measures in terms of the time coordinate $t$ how long a light ray needs to travel from its source to the observer. We will then project these quantities onto the observer's celestial sphere to construct maps. Since the photon rings only appear in the direction of the black hole, we will limit the field of view of the observer to $0\leq\Sigma\leq \pi/3$. 

In addition, it will be rather tedious to always refer to points in the maps by their latitude-longitude coordinates and thus we will now agree on the following naming conventions. The line marked by the celestial longitudes $\Psi=\pi/2$ and $\Psi=3\pi/2$ splits the maps into two equal parts and thus we will refer to it as \emph{celestial equator}. The celestial equator splits the sky into two regions. Here, we will refer to the region $0\leq\Psi<\pi/2$ and $3\pi/2<\Psi$ as \emph{southern hemisphere}. On the other hand we will refer to the region $\pi/2<\Psi<3\pi/2$ as \emph{northern hemisphere}. In addition, we will refer to the line $\Psi=0$ as \emph{meridian} and to the line $\Psi=\pi$ as \emph{antimeridian}. Similar to the celestial equator they separate the region $0<\Psi<\pi$ from the region $\pi<\Psi<2\pi$. Here, we will now refer to the former as \emph{western hemisphere} and to the latter as \emph{eastern hemisphere}.
 
\subsection{The Photon Rings and the Lens Equation}
\begin{figure}\label{fig:SLD}
\includegraphics[width=80mm]{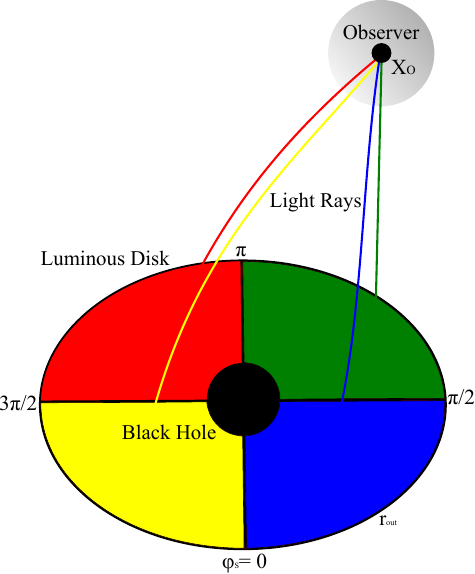}
	\caption{Schematic of the lens equation. In our case we have a luminous disk in the equatorial plane between the radius coordinates $r_{\text{in}}=r_{\text{H}}$ and $r_{\text{out}}$. We trace back light rays detected on the observer's celestial sphere at the celestial latitude $\Sigma$ and the celestial longitude $\Psi$ to their sources in the equatorial plane. For the visualisation of the lens map we now divide the luminous disk into four different quadrants and colour the different quadrants using an adapted version of the colour scheme of Bohn \emph{et al.} \cite{Bohn2015}. Here, we colour the quadrants as follows. We colour the quadrant with $0\leq\varphi_{S}\leq\pi/2$ blue, the quadrant with $\pi/2<\varphi_{S}\leq\pi$ green, the quadrant with $\pi<\varphi_{S}\leq 3\pi/2$ red, and the quadrant $3\pi/2<\varphi_{S}$ yellow.}
\end{figure}
We start with calculating the photon ring structure and the associated lens equations for the direct image, and the photon rings of first and second order. For this purpose we have to first define what we understand as a lens equation. A general-relativistic lens equation was first brought forward by Frittelli and Newman \cite{Frittelli1999}. It was later adapted to spherically symmetric and static spacetimes by Perlick \cite{Perlick2004}. It is commonly defined as a map from the latitude-longitude coordinates on the celestial sphere of an observer to a set of spacetime coordinates. In our case the source surface is given by a luminous disk in the equatorial plane. Here, we assume that the boundaries of this luminous disk are given by $r_{\text{in}}=r_{\text{H}}$ and $r_{\text{out}}$ and as a consequence in our case the spacetime coordinates of the lens map are given by the radius coordinate $r_{S}$ and the spacetime longitude $\varphi_{S}$ of sources in the luminous disk. Thus, in our case the lens equation maps an image observed at the latitude-longitude coordinates $\Sigma$ and $\Psi$ on the celestial sphere of the observer to its source at the spacetime coordinates $r_{S}$ and $\varphi_{S}$ in the luminous disk. Therefore, the lens map reads
\begin{eqnarray}
\left(\Sigma,\Psi\right)\rightarrow\left(r_{S}\left(\Sigma,\Psi\right),\varphi_{S}\left(\Sigma,\Psi\right)\right).
\end{eqnarray} 
In the following we will now briefly outline how we evaluate the lens equation for the direct image and the photon rings of the first and second order. As already mentioned above we assume that our observer detects the light rays at an event marked by the coordinates $x_{O}=(t_{O},r_{O},\vartheta_{O},\varphi_{O})$. In addition, we know that our light sources are located in the equatorial plane and thus we have $\vartheta_{S}=\pi/2$. We also know that the sources have to be located between $r_{\text{in}}=r_{\text{H}}$ and $r_{\text{out}}$. Now the only quantity we have to calculate is the Mino parameter $\lambda_{S}$ at which each light ray was emitted by its source. Since the Mino parameter is defined up to an affine transformation we can now without loss of generality choose $\lambda_{O}=0$. In addition, we agreed on the convention that when we follow light rays along their trajectories back into the past the Mino parameter is negative and thus we have $\lambda_{S}<\lambda_{O}=0$. Now we use these boundary conditions to rewrite (\ref{eq:Intx}) as
\begin{eqnarray}
&\hspace{-0.6cm}\lambda_{S}=\int_{\cos\vartheta_{i}...}^{...0}\frac{\text{d}x'}{\sqrt{-\omega_{\text{p}}^2 E_{\text{C}}^2 x'^4+\left(2\omega_{\text{p}}^2  E_{\text{C}}^2-K\right)x'^2+K-\omega_{\text{p}}^2E_{\text{C}}^2-L_{z}^2}},
\end{eqnarray}
where here for the explicit integration we have to rewrite $L_{z}$ and $K$ in terms of the latitude-longitude coordinates on the observer's celestial sphere using (\ref{eq:CoMLz}) and (\ref{eq:CoMK}). The dots in the limits of the integral shall again indicate that we have to split the integral at potential turning points into separate terms and that we then have to evaluate each term separately. In addition, the sign of the root in the denominator has to be chosen according to the direction of the $x=\cos\vartheta$ motion along each section of the trajectory (note that since here we integrate over $x'=\cos\vartheta'$ the sign of the root is always the opposite of the sign of the $\vartheta$ motion). Now for light rays travelling in vacuum or through a homogeneous plasma we rewrite this integral in terms of elementary functions, while for light rays travelling through the inhomogeneous plasma described by the plasma distribution given by Eq.~(\ref{eq:PlasmaEn}) we rewrite it in terms of Legendre's elliptic integral of the first kind. Then we insert the calculated $\lambda_{S}$ to calculate $r_{S}\left(\Sigma,\Psi\right)$ and $\varphi_{S}\left(\Sigma,\Psi\right)$ to obtain the lens equation. 

\begin{figure*}\label{fig:DI}
  \begin{tabular}{cc}
    \hspace{-0.5cm}Vacuum & \hspace{0.5cm} Homogeneous Plasma ($\omega_{\text{p}}=0$)\\
\\
    \hspace{-0.5cm}\includegraphics[width=80mm]{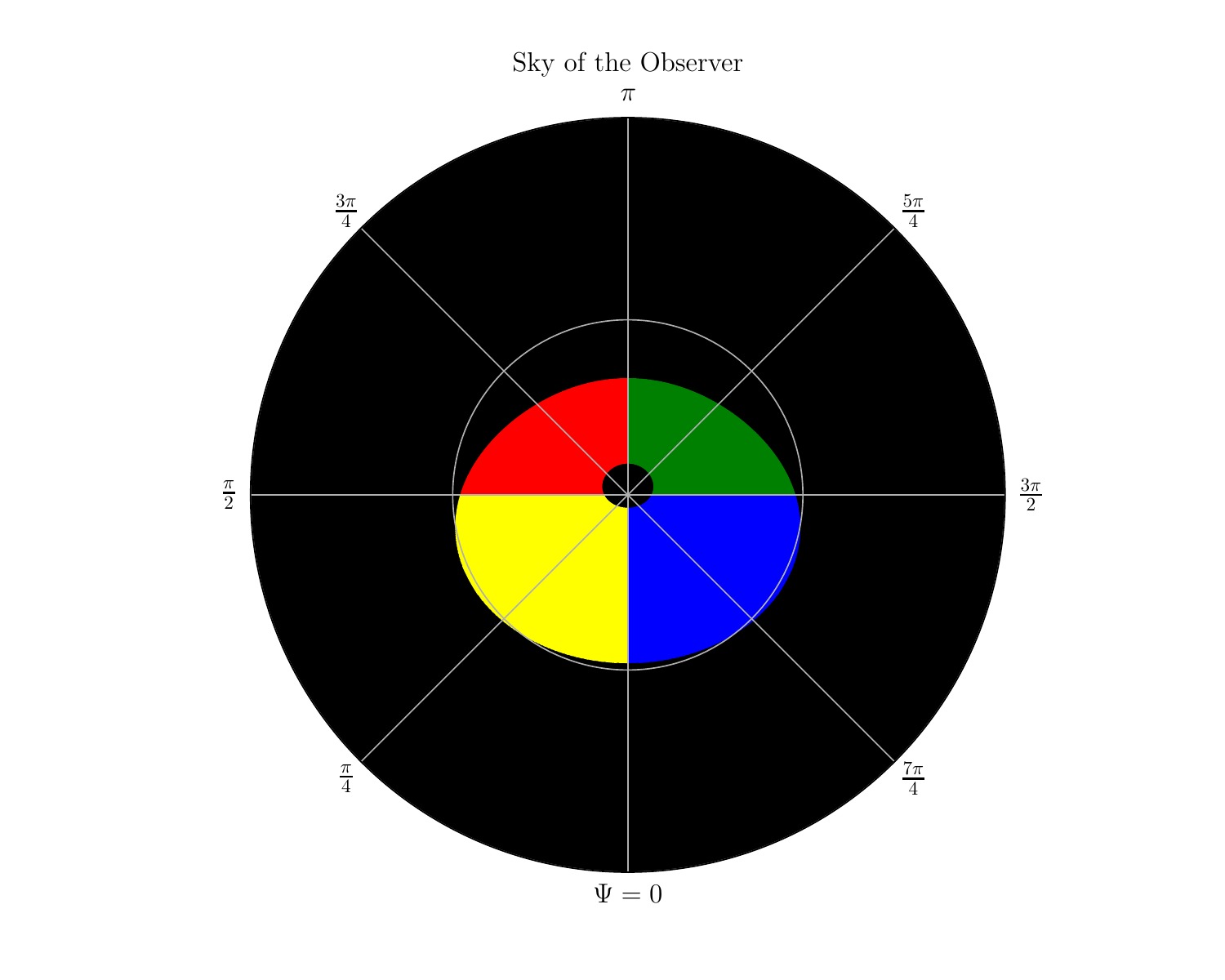} &   \hspace{0.5cm}\includegraphics[width=80mm]{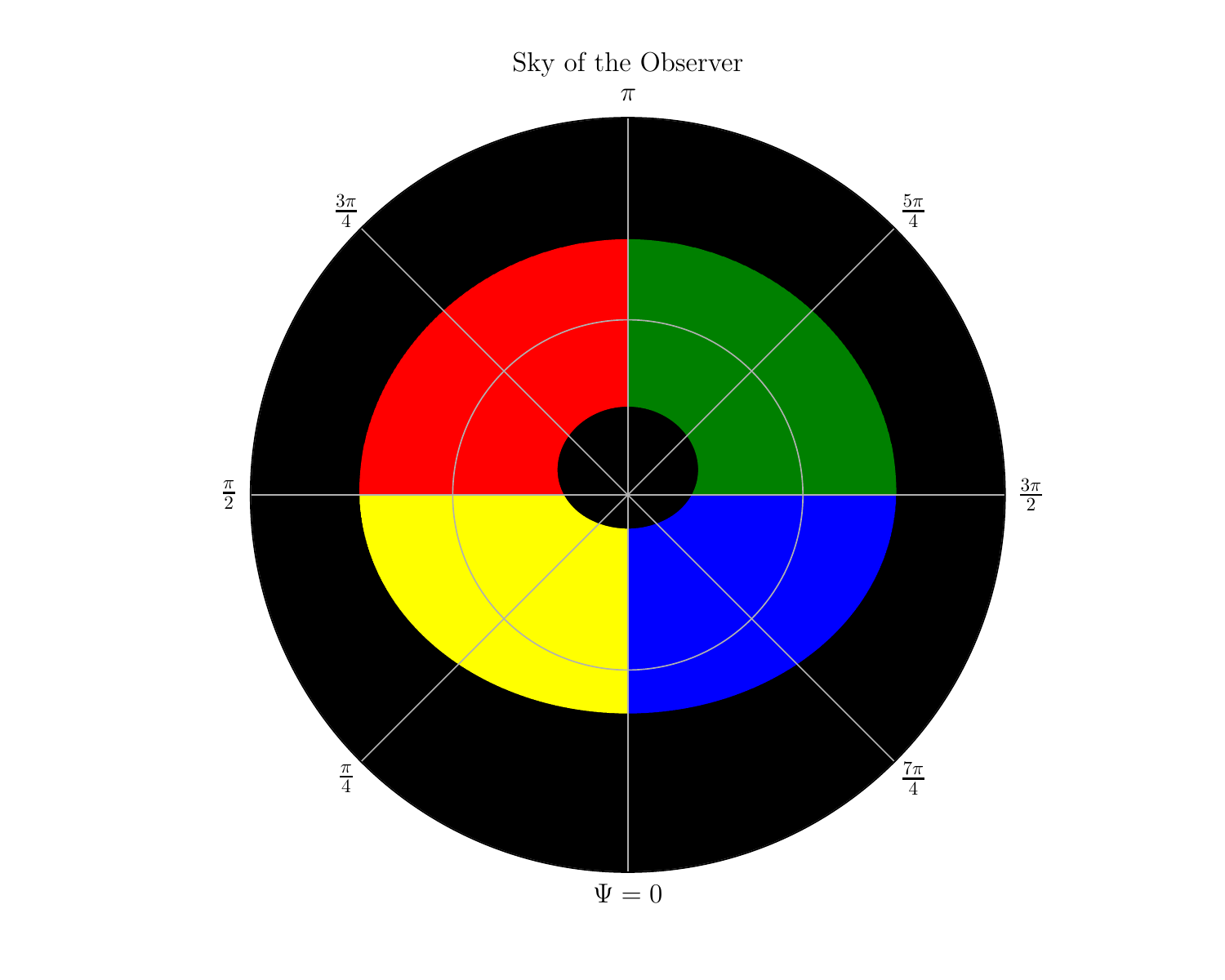} \\
\\
    \hspace{-0.5cm}Inhomogeneous Plasma with $\omega_{\text{p}}=m$ & \hspace{0.5cm} Inhomogeneous Plasma with $\omega_{\text{p}}=2m$\\
\\
    \hspace{-0.5cm}\includegraphics[width=80mm]{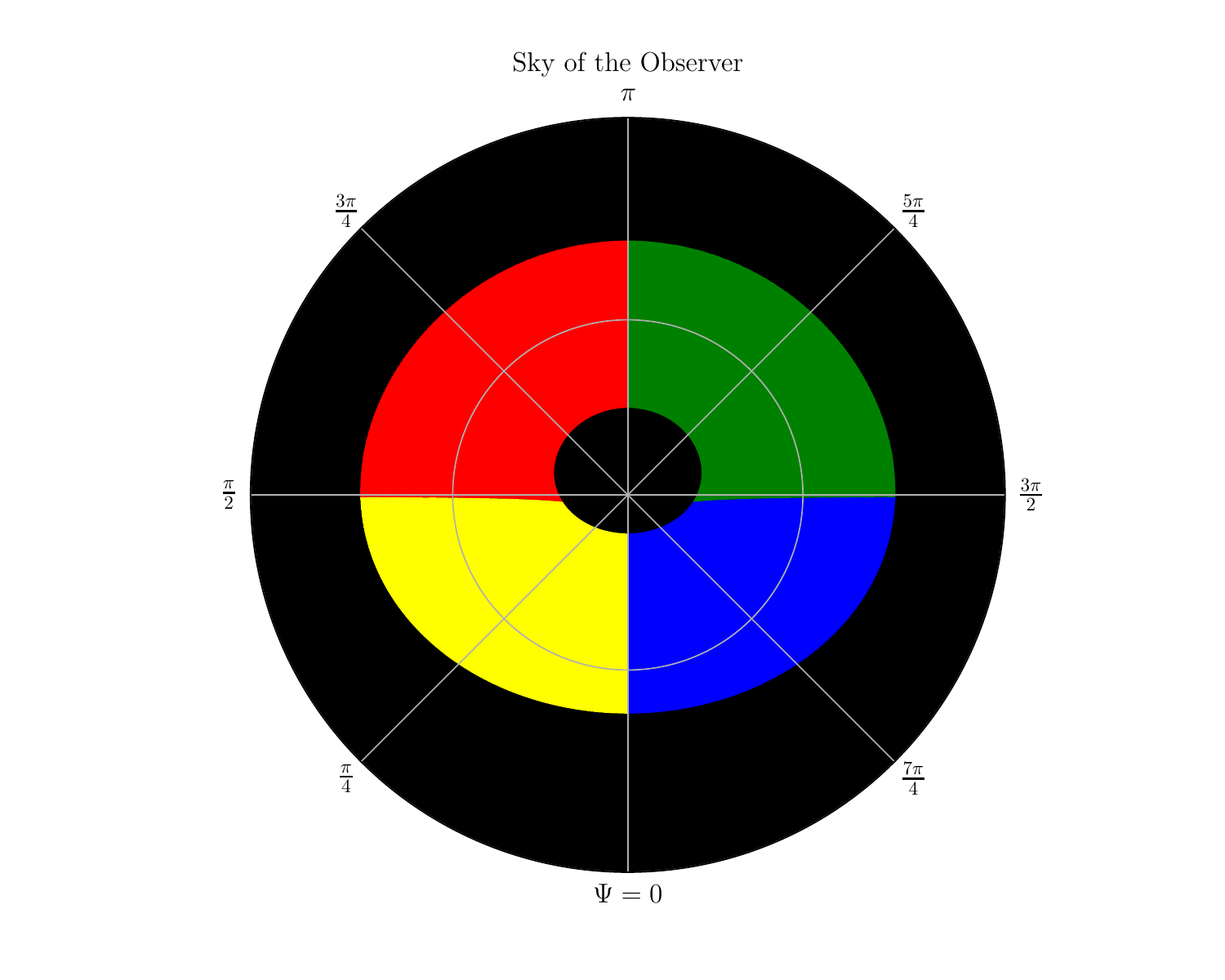} &   \hspace{0.5cm}\includegraphics[width=80mm]{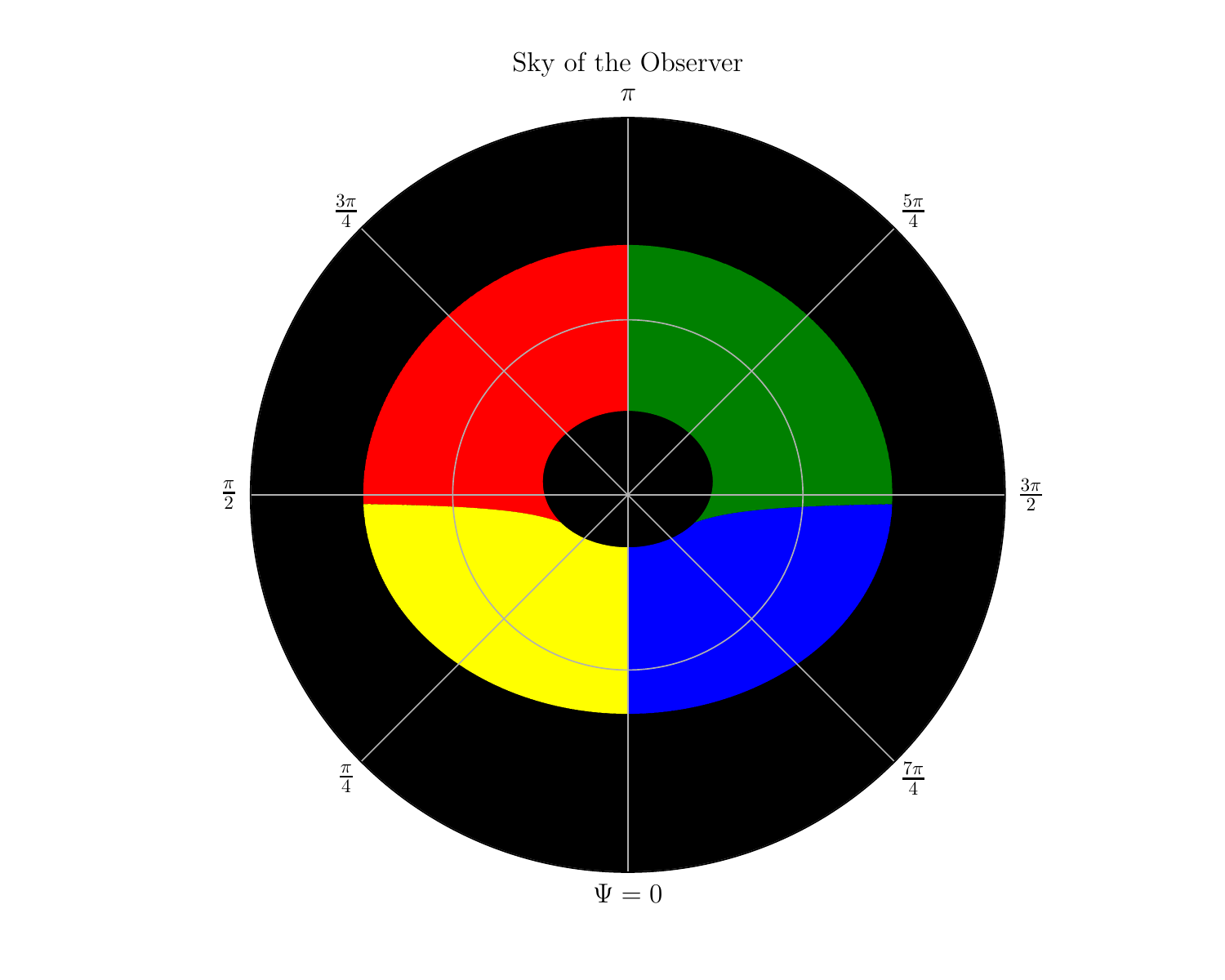} \\
    \hspace{-0.5cm}Inhomogeneous Plasma with $\omega_{\text{p}}=3m$ & \hspace{0.5cm} Inhomogeneous Plasma with $\omega_{\text{p}}=4m$\\
    \\
    \hspace{-0.5cm}\includegraphics[width=80mm]{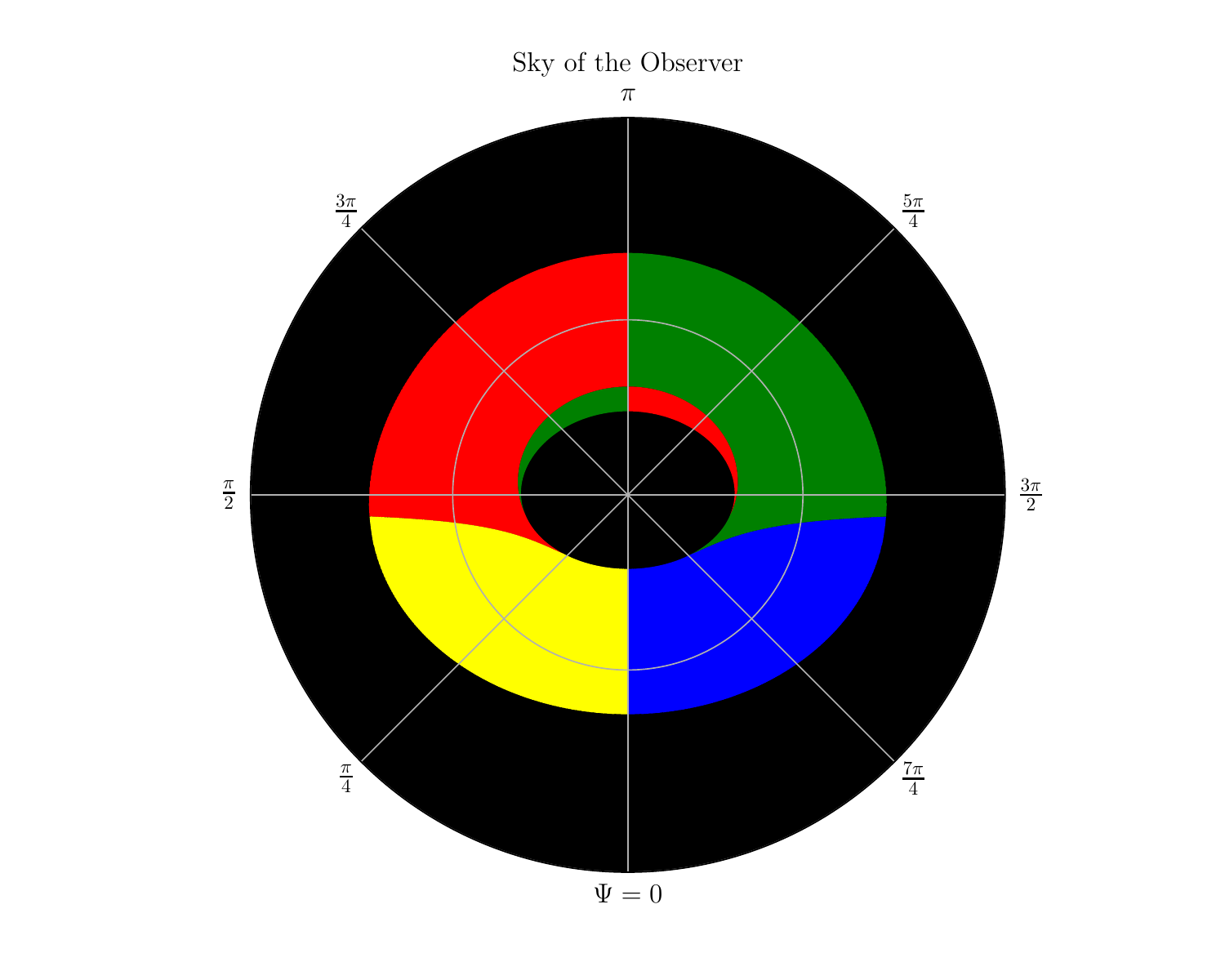} &   \hspace{0.5cm}\includegraphics[width=80mm]{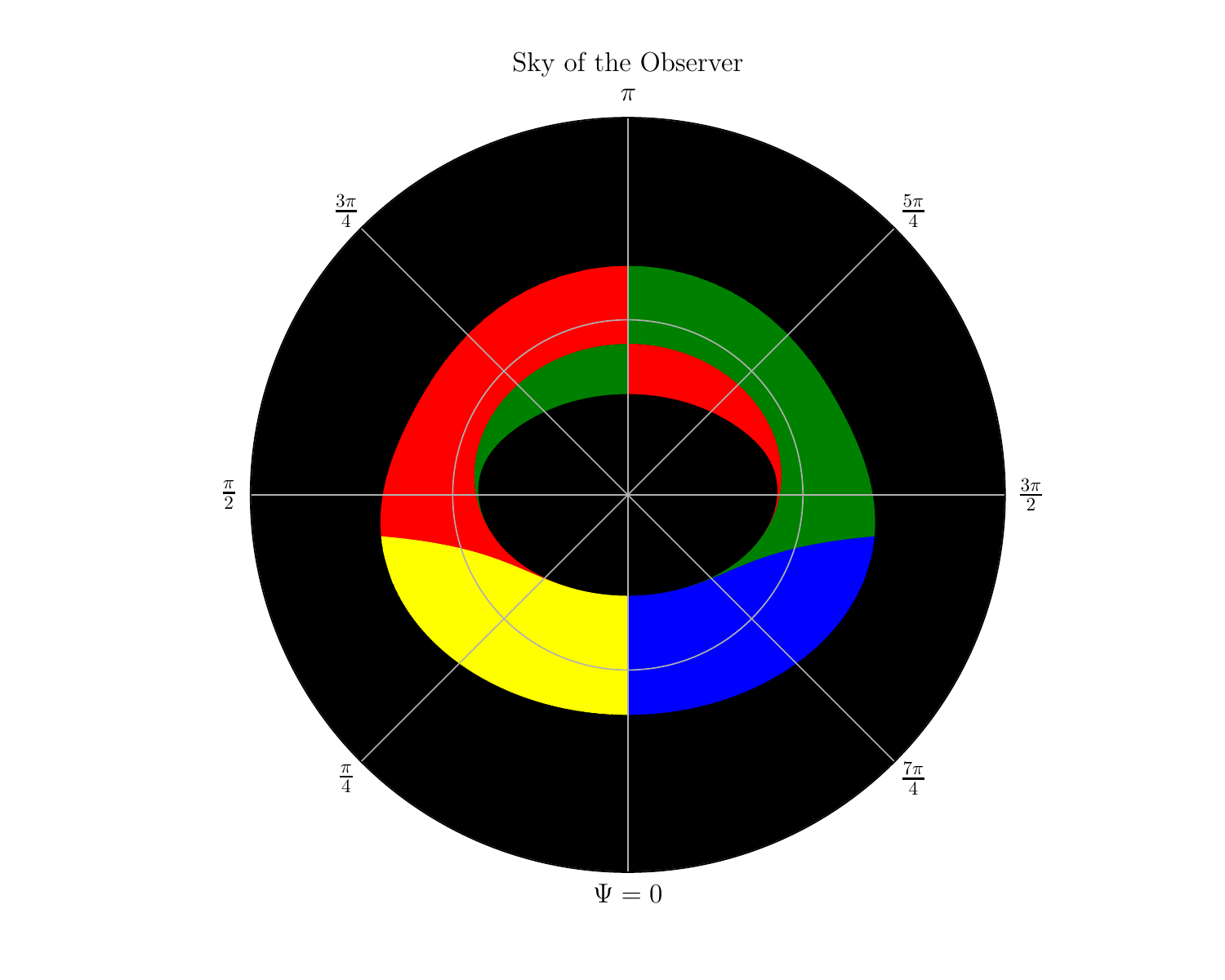} \\
  \end{tabular}
	\caption{Lens maps for the direct images in the Schwarzschild spacetime for light rays travelling in vacuum (top left panel), through a homogeneous plasma (top right panel), and through an inhomogeneous plasma described by the distribution $E_{\text{pl}}(r,\vartheta)$ given by (\ref{eq:PlasmaEn}) with $\omega_{\text{p}}=m$ (middle left panel), $\omega_{\text{p}}=2m$ (middle right panel), $\omega_{\text{p}}=3m$ (bottom left panel), and $\omega_{\text{p}}=4m$ (bottom right panel). The observer is located at $r_{O}=40m$ and $\vartheta_{O}=\pi/4$ and the luminous disk is located in the equatorial plane between $r_{\text{in}}=2m$ and $r_{\text{out}}=20m$. For the light rays travelling through one of the plasmas the energy measured at the position of the observer is $E_{O}=\sqrt{53/50}E_{\text{C}}$.}
\end{figure*}

\begin{figure*}\label{fig:PhotonRing1E1}
  \begin{tabular}{cc}
    \hspace{-0.5cm}Vacuum & \hspace{0.5cm} Homogeneous Plasma ($\omega_{\text{p}}=0$)\\
\\
    \hspace{-0.5cm}\includegraphics[width=80mm]{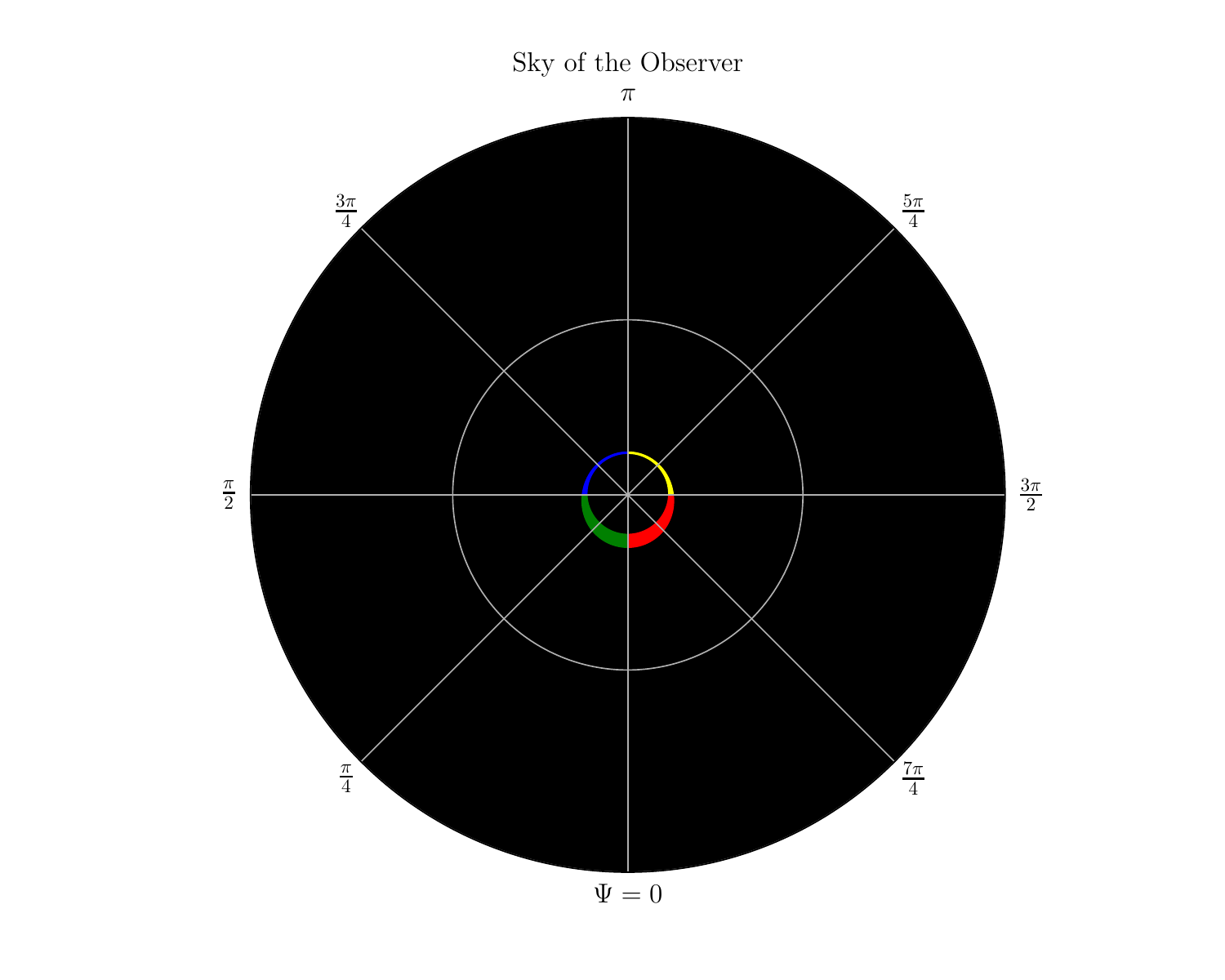} &   \hspace{0.5cm}\includegraphics[width=80mm]{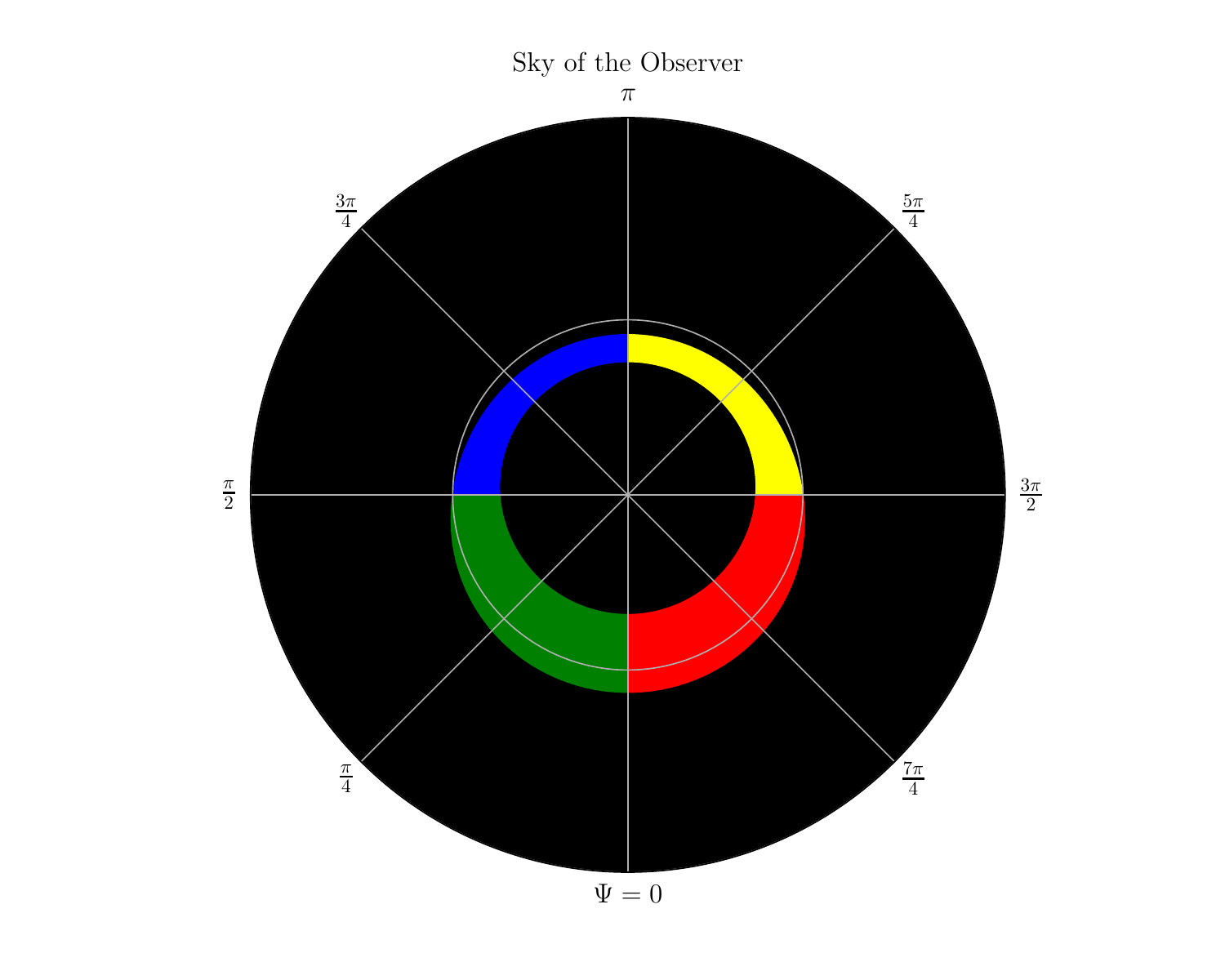} \\
\\
    \hspace{-0.5cm}Inhomogeneous Plasma with $\omega_{\text{p}}=m$ & \hspace{0.5cm} Inhomogeneous Plasma with $\omega_{\text{p}}=2m$\\
\\
    \hspace{-0.5cm}\includegraphics[width=80mm]{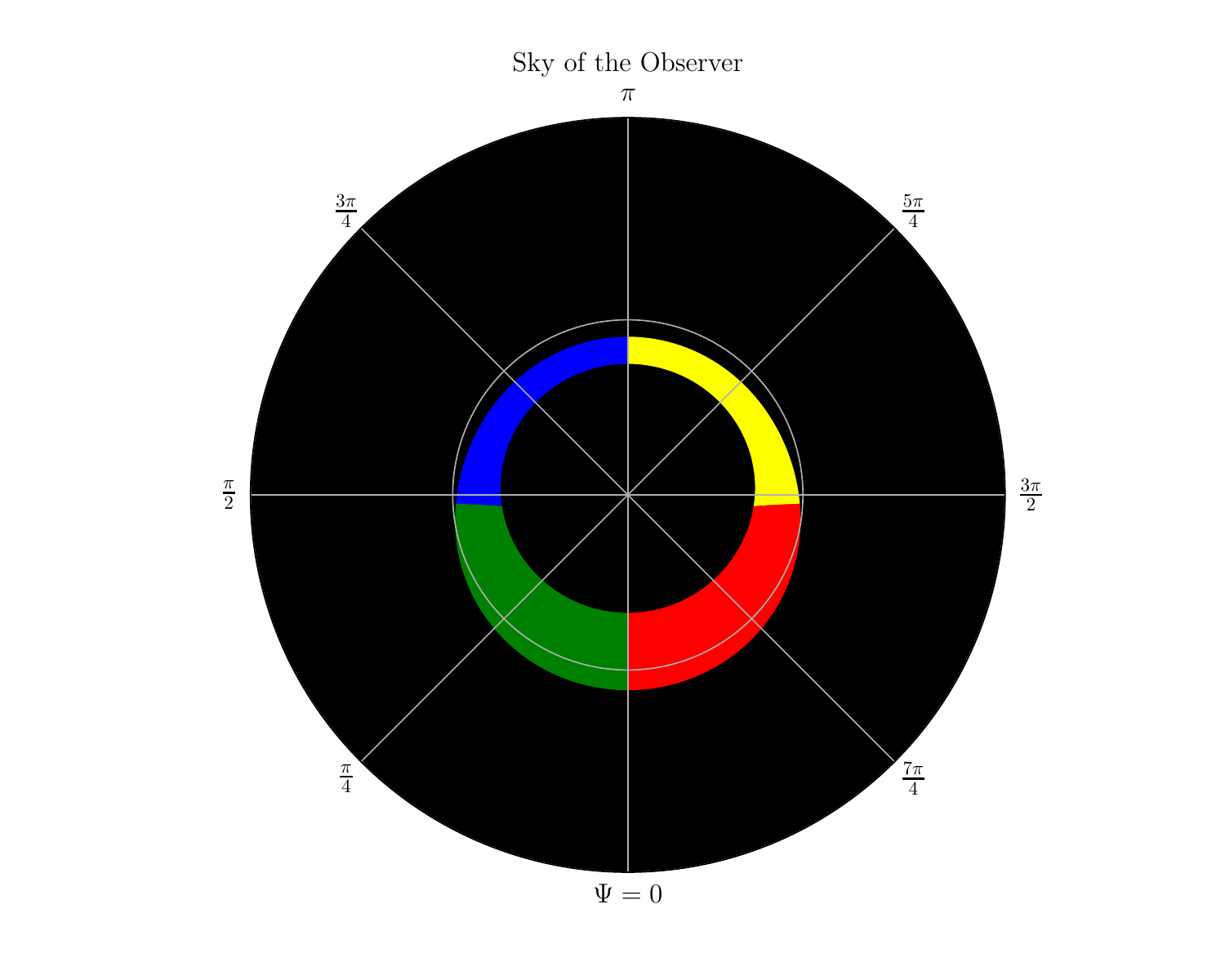} &   \hspace{0.5cm}\includegraphics[width=80mm]{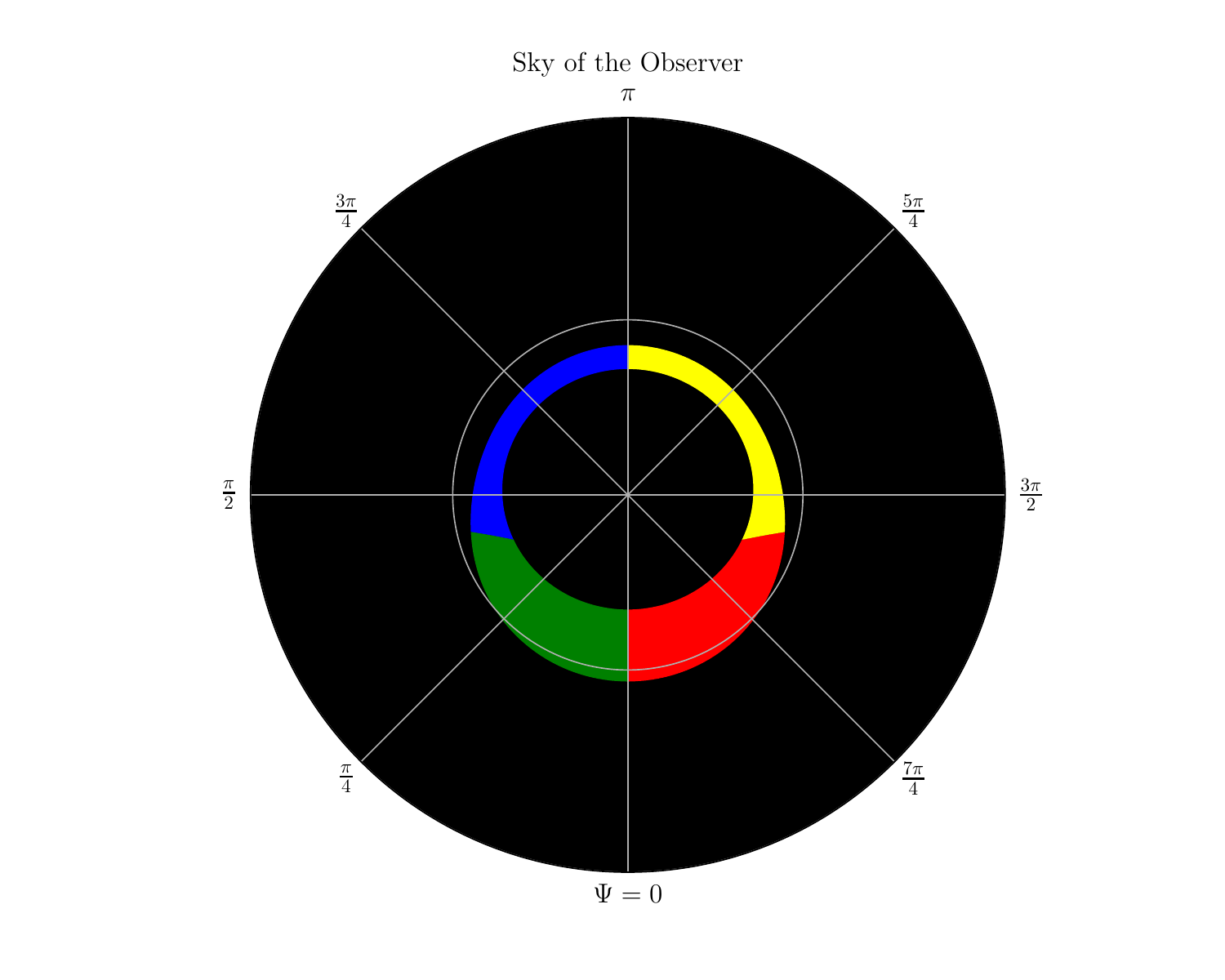} \\
    \hspace{-0.5cm}Inhomogeneous Plasma with $\omega_{\text{p}}=3m$ & \hspace{0.5cm} Inhomogeneous Plasma with $\omega_{\text{p}}=4m$\\
    \\
    \hspace{-0.5cm}\includegraphics[width=80mm]{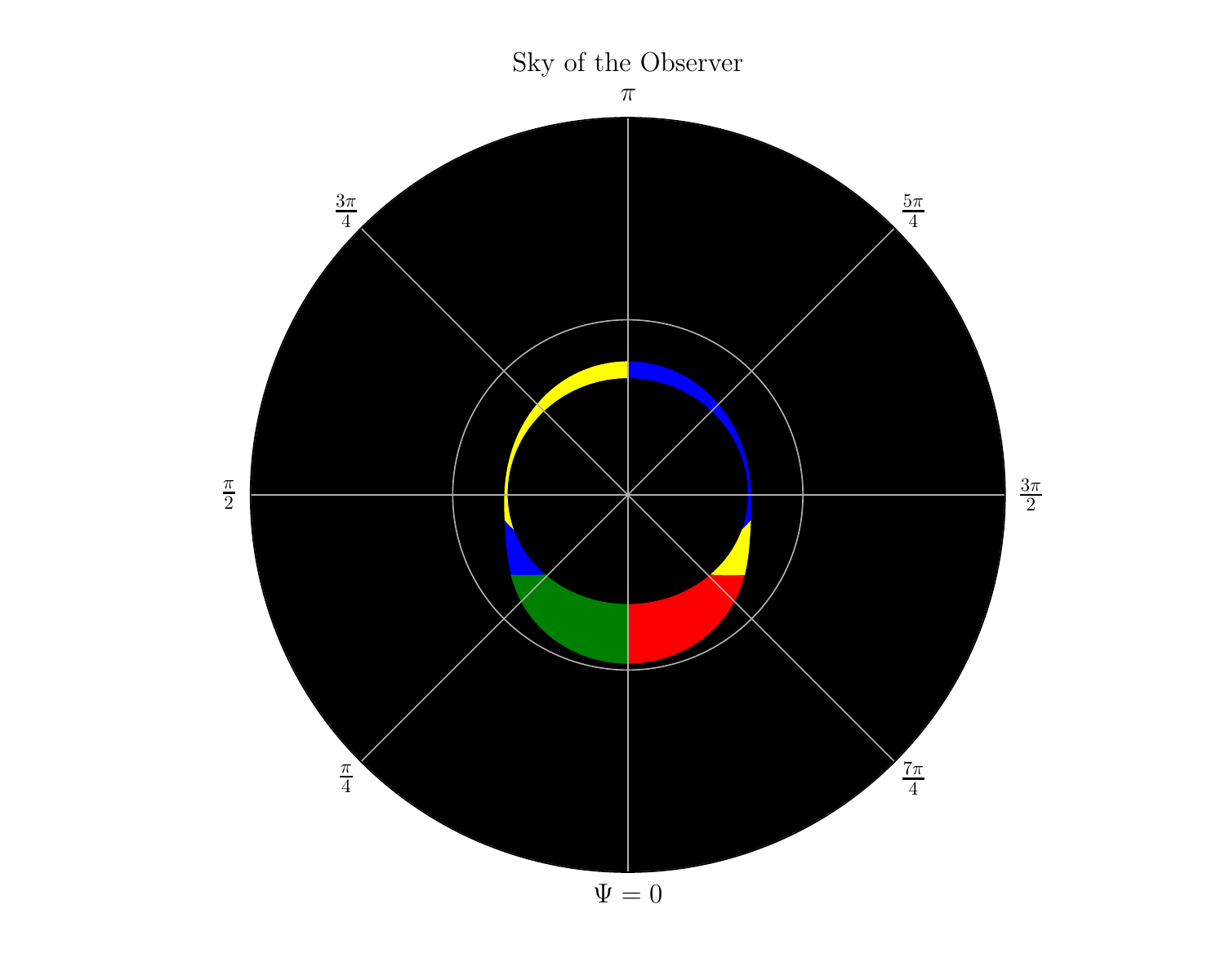} &   \hspace{0.5cm}\includegraphics[width=80mm]{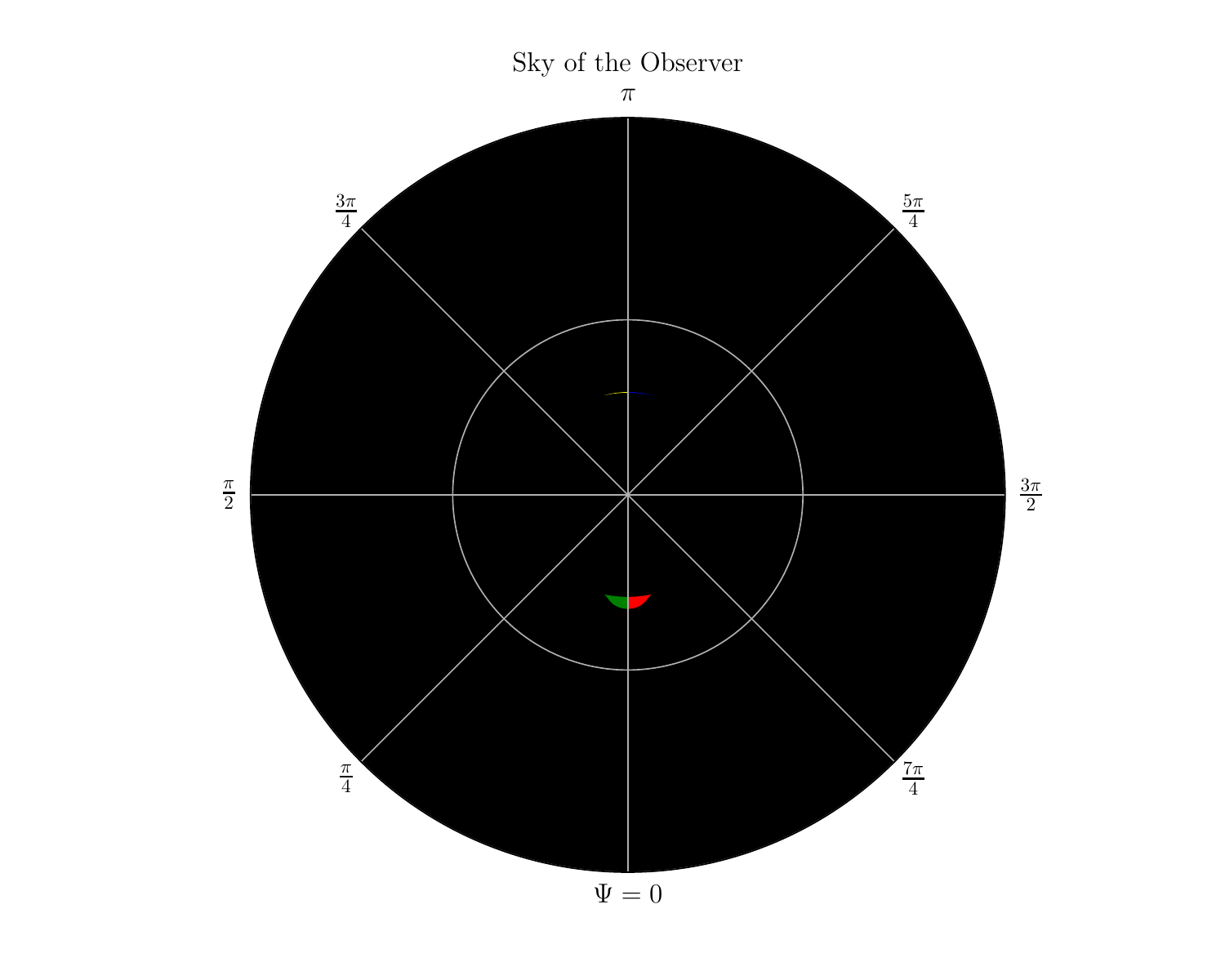} \\
  \end{tabular}
	\caption{Lens maps for the first-order photon rings in the Schwarzschild spacetime for light rays travelling in vacuum (top left panel), through a homogeneous plasma (top right panel), and through an inhomogeneous plasma described by the distribution $E_{\text{pl}}(r,\vartheta)$ given by (\ref{eq:PlasmaEn}) with $\omega_{\text{p}}=m$ (middle left panel), $\omega_{\text{p}}=2m$ (middle right panel), $\omega_{\text{p}}=3m$ (bottom left panel), and $\omega_{\text{p}}=4m$ (bottom right panel). The observer is located at $r_{O}=40m$ and $\vartheta_{O}=\pi/4$ and the luminous disk is located in the equatorial plane between $r_{\text{in}}=2m$ and $r_{\text{out}}=20m$. For the light rays travelling through one of the plasmas the energy measured at the position of the observer is $E_{O}=\sqrt{53/50}E_{\text{C}}$.}
\end{figure*}

\begin{figure*}\label{fig:PhotonRing2E1}
  \begin{tabular}{cc}
    \hspace{-0.5cm}Vacuum & \hspace{0.5cm} Homogeneous Plasma ($\omega_{\text{p}}=0$)\\
\\
    \hspace{-0.5cm}\includegraphics[width=80mm]{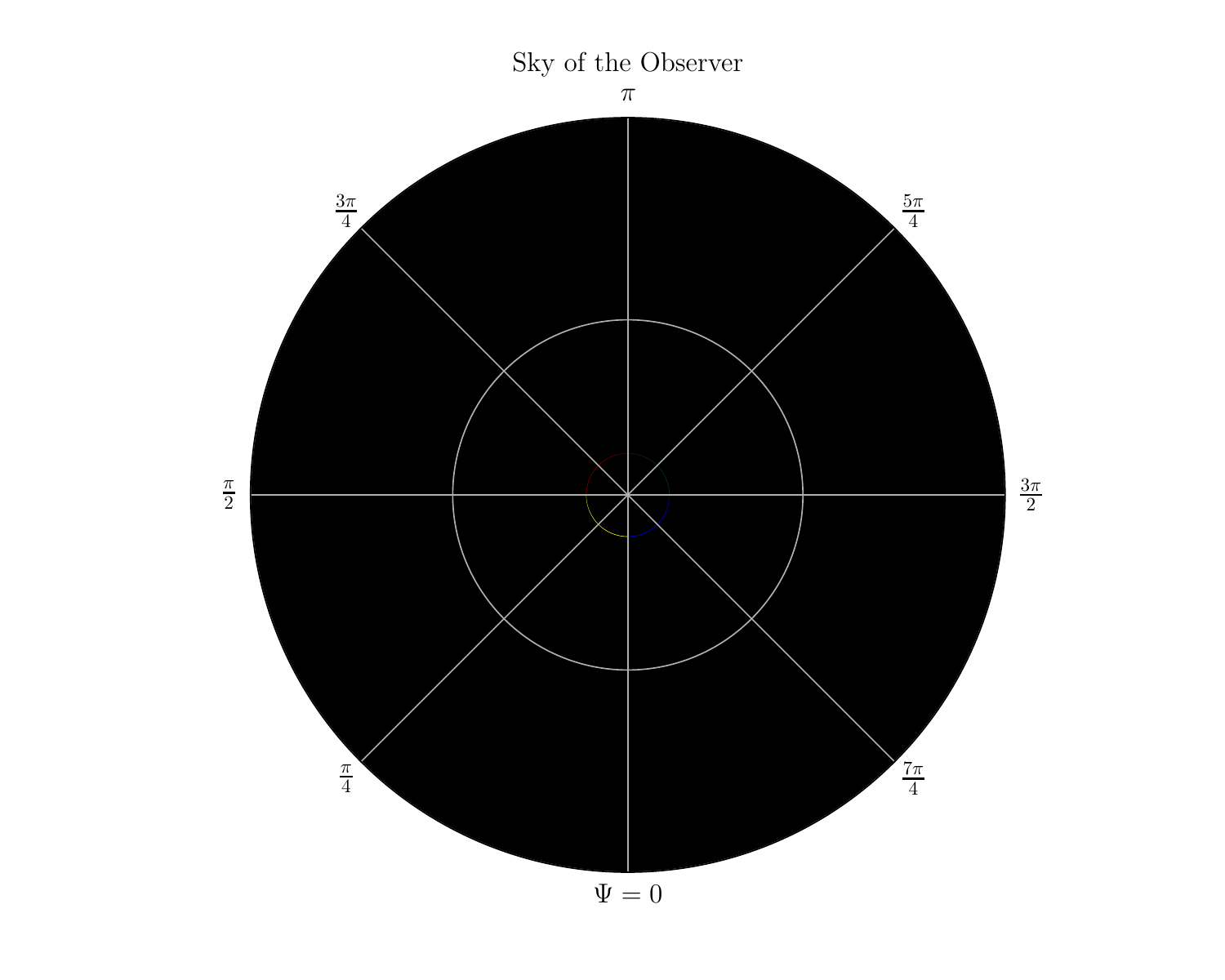} &   \hspace{0.5cm}\includegraphics[width=80mm]{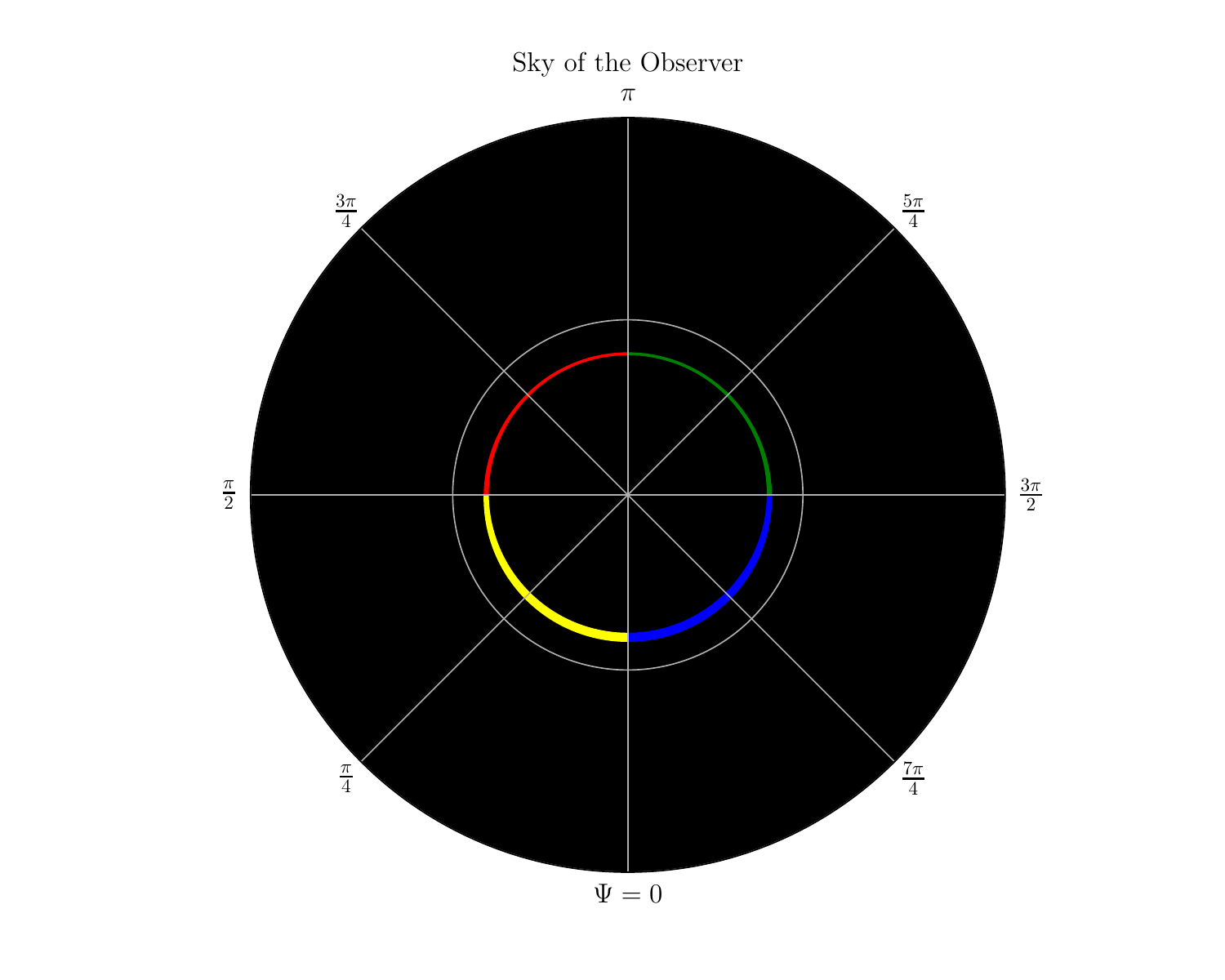} \\
\\
    \hspace{-0.5cm}Inhomogeneous Plasma with $\omega_{\text{p}}=m$ & \hspace{0.5cm} Inhomogeneous Plasma with $\omega_{\text{p}}=2m$\\
\\
    \hspace{-0.5cm}\includegraphics[width=80mm]{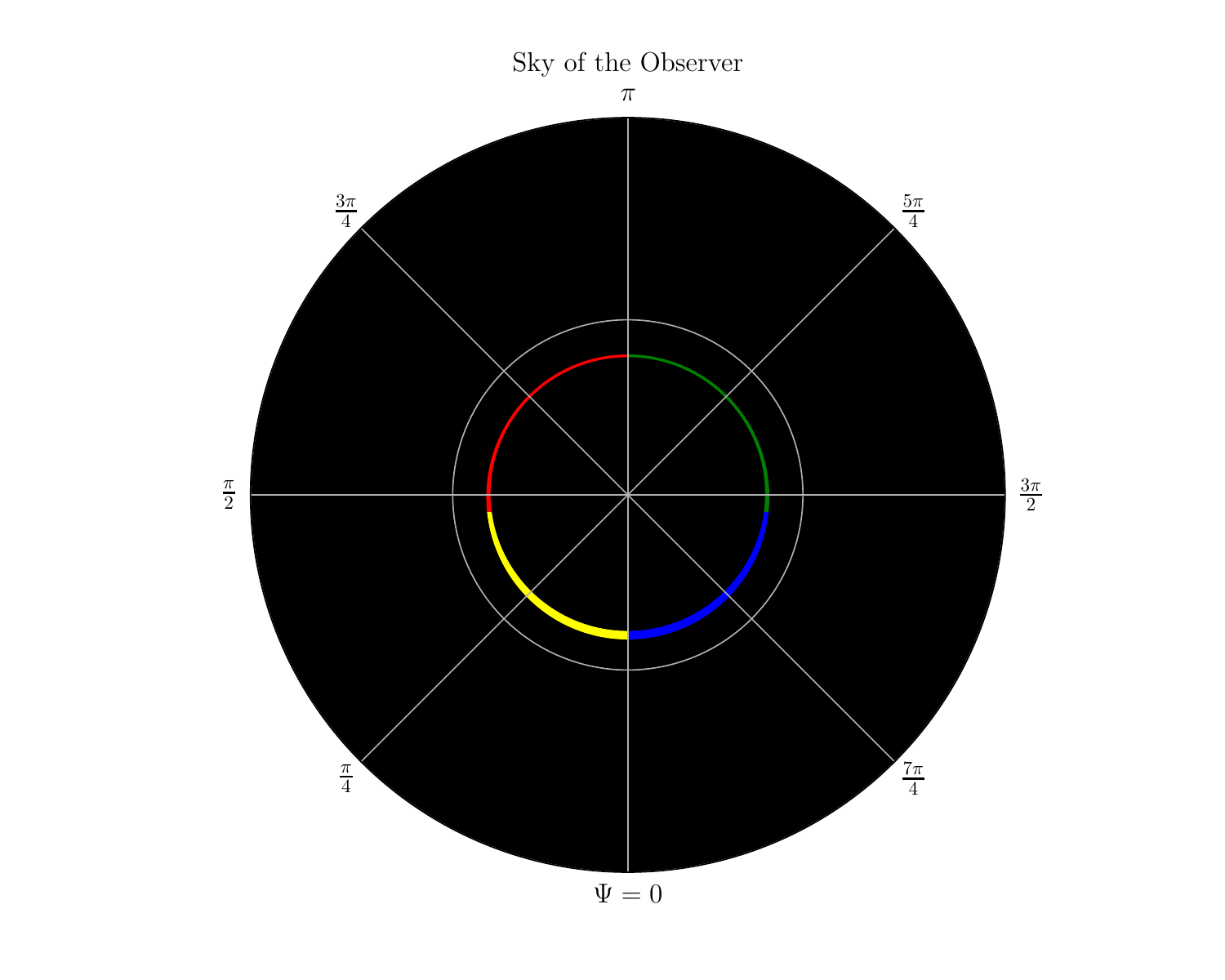} &   \hspace{0.5cm}\includegraphics[width=80mm]{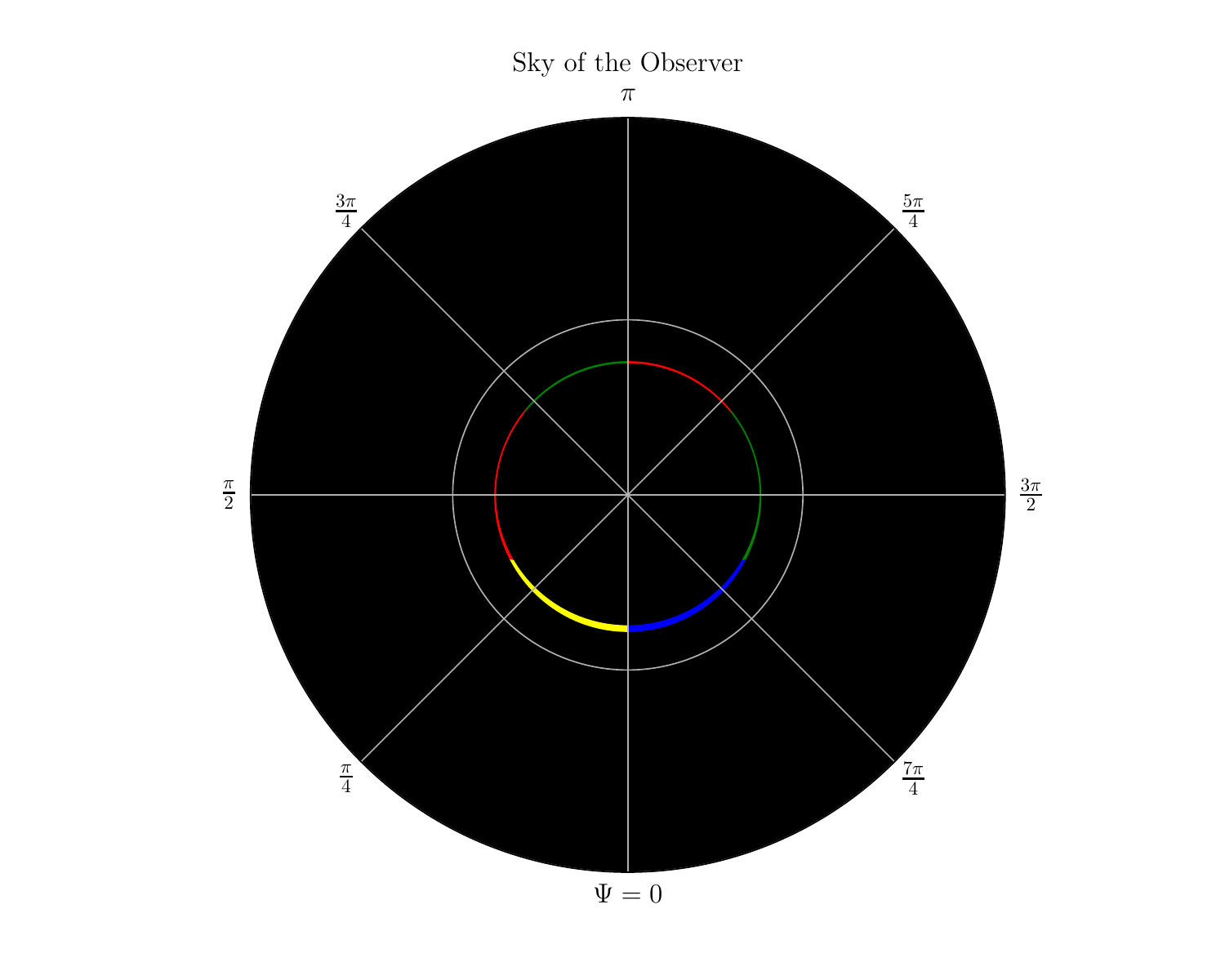} \\
    \hspace{-0.5cm}Inhomogeneous Plasma with $\omega_{\text{p}}=3m$ & \hspace{0.5cm} Inhomogeneous Plasma with $\omega_{\text{p}}=4m$\\
    \\
    \hspace{-0.5cm}\includegraphics[width=80mm]{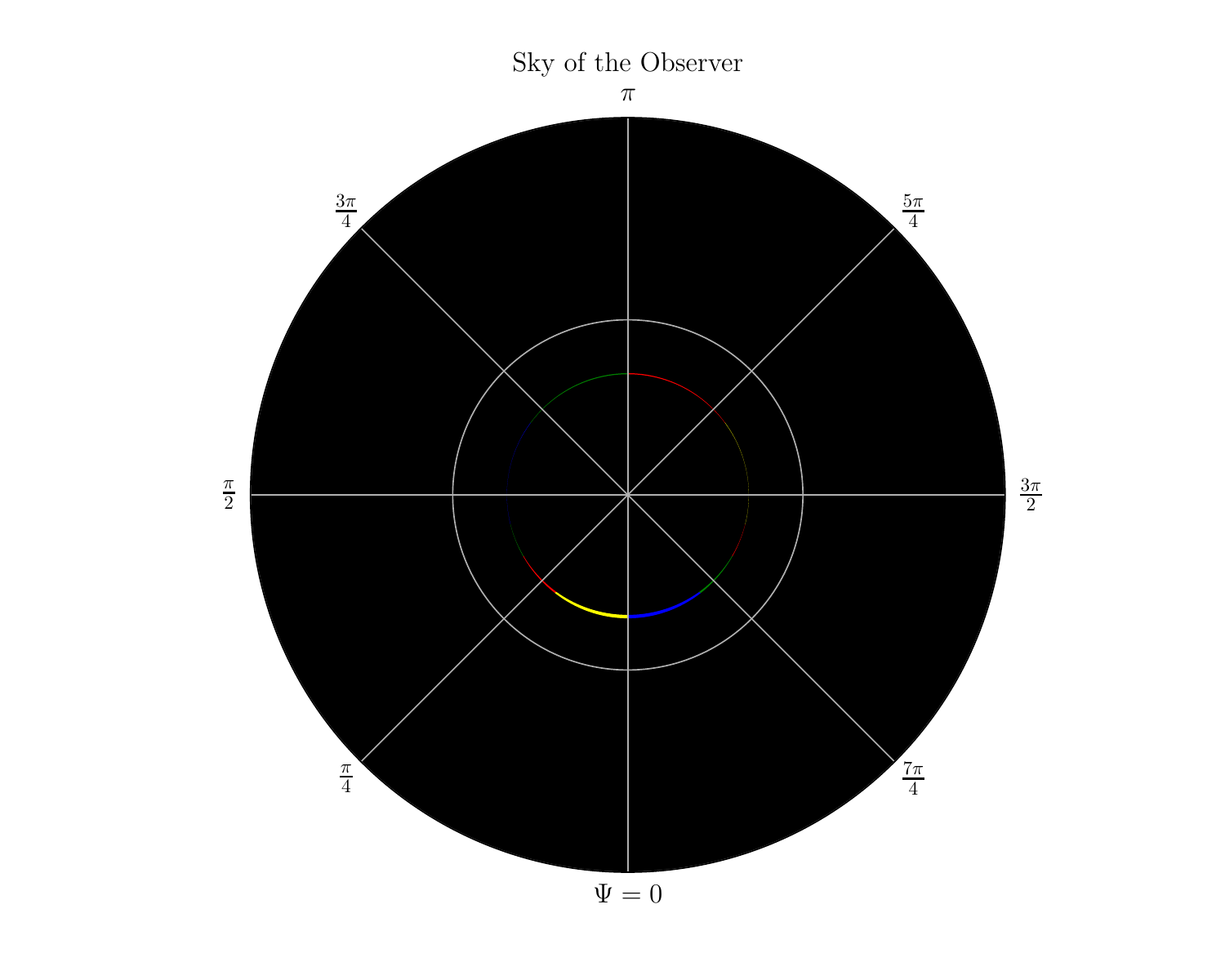} &   \hspace{0.5cm}\includegraphics[width=80mm]{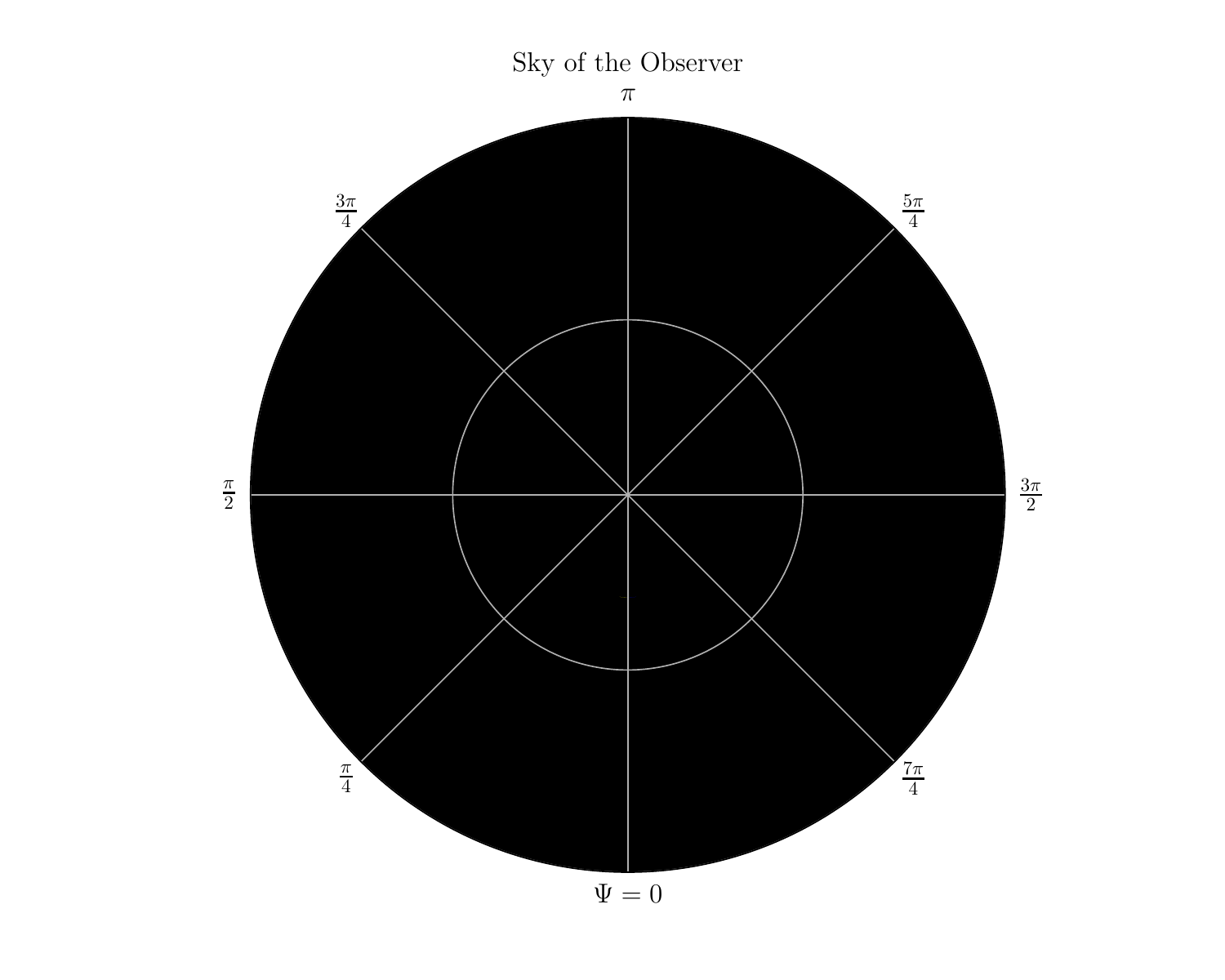} \\
  \end{tabular}
	\caption{Lens maps for the second-order photon rings in the Schwarzschild spacetime for light rays travelling in vacuum (top left panel), through a homogeneous plasma (top right panel), and through an inhomogeneous plasma described by the distribution $E_{\text{pl}}(r,\vartheta)$ given by (\ref{eq:PlasmaEn}) with $\omega_{\text{p}}=m$ (middle left panel), $\omega_{\text{p}}=2m$ (middle right panel), $\omega_{\text{p}}=3m$ (bottom left panel), and $\omega_{\text{p}}=4m$ (bottom right panel). The observer is located at $r_{O}=40m$ and $\vartheta_{O}=\pi/4$ and the luminous disk is located in the equatorial plane between $r_{\text{in}}=2m$ and $r_{\text{out}}=20m$. For the light rays travelling through one of the plasmas the energy measured at the position of the observer is $E_{O}=\sqrt{53/50}E_{\text{C}}$.}
\end{figure*}

\begin{figure*}\label{fig:DIIHP2m}
  \begin{tabular}{cc}
    \hspace{-0.5cm} $E_{O}=\sqrt{53/50}E_{\text{C}}$ & \hspace{0.5cm$E_{O}=\sqrt{15/14}E_{\text{C}}$} \\
\\
    \hspace{-0.5cm}\includegraphics[width=80mm]{Schwarzschild_Metric_Photon_Rings_IPlasma_ULR_Calculations_IHP1_IHP2.pdf} &   \hspace{0.5cm}\includegraphics[width=80mm]{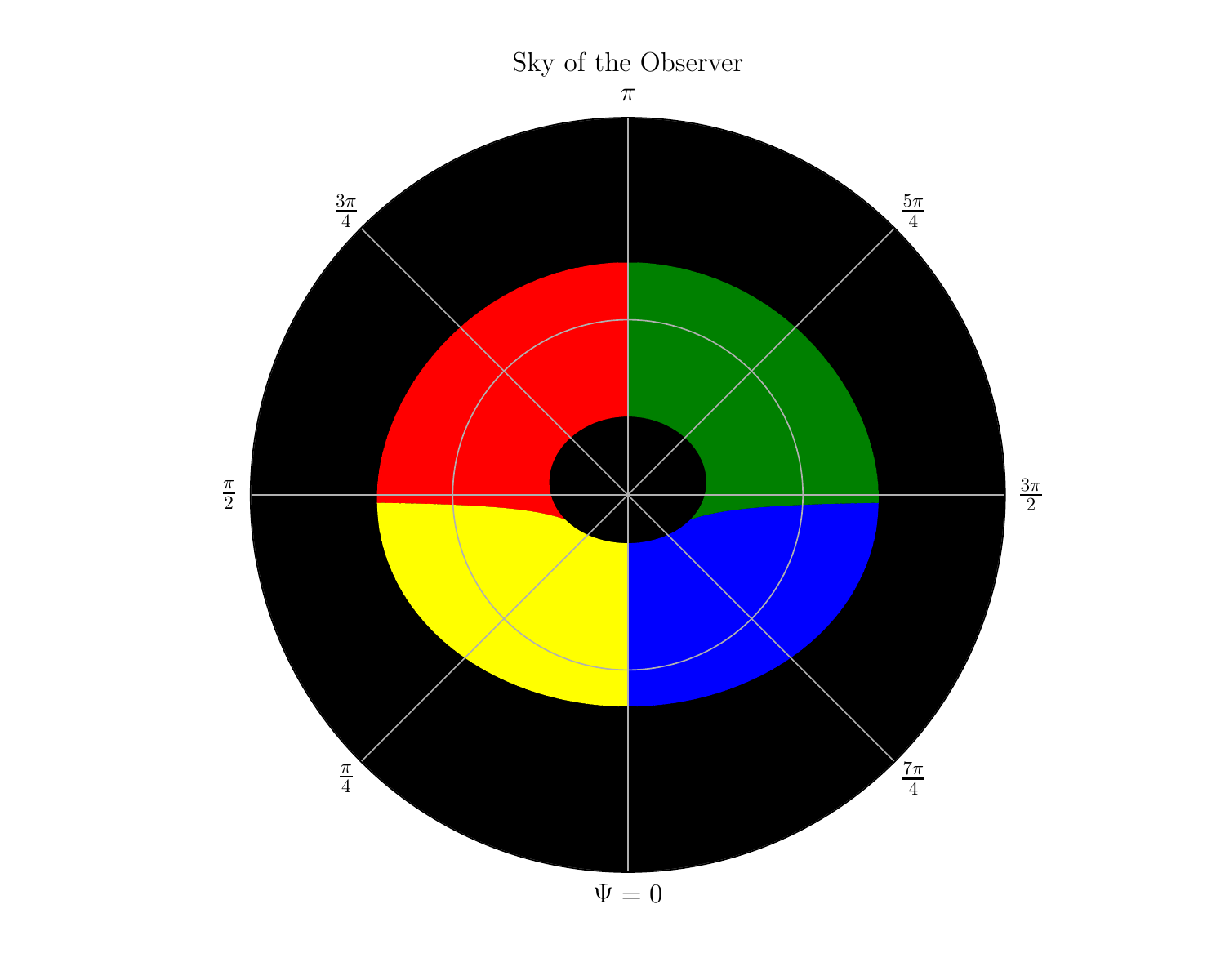} \\
\\
    \hspace{-0.5cm} $E_{O}=\sqrt{5}E_{\text{C}}/2$ & \hspace{0.5cm} $E_{O}=\sqrt{15/8}E_{\text{C}}$\\
\\
    \hspace{-0.5cm}\includegraphics[width=80mm]{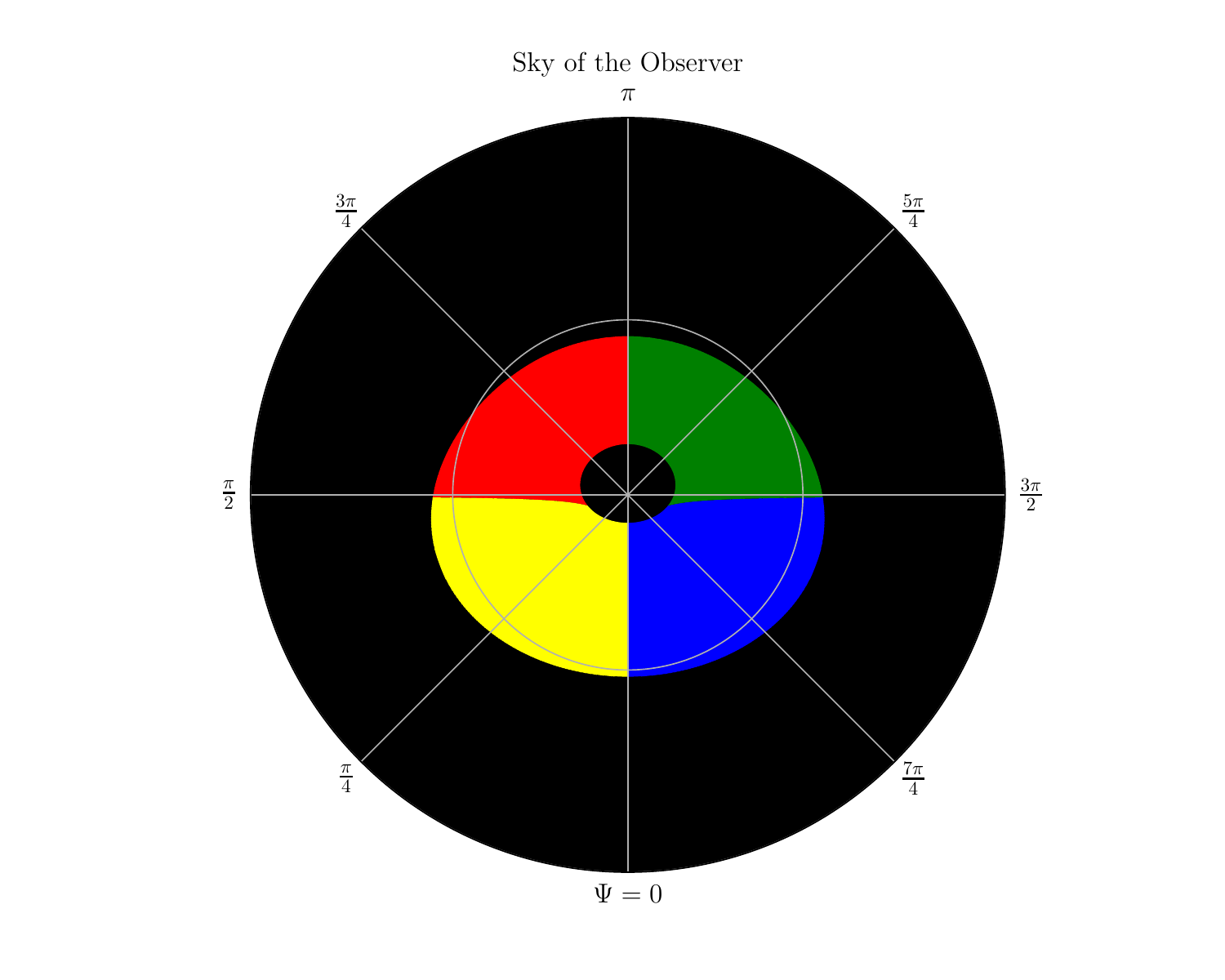} &   \hspace{0.5cm}\includegraphics[width=80mm]{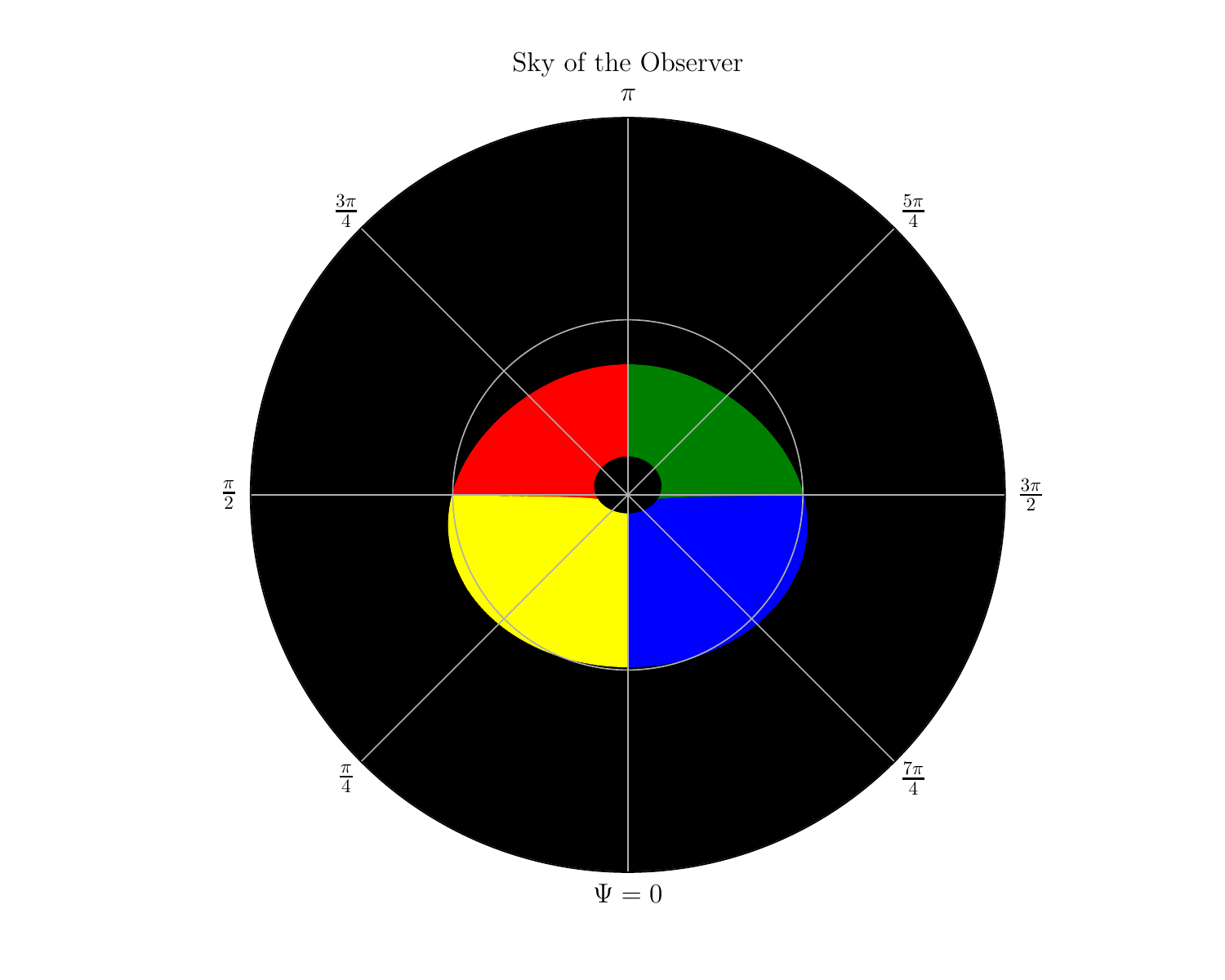} \\
    \hspace{-0.5cm}$E_{O}=\sqrt{2}E_{\text{C}}$ & \hspace{0.5cm} $E_{O}=\sqrt{5}E_{\text{C}}$\\
    \\
    \hspace{-0.5cm}\includegraphics[width=80mm]{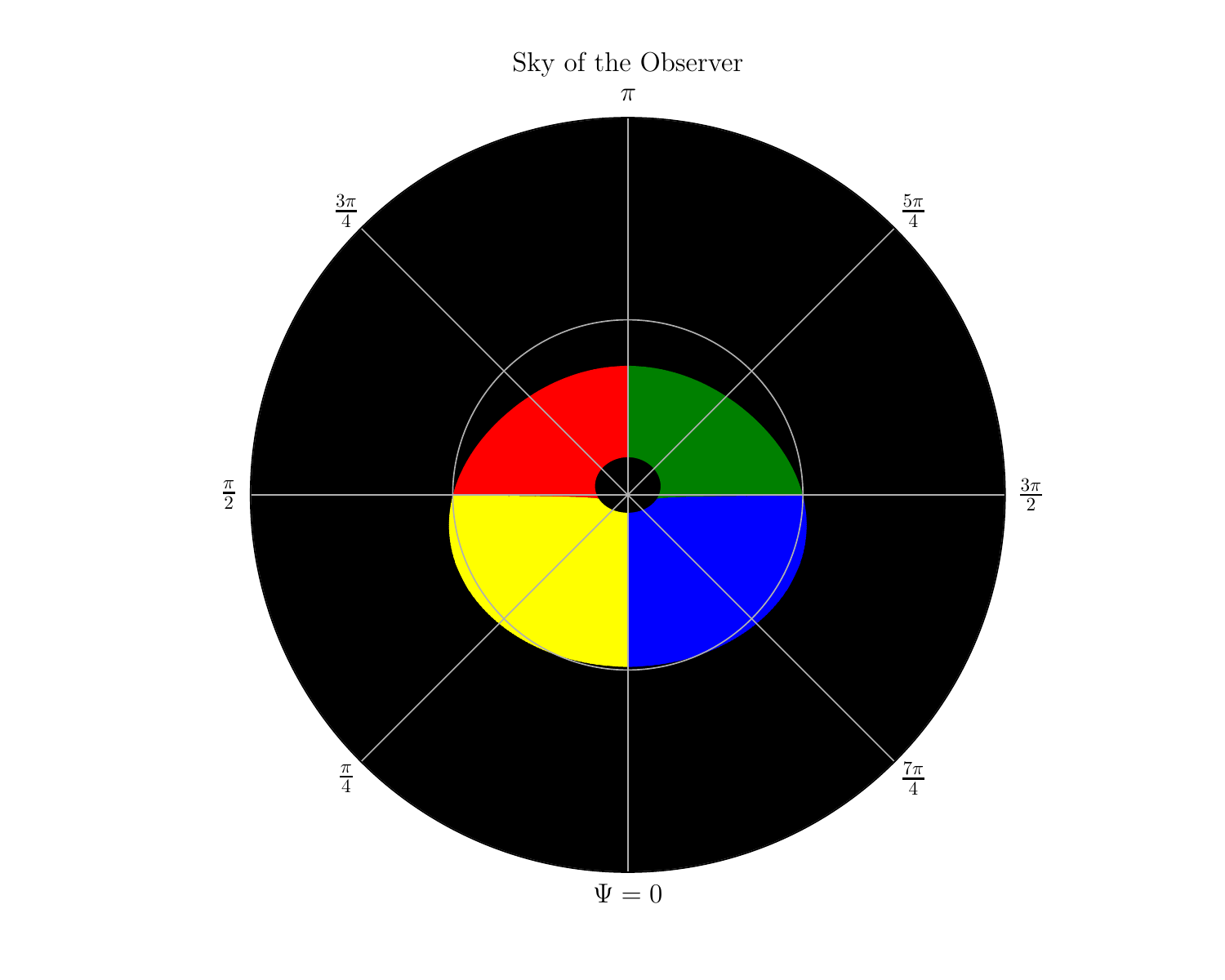} &   \hspace{0.5cm}\includegraphics[width=80mm]{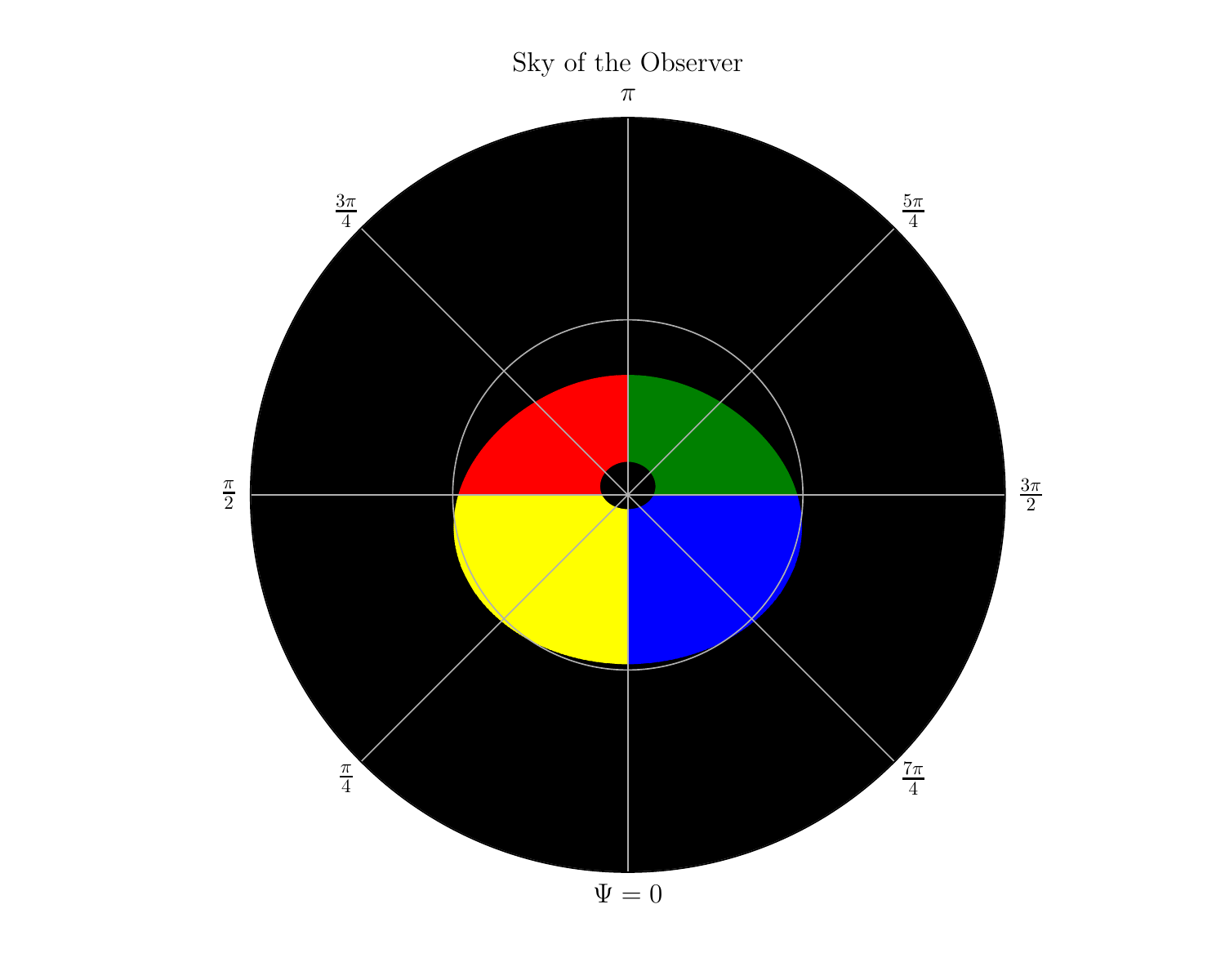} \\
  \end{tabular}
	\caption{Lens maps for the direct images in the Schwarzschild spacetime for light rays with $E_{O}=\sqrt{53/50}E_{\text{C}}$ (top left panel), $E_{O}=\sqrt{15/14}E_{\text{C}}$ (top right panel), $E_{O}=\sqrt{5}E_{\text{C}}/2$ (middle left panel), $E_{O}=\sqrt{15/8}E_{\text{C}}$ (middle right panel), $E_{O}=\sqrt{2}E_{\text{C}}$ (bottom left panel), and $E_{O}=\sqrt{5}E_{\text{C}}$ (bottom right panel) travelling through an inhomogeneous plasma described by the distribution $E_{\text{pl}}(r,\vartheta)$ given by (\ref{eq:PlasmaEn}) with $\omega_{\text{p}}=2m$. The observer is located at $r_{O}=40m$ and $\vartheta_{O}=\pi/4$ and the luminous disk is located in the equatorial plane between $r_{\text{in}}=2m$ and $r_{\text{out}}=20m$.}
\end{figure*}

\begin{figure*}\label{fig:PR1IHP2m}
  \begin{tabular}{cc}
    \hspace{-0.5cm} $E_{O}=\sqrt{53/50}E_{\text{C}}$ & \hspace{0.5cm}$E_{O}=\sqrt{15/14}E_{\text{C}}$ \\
\\
    \hspace{-0.5cm}\includegraphics[width=80mm]{Schwarzschild_Metric_Photon_Rings_IPlasma_ULR_Calculations_IHP3_IHP4.pdf} &   \hspace{0.5cm}\includegraphics[width=80mm]{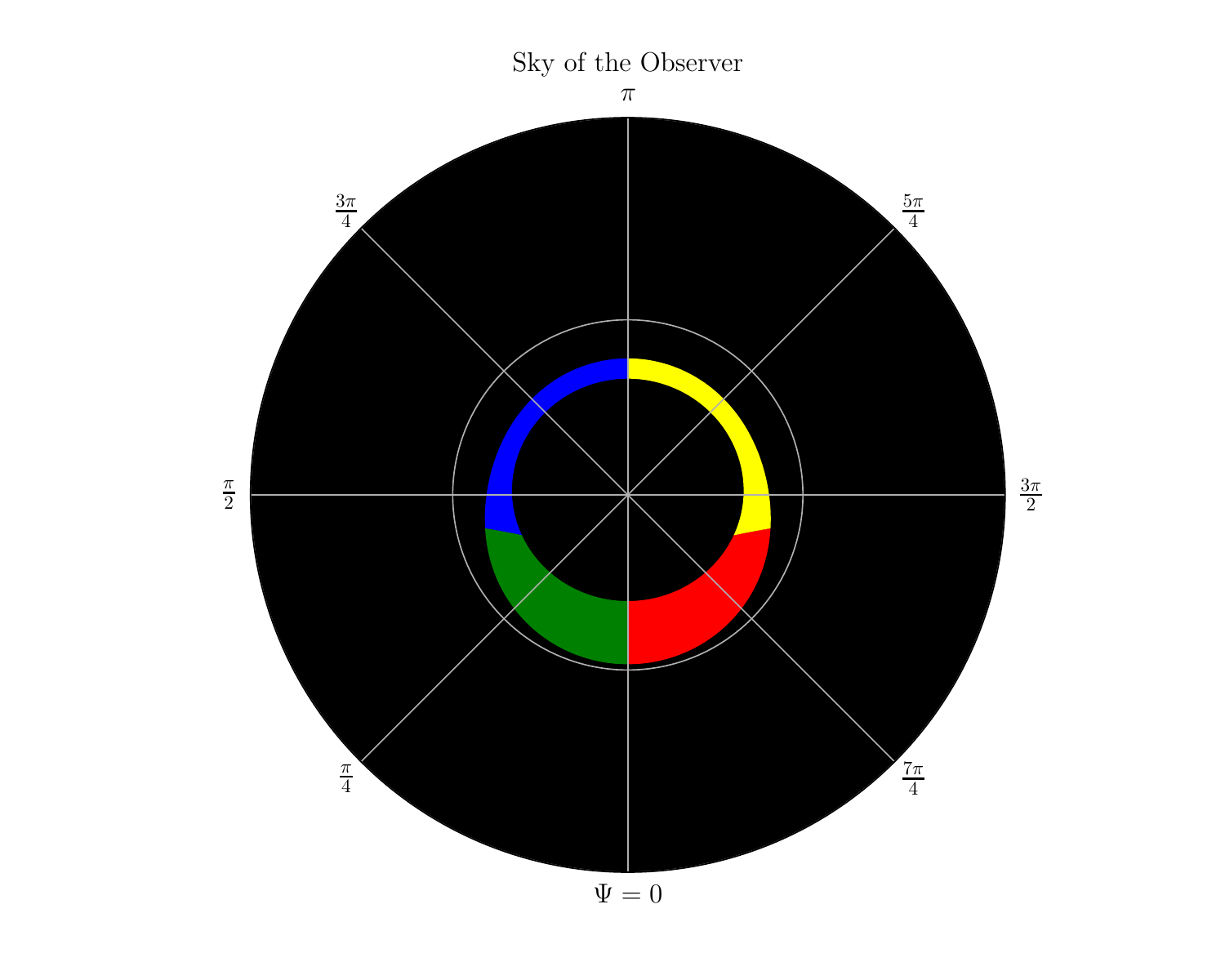} \\
\\
    \hspace{-0.5cm}$E_{O}=\sqrt{5}E_{\text{C}}/2$ & \hspace{0.5cm} $E_{O}=\sqrt{15/8}E_{\text{C}}$\\
\\
    \hspace{-0.5cm}\includegraphics[width=80mm]{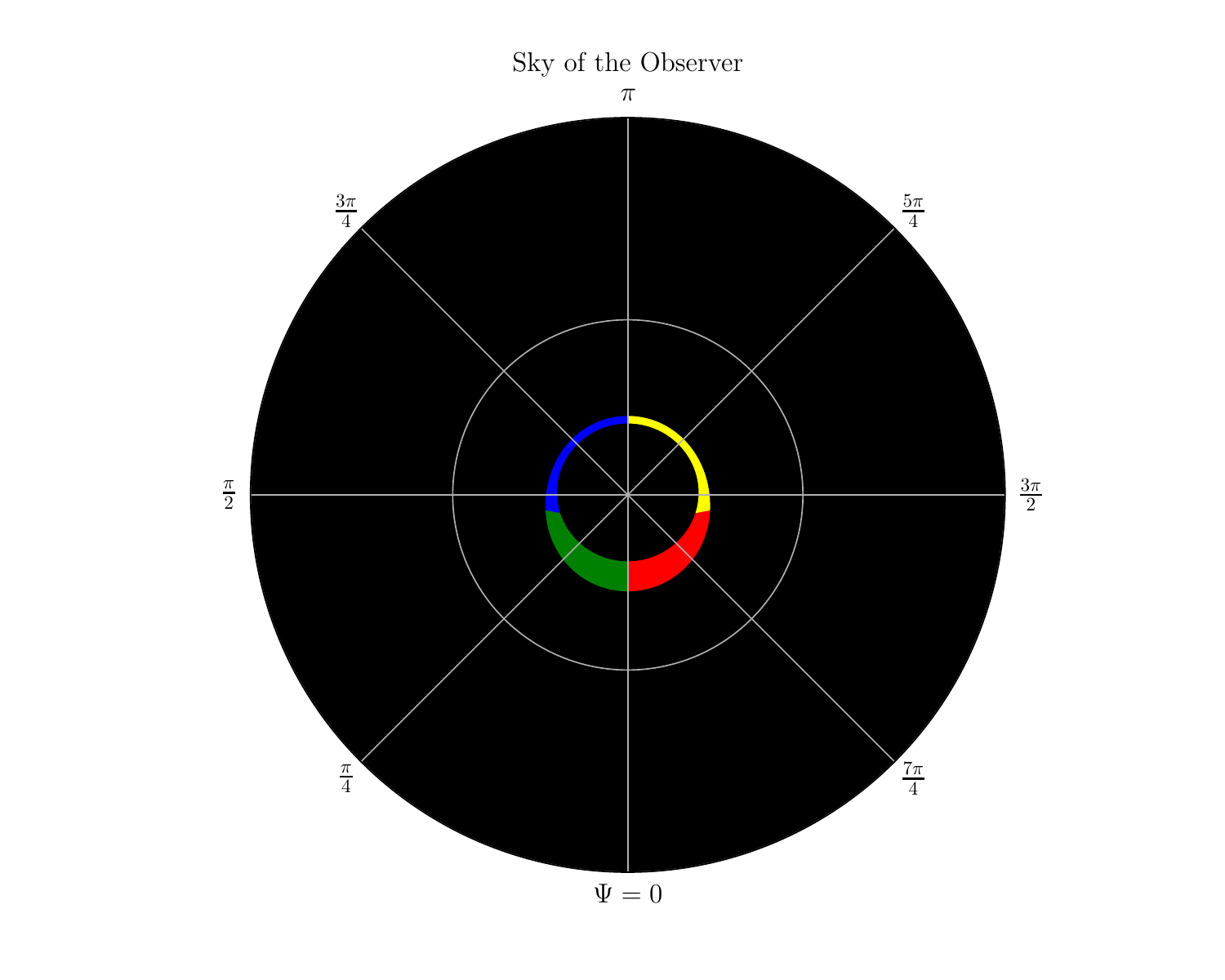} &   \hspace{0.5cm}\includegraphics[width=80mm]{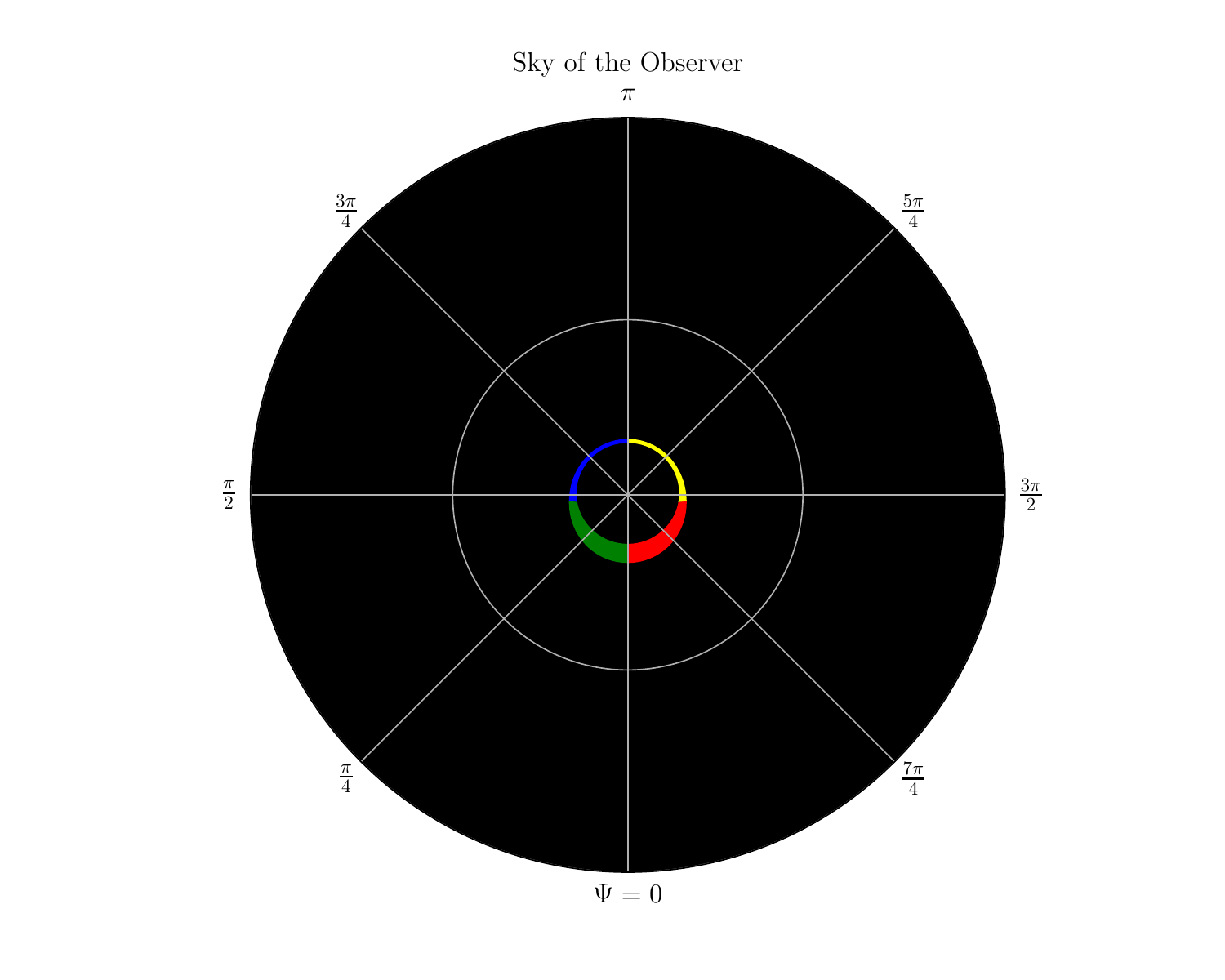} \\
    \hspace{-0.5cm}$E_{O}=\sqrt{2}E_{\text{C}}$ & \hspace{0.5cm} $E_{O}=\sqrt{5}E_{\text{C}}$\\
    \\
    \hspace{-0.5cm}\includegraphics[width=80mm]{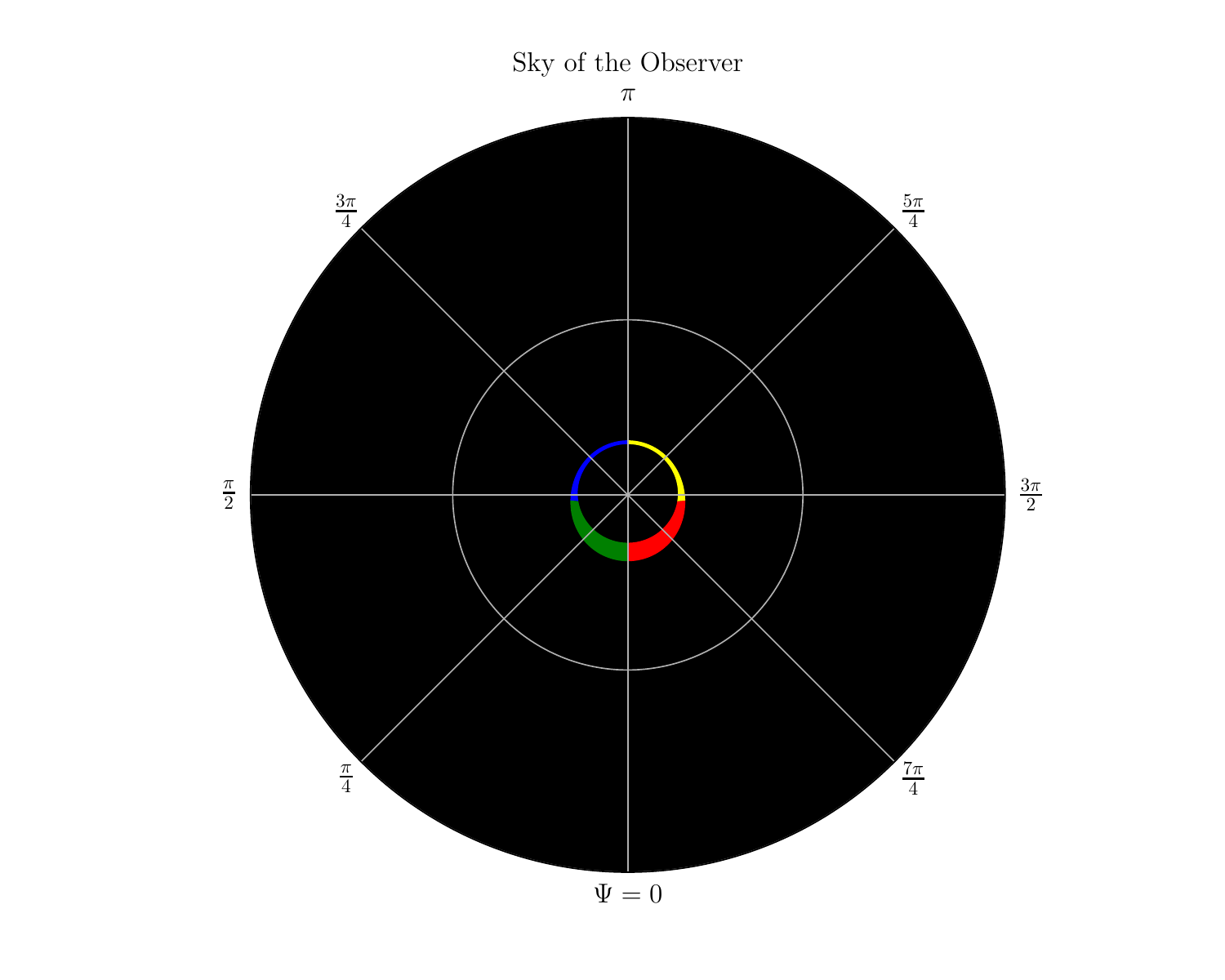} &   \hspace{0.5cm}\includegraphics[width=80mm]{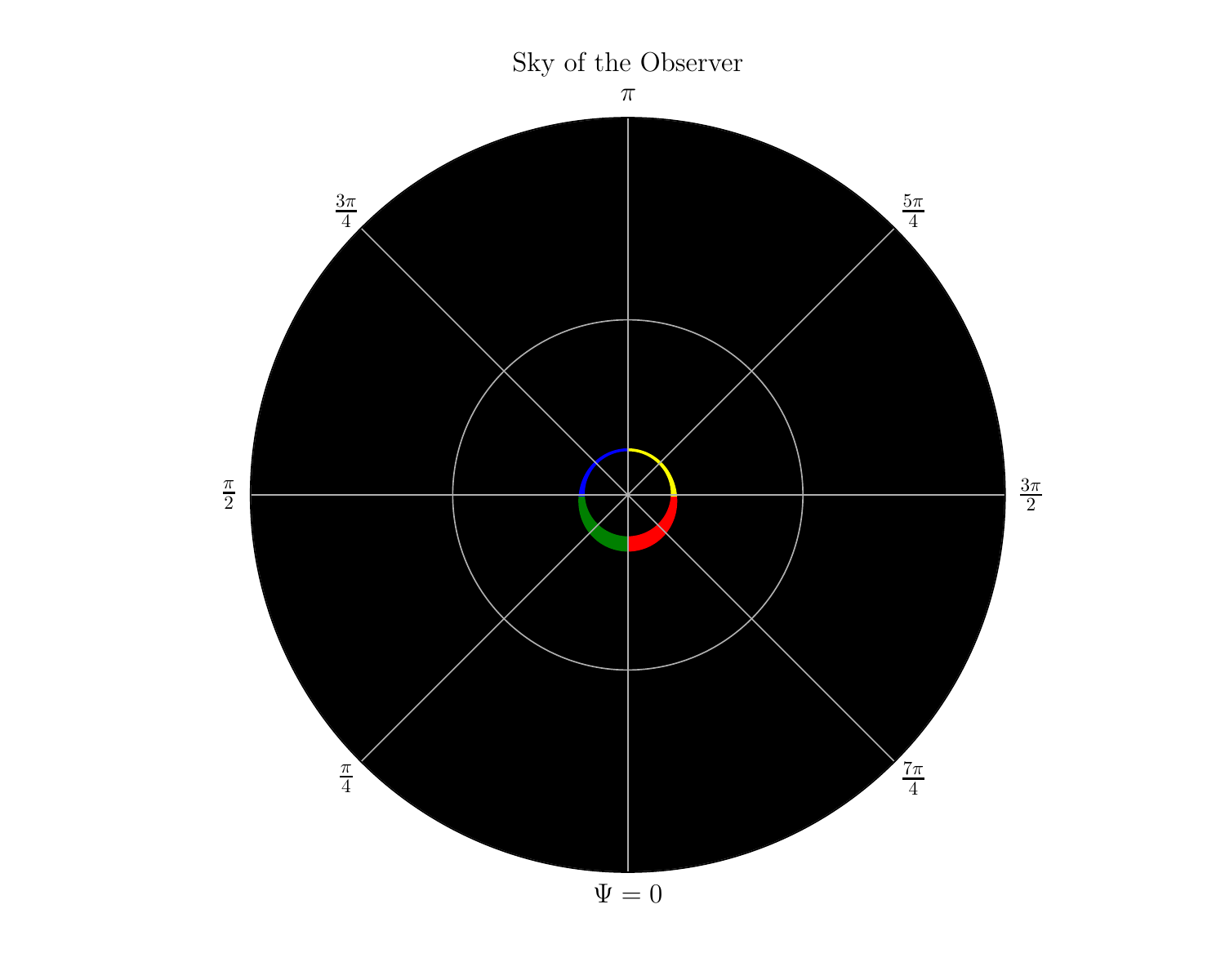} \\
  \end{tabular}
	\caption{Lens maps for the first-order photon rings in the Schwarzschild spacetime for light rays with $E_{O}=\sqrt{53/50}E_{\text{C}}$ (top left panel), $E_{O}=\sqrt{15/14}E_{\text{C}}$ (top right panel), $E_{O}=\sqrt{5}E_{\text{C}}/2$ (middle left panel), $E_{O}=\sqrt{15/8}E_{\text{C}}$ (middle right panel), $E_{O}=\sqrt{2}E_{\text{C}}$ (bottom left panel), and $E_{O}=\sqrt{5}E_{\text{C}}$ (bottom right panel) travelling through an inhomogeneous plasma described by the distribution $E_{\text{pl}}(r,\vartheta)$ given by (\ref{eq:PlasmaEn}) with $\omega_{\text{p}}=2m$. The observer is located at $r_{O}=40m$ and $\vartheta_{O}=\pi/4$ and the luminous disk is located in the equatorial plane between $r_{\text{in}}=2m$ and $r_{\text{out}}=20m$.}
\end{figure*}

\begin{figure*}\label{fig:PR2IHP2m}
  \begin{tabular}{cc}
    \hspace{-0.5cm} $E_{O}=\sqrt{53/50}E_{\text{C}}$ & \hspace{0.5cm}$E_{O}=\sqrt{15/14}E_{\text{C}}$  \\
\\
    \hspace{-0.5cm}\includegraphics[width=80mm]{Schwarzschild_Metric_Photon_Rings_IPlasma_ULR_Calculations_IHP5_IHP6.pdf} &   \hspace{0.5cm}\includegraphics[width=80mm]{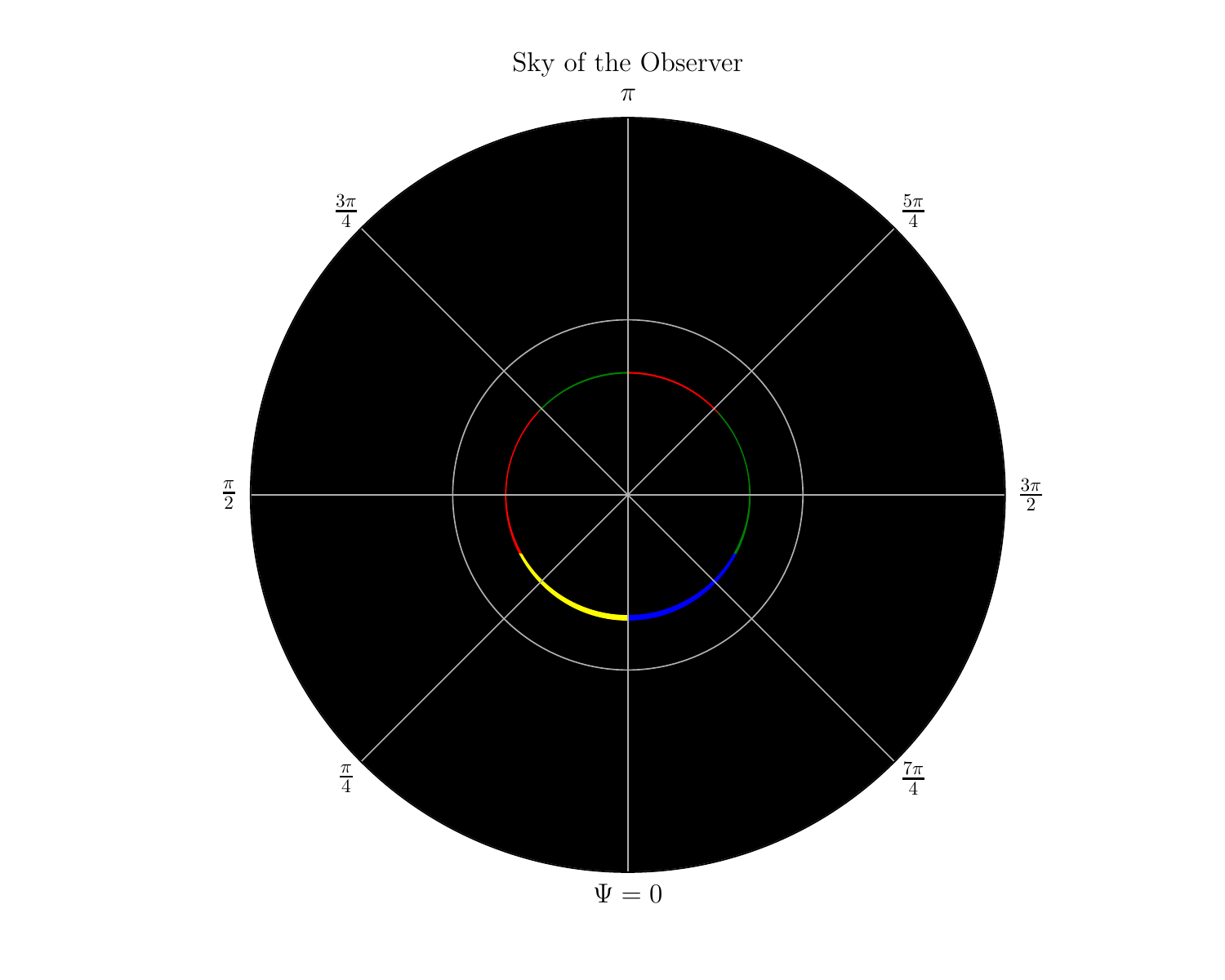} \\
\\
    \hspace{-0.5cm}$E_{O}=\sqrt{5}E_{\text{C}}/2$ & \hspace{0.5cm} $E_{O}=\sqrt{15/8}E_{\text{C}}$\\
\\
    \hspace{-0.5cm}\includegraphics[width=80mm]{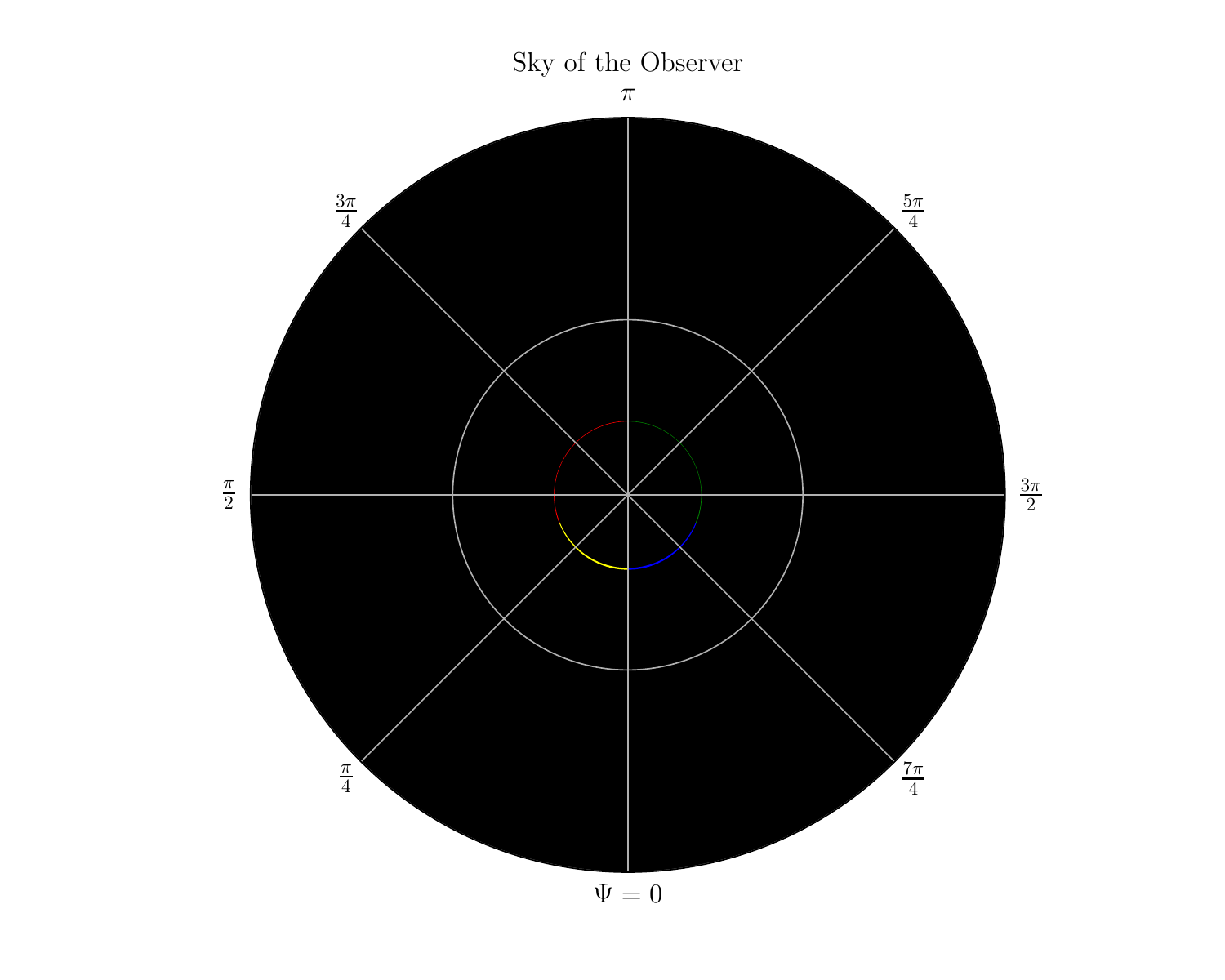} &   \hspace{0.5cm}\includegraphics[width=80mm]{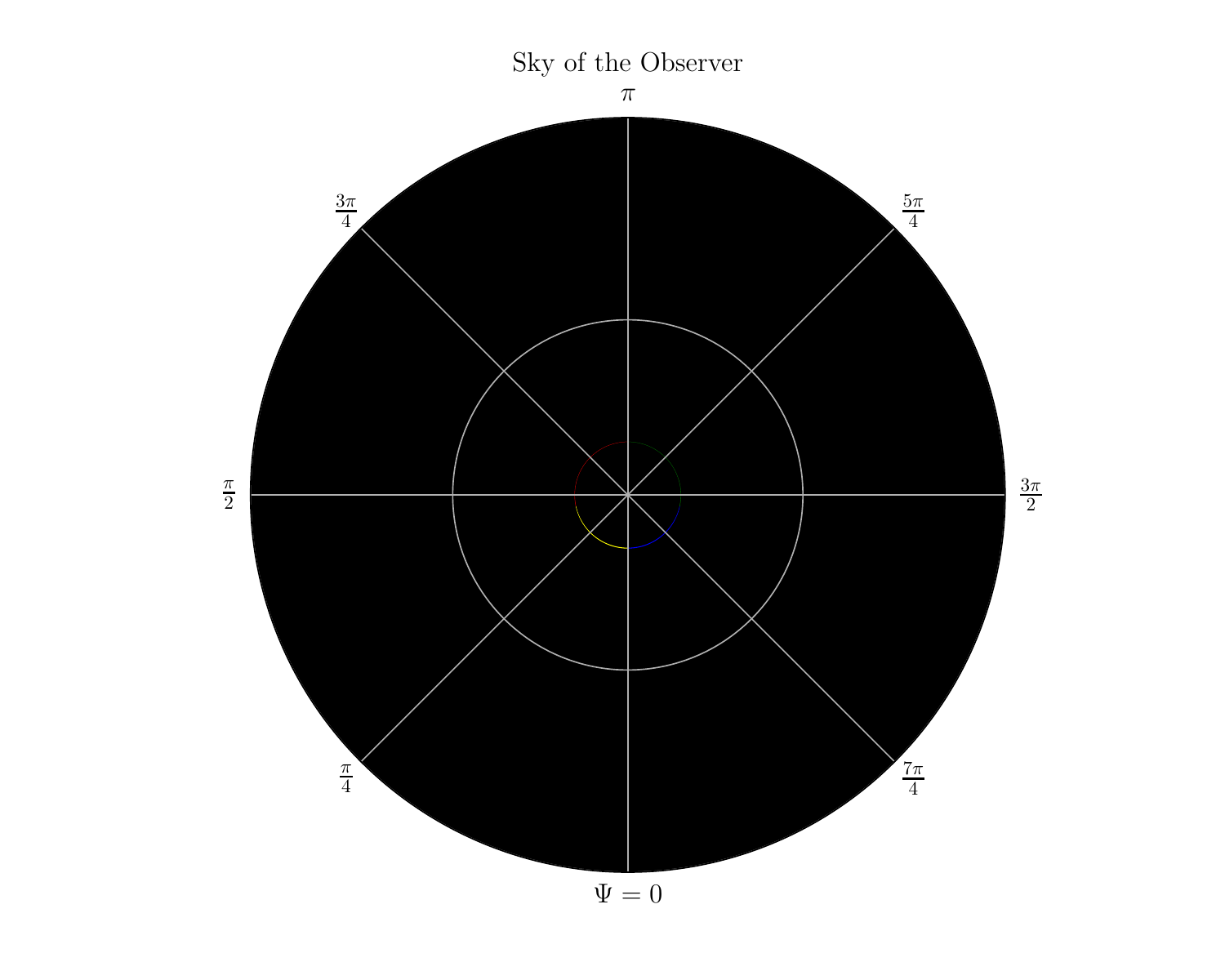} \\
    \hspace{-0.5cm}$E_{O}=\sqrt{2}E_{\text{C}}$ & \hspace{0.5cm} $E_{O}=\sqrt{5}E_{\text{C}}$\\
    \\
    \hspace{-0.5cm}\includegraphics[width=80mm]{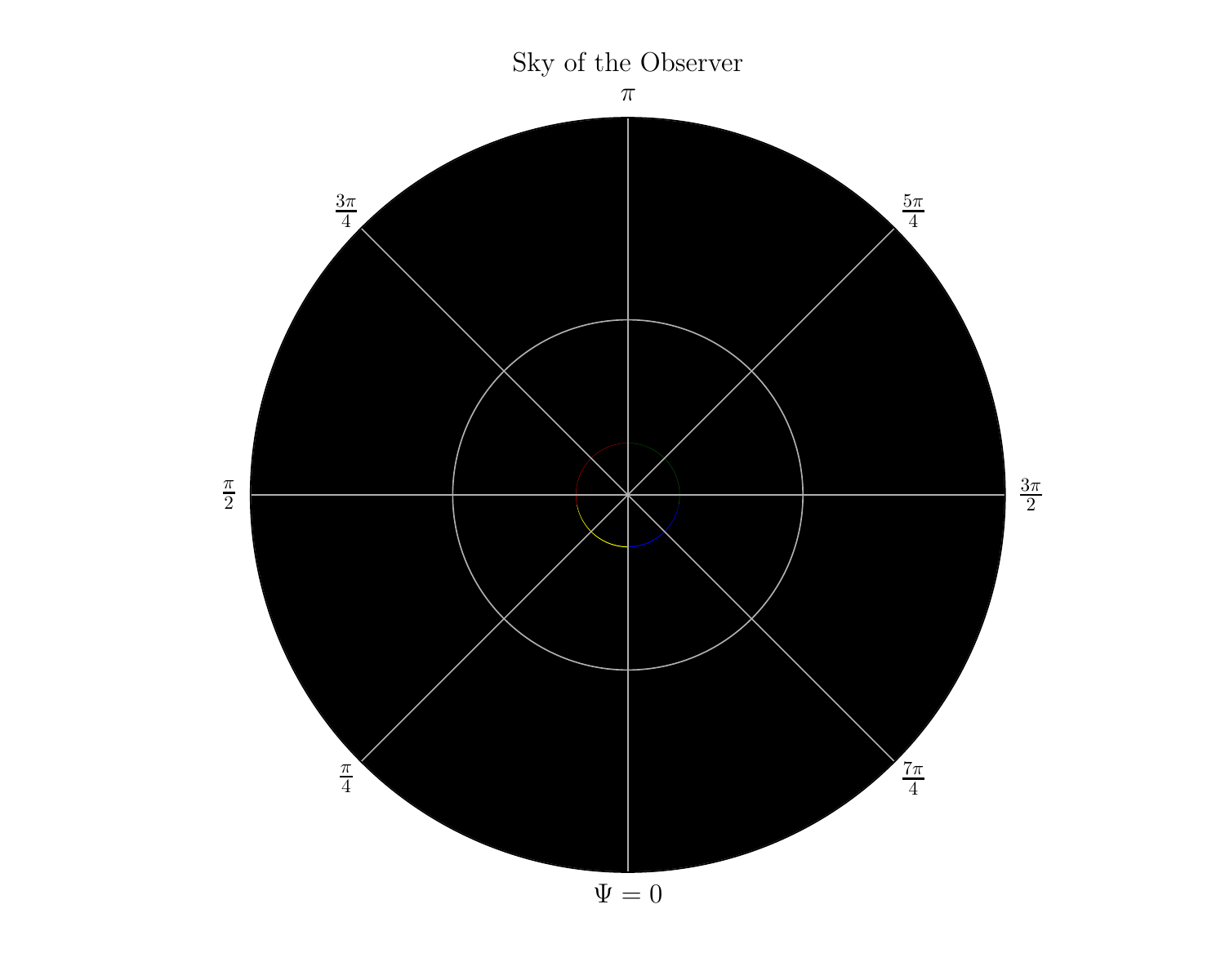} &   \hspace{0.5cm}\includegraphics[width=80mm]{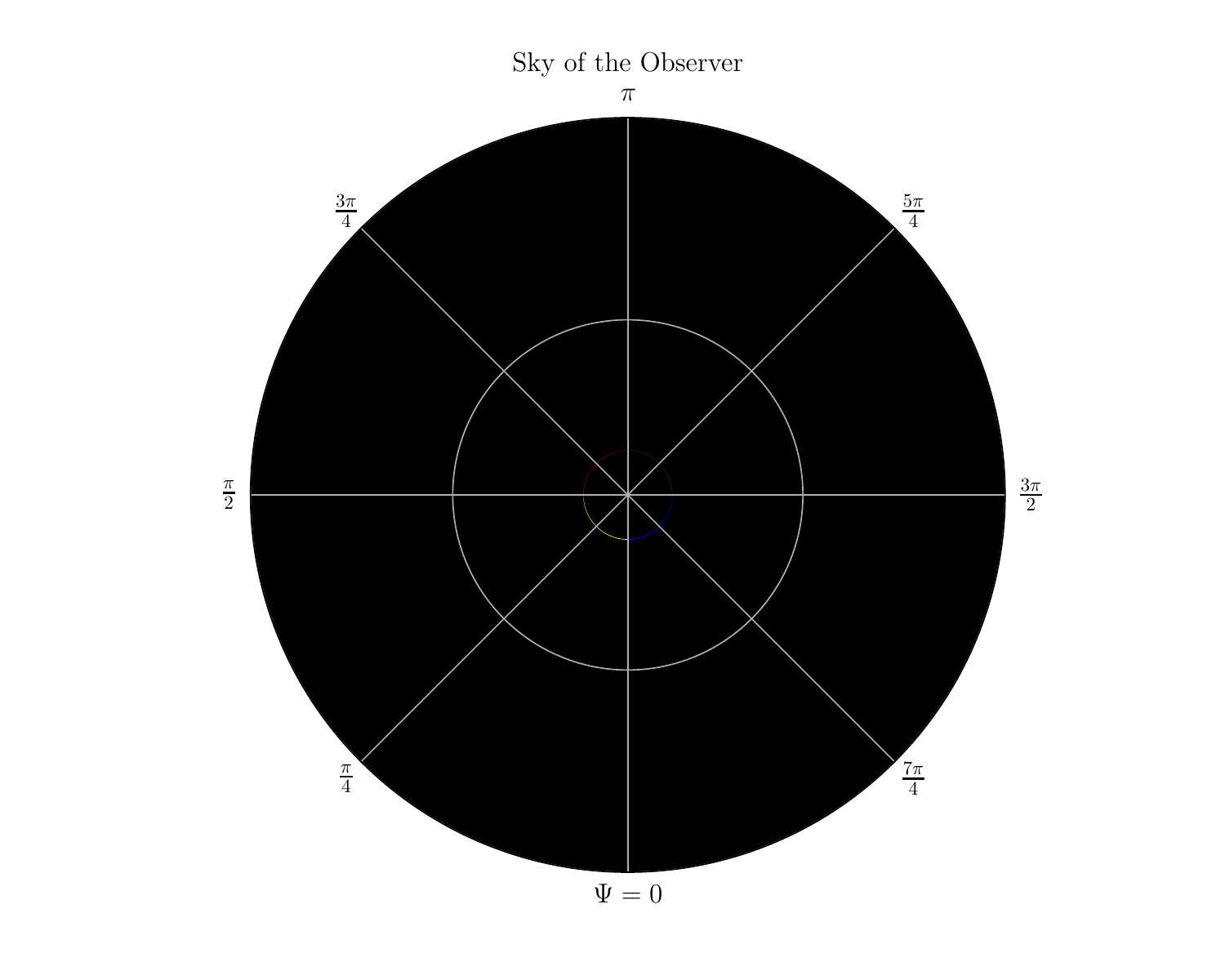} \\
  \end{tabular}
	\caption{Lens maps for the second-order photon rings in the Schwarzschild spacetime for light rays with $E_{O}=\sqrt{53/50}E_{\text{C}}$ (top left panel), $E_{O}=\sqrt{15/14}E_{\text{C}}$ (top right panel), $E_{O}=\sqrt{5}E_{\text{C}}/2$ (middle left panel), $E_{O}=\sqrt{15/8}E_{\text{C}}$ (middle right panel), $E_{O}=\sqrt{2}E_{\text{C}}$ (bottom left panel), and $E_{O}=\sqrt{5}E_{\text{C}}$ (bottom right panel) travelling through an inhomogeneous plasma described by the distribution $E_{\text{pl}}(r,\vartheta)$ given by (\ref{eq:PlasmaEn}) with $\omega_{\text{p}}=2m$. The observer is located at $r_{O}=40m$ and $\vartheta_{O}=\pi/4$ and the luminous disk is located in the equatorial plane between $r_{\text{in}}=2m$ and $r_{\text{out}}=20m$.}
\end{figure*}

For a fast evaluation the calculation of the lens equation (and also the evaluation of the redshift and the travel time in the next two subsections) was implemented in the programming language Julia \cite{Bezanson2017}. Here, using the angular radius of the shadow the calculations were split into two parts. The first part covers the part of the luminous disk between the outer boundary and the radius coordinate of the photon sphere. The second part covers the part of the luminous disk between the photon sphere and the event horizon. 

For the visualisation of the lens equation we will now use the same colour coding as used by Bohn \emph{et al.} \cite{Bohn2015}, however, we adapt it to the luminous disk as illustrated in Fig.~1. We first split the luminous disk into four different quadrants and then colour the quadrant with $0\leq\varphi_{S}\leq\pi/2$ blue, the quadrant with $\pi/2<\varphi_{S}\leq\pi$ green, the quadrant with $\pi<\varphi_{S}\leq 3\pi/2$ red, and the quadrant $3\pi/2<\varphi_{S}$ yellow. Then we plot the calculated lens equations as maps on the observer's celestial sphere. In the following we discuss two different examples. The first are the direct images and the photon rings of first and second order for light rays travelling in vacuum and light rays with $E_{O}=\sqrt{53/50}E_{\text{C}}$ travelling through a homogeneous plasma, and through the inhomogeneous plasma distribution given by Eq.~(\ref{eq:PlasmaEn}) for four different values of $\omega_{\text{p}}$. The second are light rays with six different energies $E_{O}$ travelling through the inhomogeneous plasma distribution given by (\ref{eq:PlasmaEn}) with $\omega_{\text{p}}=2m$. 

We start our discussion with the first case. The lens maps for the direct image are shown in Fig.~2. The lens maps for the photon rings of the first and second order on the other hand are shown in Fig.~3 and Fig.~4, respectively. The different panels in these three figures show the lens maps for light rays travelling in vacuum (top left panel), through a homogeneous plasma ($\omega_{\text{p}}=0$; top right panel), and through the inhomogeneous plasma distribution described by Eq.~(\ref{eq:PlasmaEn}) with $\omega_{\text{p}}=m$ (middle left panel), $\omega_{\text{p}}=2m$ (middle right panel), $\omega_{\text{p}}=3m$ (bottom left panel), and $\omega_{\text{p}}=4m$ (bottom right panel). Please note that here and in all subsequent maps the gray circles mark $\Sigma=\pi/6$.

We start with the direct image for light rays in vacuum. It is shown in the top left panel of Fig.~2. At the centre we see what is commonly referred to as the \emph{inner shadow} of the black hole (please note that in the following due to the lack of a better term we will refer to all shadow areas as inner shadow even if they are already very close to the conventional shadow). In addition, we can easily see that on the celestial sphere of the observer the four quadrants of the luminous disk are clearly separated from each other by the meridian, the antimeridian, and the celestial equator. Here, the area covered by the image of the luminous disk on the southern hemisphere is much larger than on the northern hemisphere. 

The top right panel shows the lens map of the direct image for light rays travelling in a homogeneous plasma. In the map the four quadrants are still clearly separated by the meridian, the antimeridian, and the celestial equator, however, due to the presence of the plasma the structure and the size of the image are different. In the presence of a homogeneous plasma the size of the inner shadow increases and also the image structure extends to higher latitudes. However, the most drastic change is that now the parts of the images on the southern and northern hemispheres cover roughly the same apparent area. In addition, on the northern hemisphere the direct image can already be observed at latitudes much further away from the centre of the map than on the southern hemisphere. 

When we now turn on the plasma parameter $\omega_{\text{p}}$ and set it to $\omega_{\text{p}}=m$ (middle left panel of Fig. 2) we can easily see that the observed features are slightly smaller than for the homogeneous plasma but still much larger than for the direct image created by light rays travelling in vacuum. In addition, we see, that the transitions between the yellow and the red areas on the western hemisphere and the transition between the blue and the green areas on the eastern hemisphere are not aligned with the celestial equator (marked by $\Psi=\pi/2$ on the western hemisphere and $\Psi=3\pi/2$ on the eastern hemisphere) anymore. Instead they are slightly shifted towards the southern hemisphere. When we now increase the plasma parameter to $\omega_{\text{p}}=2m$ (shown in the middle right panel of Fig.~2) this effect increases. When we then increase the plasma parameter first to $\omega_{\text{p}}=3m$ and then to $\omega_{\text{p}}=4m$ the effect again gets stronger. In addition, on the northern hemisphere we observe another effect. On the western hemisphere close to the inner shadow we find images of sources from the green quadrant of the luminous disk. On the other hand on the eastern hemisphere close to the inner shadow we find images of sources from the red quadrant of the luminous disk. Thus these are images of second order. Now one could mean that at the antimeridian we may have a discontinuity, however, a look into the raw data indicates that the transition is still smooth and for light rays crossing $\vartheta=0$ we still have $\varphi_{S}=\pi$. 

Now we turn to the photon rings of first order in Fig.~3. First of all the area covered by the images shrinks. In addition, the dark area in the center of each maps becomes more circular. Otherwise we basically observe the same effects and the only difference is that the sources are located on different quadrants in the luminous disk. In addition, for $\omega_{\text{p}}=3m$ (bottom left panel of Fig.~3) we see that now the transition between images of sources from the blue and the yellow quadrants on the western and eastern hemispheres occur on the southern hemisphere. Furthermore, when we increase the plasma parameter to $\omega_{\text{p}}=4m$ large parts of the first-order photon ring disappear. 

When we now turn to the second order photon rings we again see basically the same effects. Again the area covered with images shrinks. In addition, compared to the first order photon rings the second order photon rings become more circular. In addition, we now already see the transitions between the images of sources in the red quadrant to images from the green quadrant, and from the green quadrant to the red quadrant on the western and eastern hemisphere, respectively, for $\omega_{\text{p}}=2m$ (middle right panel), and for $\omega_{\text{p}}=3m$ we have now in total three additional transitions, however, this time not only between images from sources on the red and green quadrants but also between images of sources on the red and the yellow quadrants. Furthermore, for $\omega_{\text{p}}=4m$ except for a small spot at the meridian (when one zooms in it can be seen roughly in the middle between $\Sigma=0$ and $\Sigma=\pi/6$) the photon ring of second order nearly completely disappeared.

Now we turn to the direct images and the photon rings of first and second order shown in Figs.~5 to 7. In these figures the single panels show direct images and first- and second-order photon rings for light rays travelling through an inhomogeneous plasma described by the distribution given by Eq.~(\ref{eq:PlasmaEn}) with $\omega_{\text{p}}=2m$ for $E_{O}=\sqrt{53/50}E_{\text{C}}$ (top left panel), $E_{O}=\sqrt{15/14}E_{\text{C}}$ (top right panel), $E_{O}=\sqrt{5}E_{\text{C}}/2$ (middle left panel), $E_{O}=\sqrt{15/8}E_{\text{C}}$ (middle right panel), $E_{O}=\sqrt{2}E_{\text{C}}$ (bottom left panel), and $E_{O}=\sqrt{5}E_{\text{C}}$ (bottom right panel). As we can see with increasing energy $E_{O}$ the lens maps for the direct image and the photon rings of first and second order shrink and approach the lens maps for the direct image and the photon rings of first and second order for light rays travelling in vacuum. Here, we can see in the top panels of Fig.~7 that for the second-order photon rings we have for $E_{O}=\sqrt{15/14}E_{\text{C}}$ the same transitions between images of sources on the red quadrant to images of sources on the green quadrant on the western hemisphere and from images of soucres on the green quadrant to images of sources on the red quadrant on the eastern hemisphere as for $E_{O}=\sqrt{53/50}E_{\text{C}}$. Here, the only difference is that for $E_{O}=\sqrt{15/14}E_{\text{C}}$ the transitions are located closer to the antimeridian then for $E_{O}=\sqrt{53/50}E_{\text{C}}$.

Now the most interesting question is how we can use these features to determine the properties of the plasma. Let us start with the characteristics of a homogeneous plasma. As we saw in the presence of a homogeneous plasma the sizes of the inner shadow, the direct image as well as the photon rings of first and second order increase while on the celestial sphere of the observer the boundaries of the four quadrants of the luminous disk remained aligned with the celestial equator, the meridian, and the antimeridian. With increasing photon energy now the sizes of the inner shadow, the direct image and the first- and second-order photon rings will shrink, however, the boundaries of the four quadrants will remain aligned with the celestial equator, the meridian, and the antimeridian. The consequence will be that independent of the frequency of the photons at all energies an image of the same source will remain at the same celestial longitude. In addition, for an observer at $\vartheta_{O}=\pi/4$ when we increase the photon energy the direct image will change its structure from being nearly symmetric with respect to the celestial equator to being smaller on the northern hemisphere and larger on the southern hemisphere. This effect to some degree also exists for the first and second-order photon rings but is less pronounced. Thus when we perform multifrequency observations of a black hole and we observe an inner shadow which shrinks with increasing photon energy, a structural change from a nearly equatorially symmetric direct image for photon energies close to the energy of marginally bound photons to a direct image which dominates on one hemisphere at large energies (note that this effect becomes weaker for observers close to $\vartheta=0$, $\vartheta=\pi$, or the equatorial plane), as well as images of characteristic sources remaining at the same celestial longitude across all frequencies, this will be a strong indicator that the black hole is surrounded by a homogeneous plasma. 

In the case of the inhomogeneous plasma distribution considered in this paper the situation is slightly more complex. However, luckily here we also have several more characteristic features. First of all for the inhomogeneous plasma considered in this paper the size of the inner shadow and the direct image as well as the first- and second-order photon rings still increases compared to the direct image and the first- and second-order photon rings in vacuum. However, here we have an asymmetry between the increase in size of the inner shadow and the increase in size of the direct image and the first- and second-order photon rings. In general the size of the inner shadow is larger than for a homogeneous plasma while the overall size of the direct image and the first- and second-order photon rings is smaller than for a homogeneous plasma. In addition, when we increase the plasma parameter $\omega_{\text{p}}$ the size of the inner shadow increases while the size of the direct image and the first- and second-order photon rings decreases. Furthermore, the inner shadow becomes more asymmetric along the celestial equator while the direct image and the first-order photon rings become more asymmetric along the meridian and the antimeridian. However, the most drastic difference is certainly that for $\omega_{\text{p}}=4m$ large parts of the first- and second order photon rings disappear. In addition, when we increase the plasma parameter $\omega_{\text{p}}$ the boundaries of the quadrants which were aligned with the celestial equator for light rays travelling in vacuum or through a homogeneous plasma move further away from the celestial equator, in our case on the southern hemisphere, and do not form straight lines anymore. Now when we perform multifrequency observations these effects all reduce with increasing photon energy and since the strength of these effects varies with the properties of the inhomogeneous plasma, in our case characterised by the plasma parameter $\omega_{\text{p}}$, when we record energy profiles of these changes and combine them it has the potential to allow us to accurately determine these properties. 

However, while the results of our analysis speak a relatively clear language we always have to remind ourselves that we only considered two very idealised and characteristic plasma distributions. However, there are also other profiles which can lead to very different effects. E.g., Perlick, Tsupko, and Bisnovatyi-Kogan \cite{Perlick2015} considered a Schwarzschild black hole accreting a spherically symmetric plasma and found that the angular size of the shadow generally decreases compared to the angular size of the shadow in vacuum. Thus the effects of a plasma on the observable structure can be completely different for different types of plasma distributions. In particular, we can safely assume that the idealised plasma distributions which we considered in this paper, may only describe certain parts of a real accretion disk and thus for obtaining the correct plasma profiles for different parts of an accretion disk we have to combine different models whose effects may compensate or enhance each other. 

In addition, there are also other effects which may affect the characterisation of a plasma in the accretion disk around a supermassive black hole. While in this paper we considered a luminous disk stretching from the horizon to an outer radius coordinate $r_{\text{out}}$ and thus the photon rings are projected on top of the direct image, when we have real accretion disks they may not extend to the horizon or even the photon sphere and thus they may only partially overlap or not at all. In this case using a wrong assumption for the accretion disk may strongly affect the determination of the plasma properties. Furthermore, even when the first- and second-order photon rings exist and would be potentially observable, when the sensitivity of our instruments and their angular resolution is not high enough we may not be able to observe them. Misinterpreting this as an effect of an inhomogeneous plasma then bears the risk of deriving completely wrong plasma properties. Thus even when we manage to perform multifrequency observations of the direct image as well as the photon rings determining the properties of the plasma surrounding the target black hole will still require a careful analysis of all observed features.

\subsection{The Redshift}
\begin{figure*}\label{fig:RSDI}
  \begin{tabular}{cc}
    \hspace{-0.5cm}Vacuum & \hspace{0.5cm} Homogeneous Plasma ($\omega_{\text{p}}=0$)\\
\\
    \hspace{-0.5cm}\includegraphics[width=80mm]{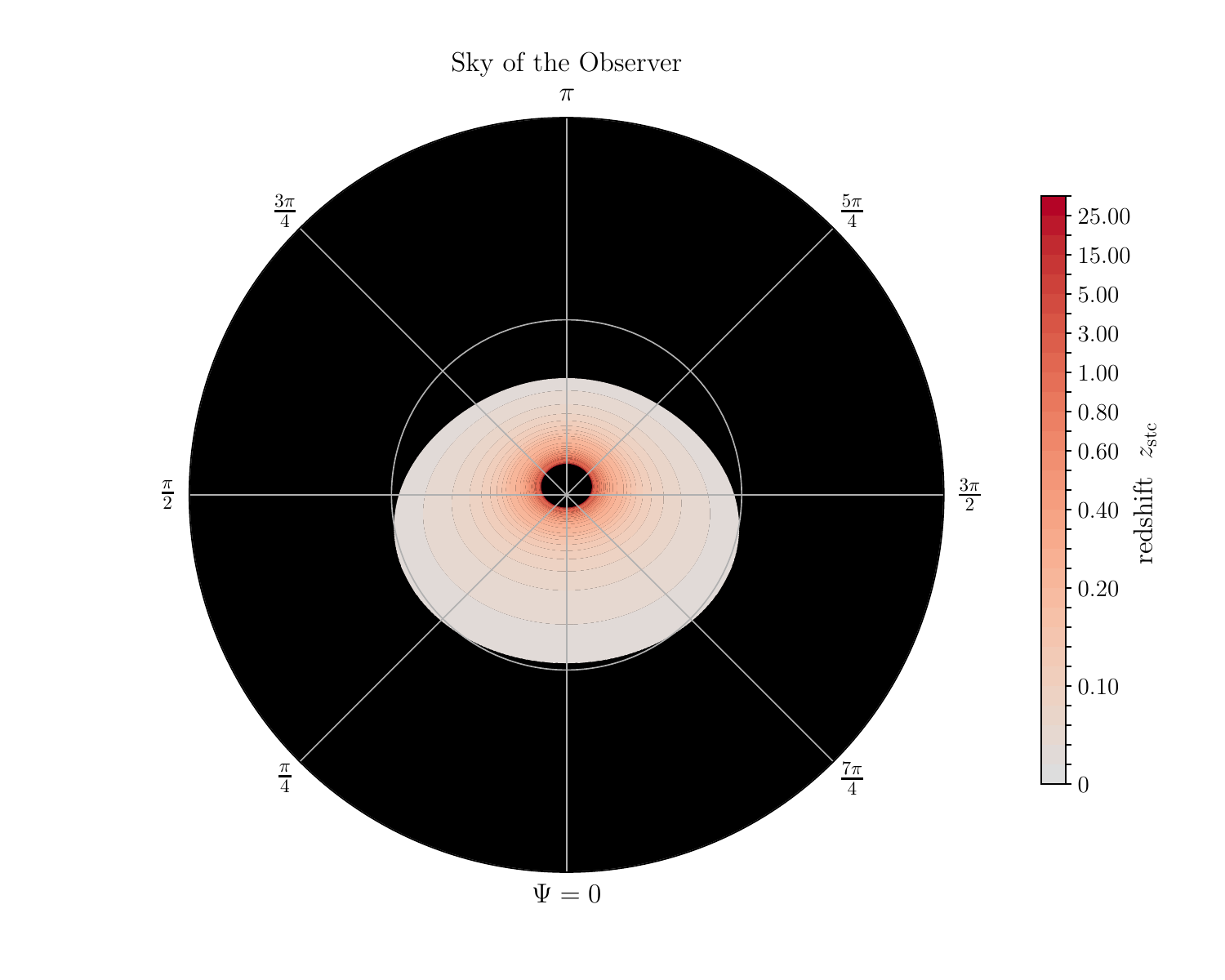} &   \hspace{0.5cm}\includegraphics[width=80mm]{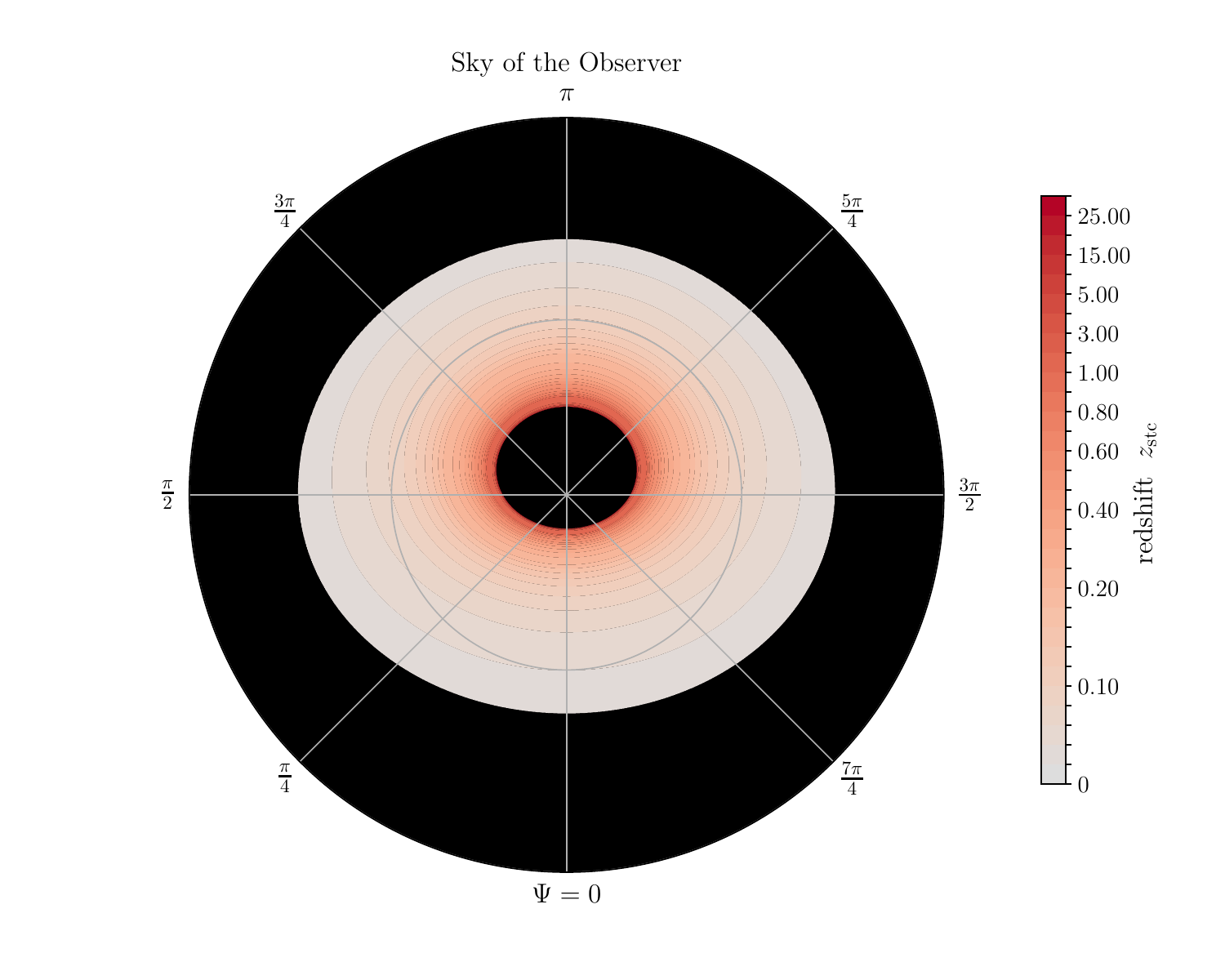} \\
\\
    \hspace{-0.5cm}Inhomogeneous Plasma with $\omega_{\text{p}}=m$ & \hspace{0.5cm} Inhomogeneous Plasma with $\omega_{\text{p}}=2m$\\
\\
    \hspace{-0.5cm}\includegraphics[width=80mm]{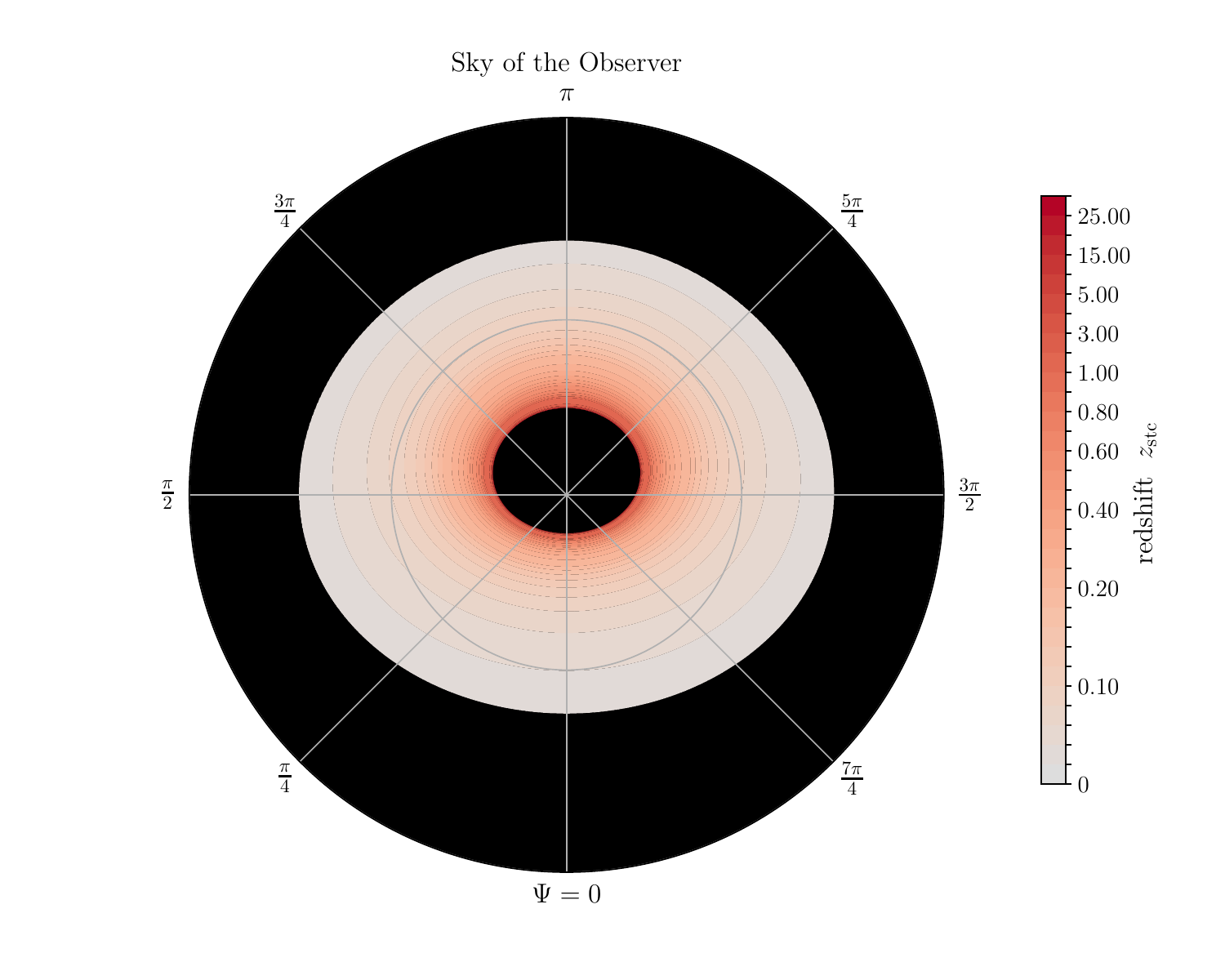} &   \hspace{0.5cm}\includegraphics[width=80mm]{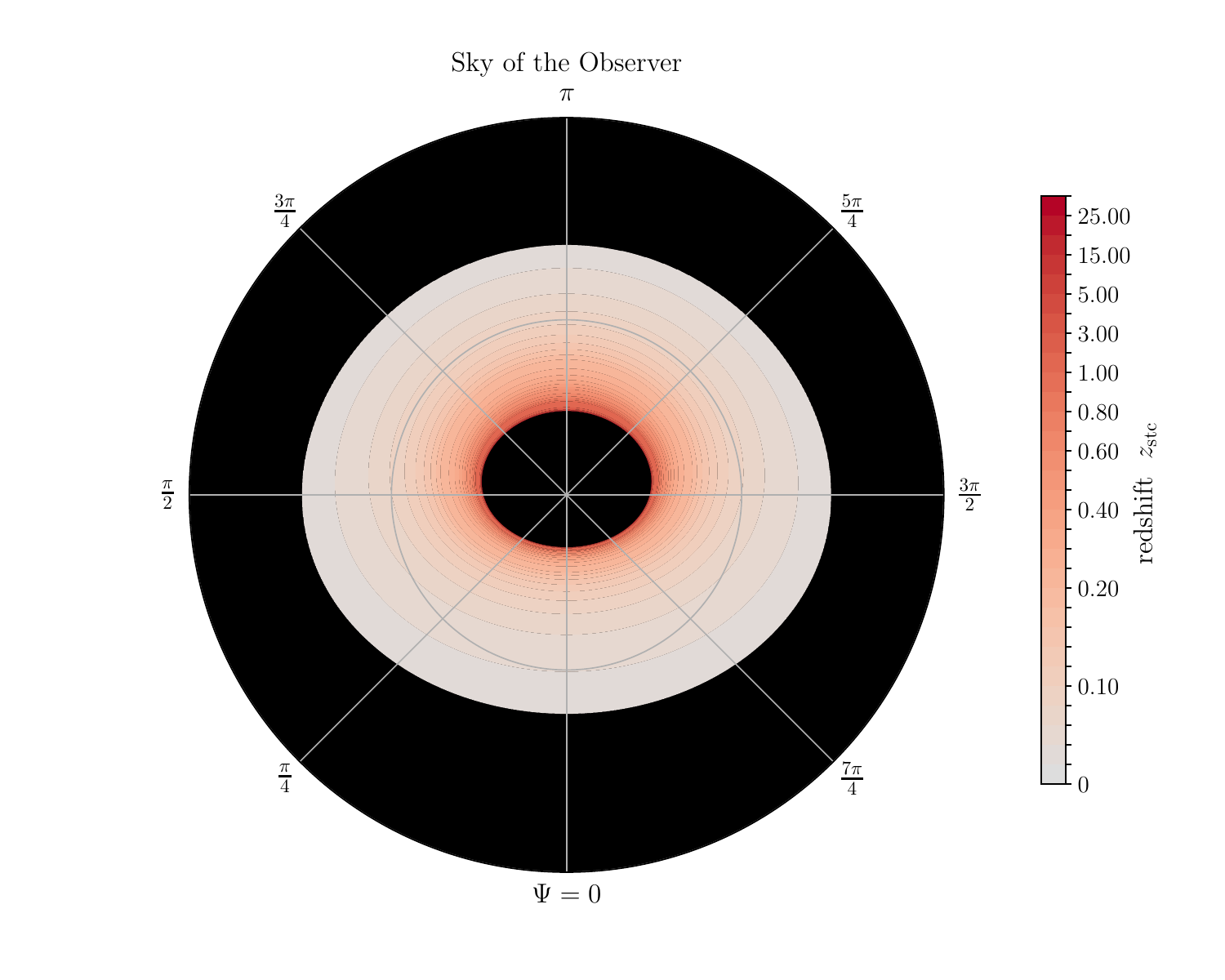} \\
    \hspace{-0.5cm}Inhomogeneous Plasma with $\omega_{\text{p}}=3m$ & \hspace{0.5cm} Inhomogeneous Plasma with $\omega_{\text{p}}=4m$\\
    \\
    \hspace{-0.5cm}\includegraphics[width=80mm]{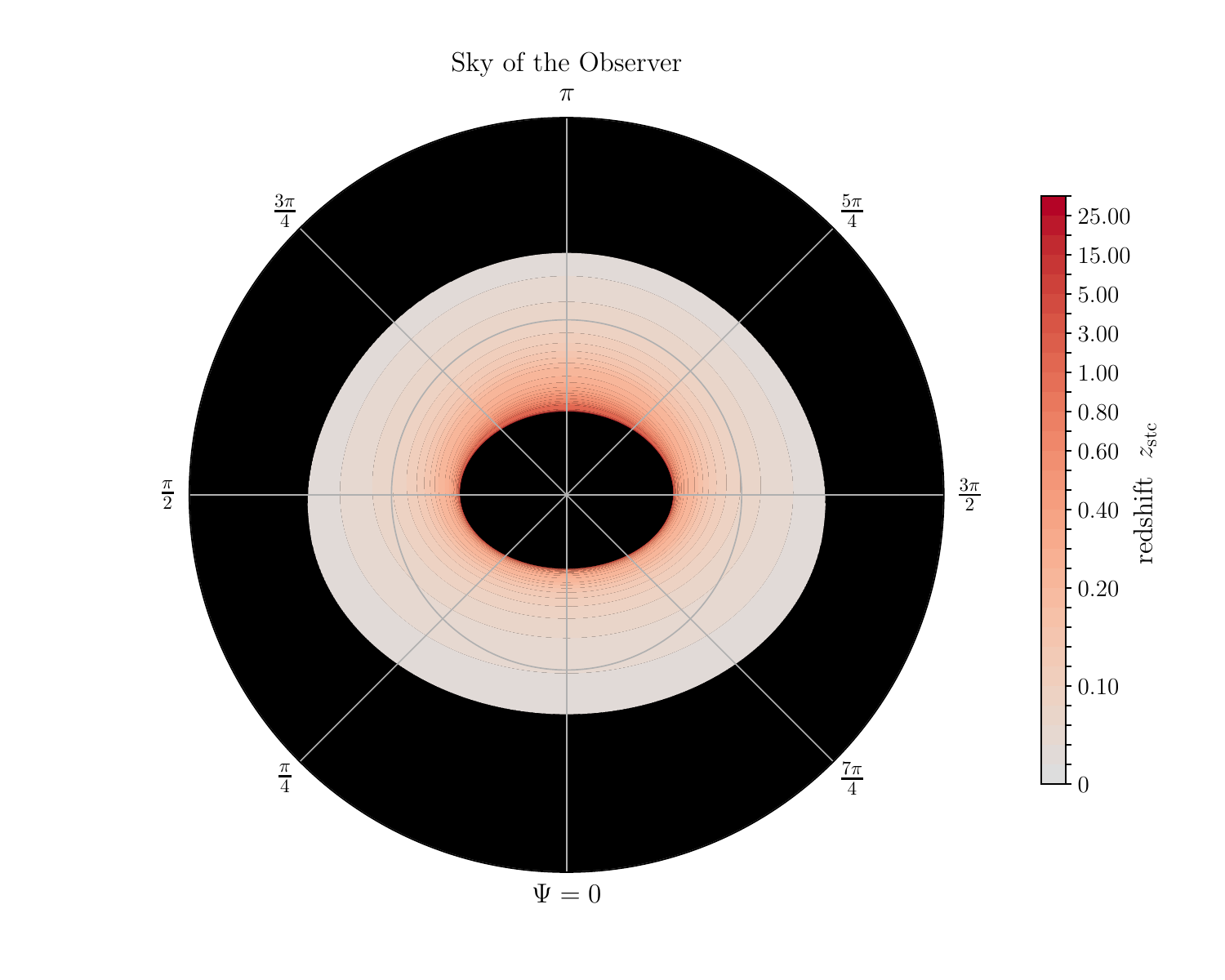} &   \hspace{0.5cm}\includegraphics[width=80mm]{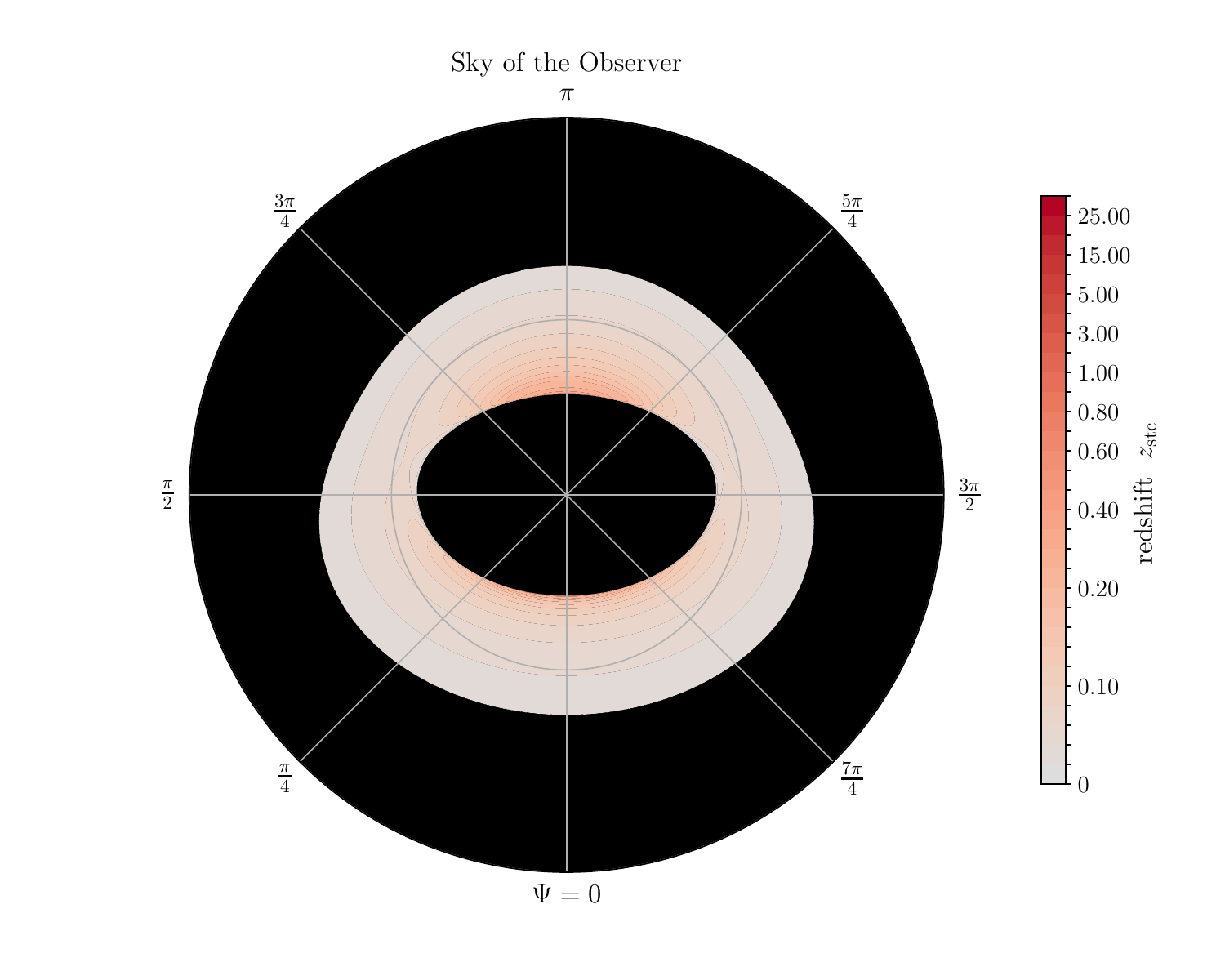} \\
  \end{tabular}
	\caption{Redshift maps for the direct images in the Schwarzschild spacetime for light rays travelling in vacuum (top left panel), through a homogeneous plasma (top right panel), and through an inhomogeneous plasma described by the distribution $E_{\text{pl}}(r,\vartheta)$ given by (\ref{eq:PlasmaEn}) with $\omega_{\text{p}}=m$ (middle left panel), $\omega_{\text{p}}=2m$ (middle right panel), $\omega_{\text{p}}=3m$ (bottom left panel), and $\omega_{\text{p}}=4m$ (bottom right panel). The observer is located at $r_{O}=40m$ and $\vartheta_{O}=\pi/4$ and the luminous disk is located in the equatorial plane between $r_{\text{in}}=2m$ and $r_{\text{out}}=20m$. For the light rays travelling through one of the plasmas the energy measured at the position of the observer is $E_{O}=\sqrt{53/50}E_{\text{C}}$.}
\end{figure*}

\begin{figure*}\label{fig:RSPhotonRing1E1}
  \begin{tabular}{cc}
    \hspace{-0.5cm}Vacuum & \hspace{0.5cm} Homogeneous Plasma ($\omega_{\text{p}}=0$)\\
\\
    \hspace{-0.5cm}\includegraphics[width=80mm]{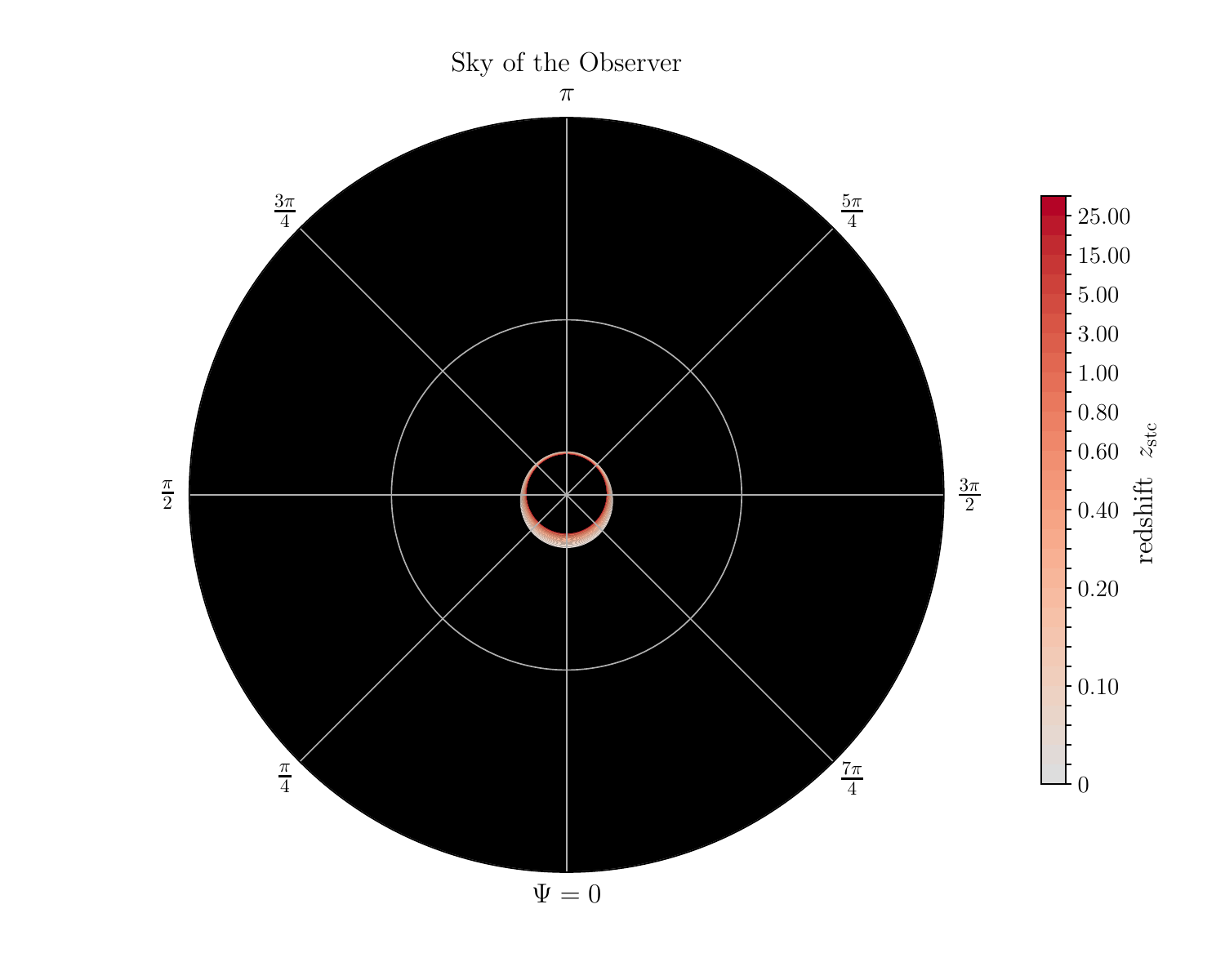} &   \hspace{0.5cm}\includegraphics[width=80mm]{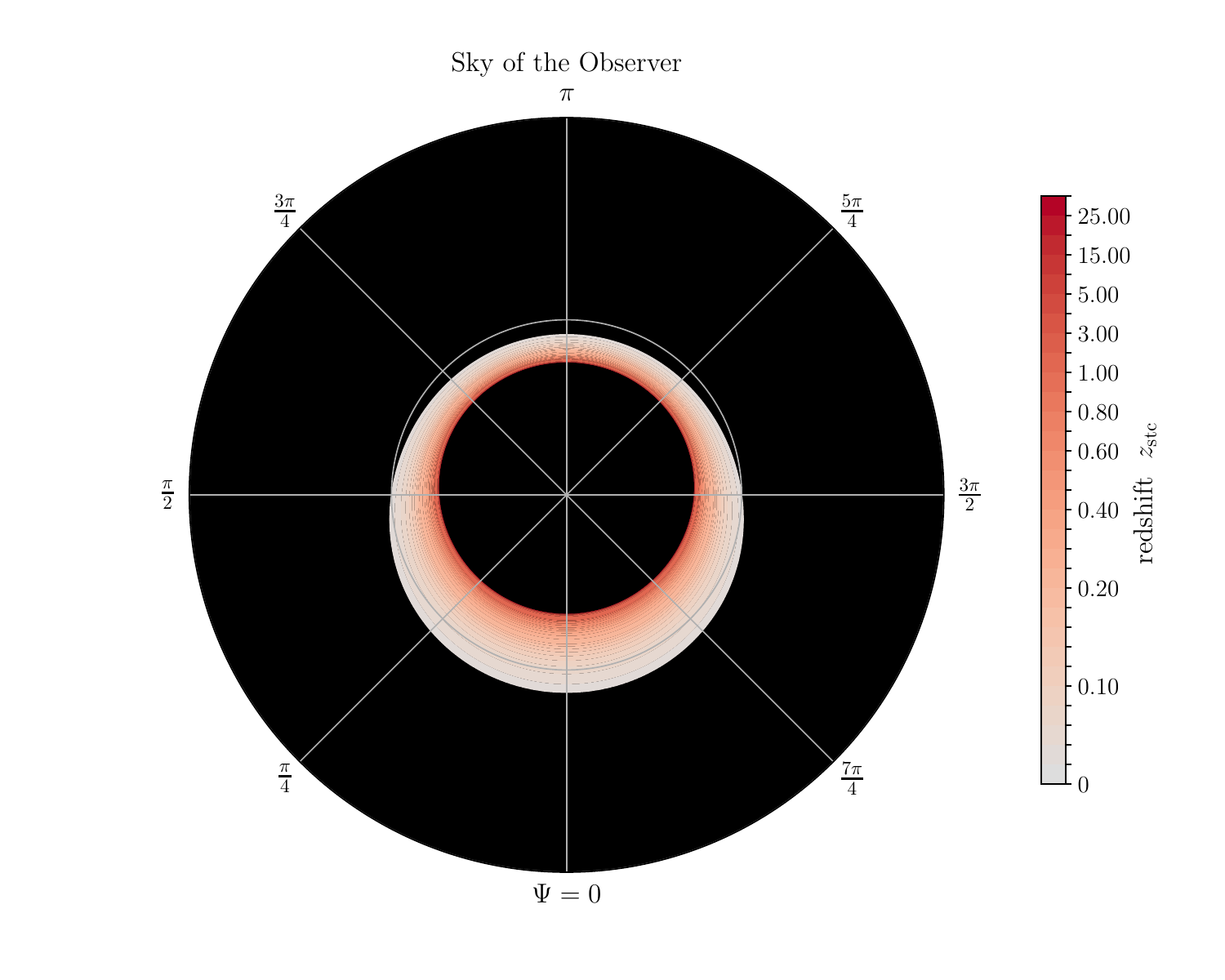} \\
\\
    \hspace{-0.5cm}Inhomogeneous Plasma with $\omega_{\text{p}}=m$ & \hspace{0.5cm} Inhomogeneous Plasma with $\omega_{\text{p}}=2m$\\
\\
    \hspace{-0.5cm}\includegraphics[width=80mm]{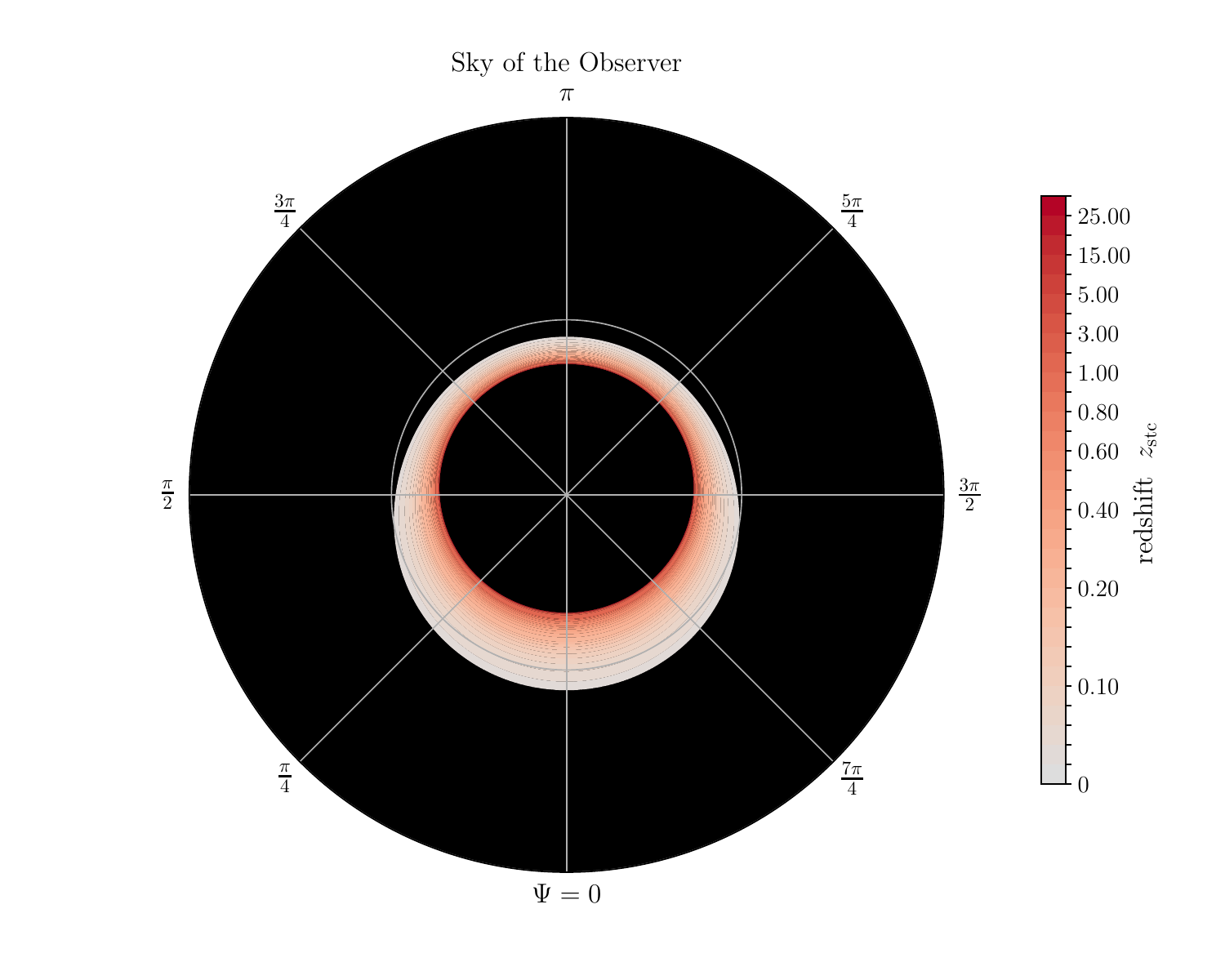} &   \hspace{0.5cm}\includegraphics[width=80mm]{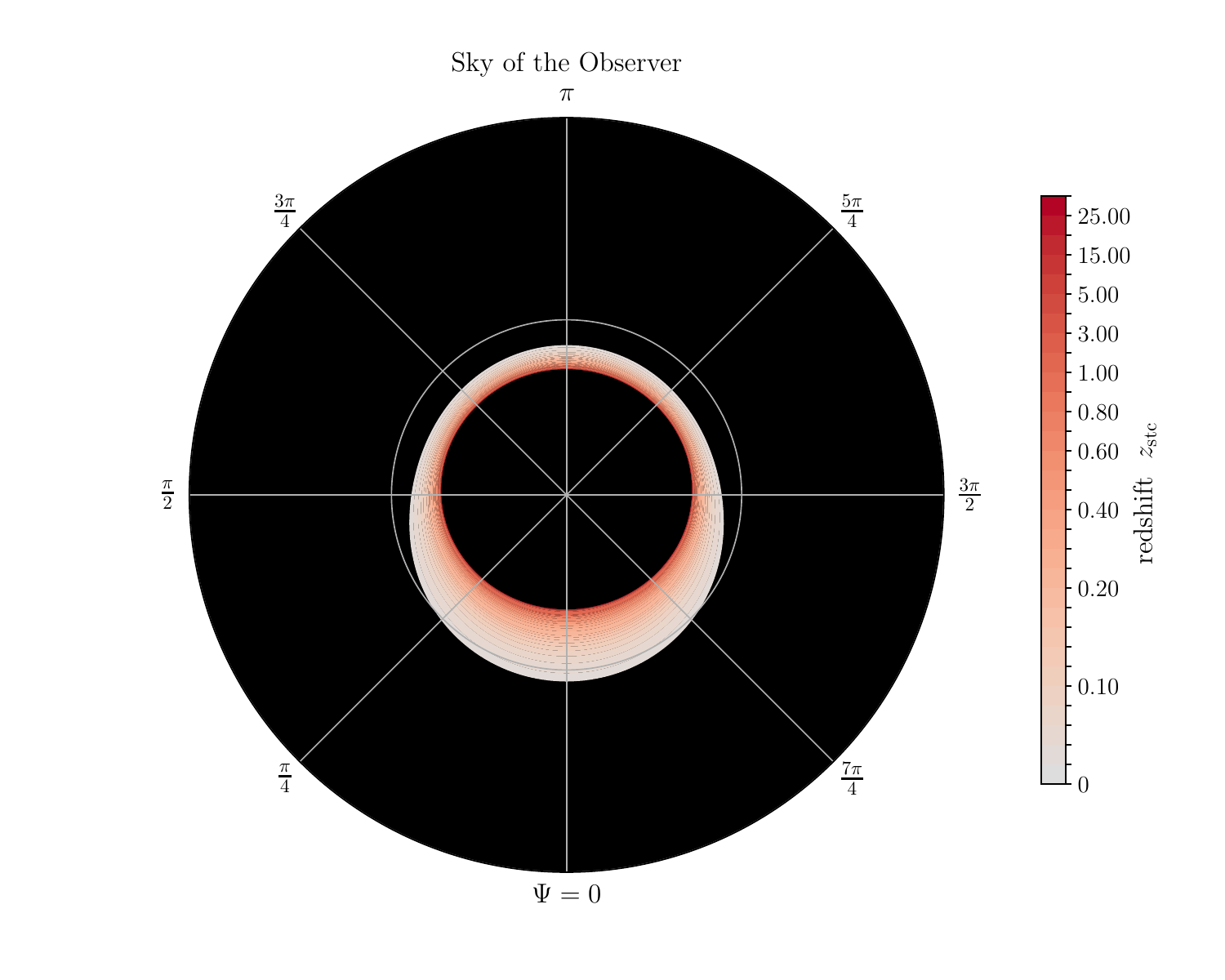} \\
    \hspace{-0.5cm}Inhomogeneous Plasma with $\omega_{\text{p}}=3m$ & \hspace{0.5cm} Inhomogeneous Plasma with $\omega_{\text{p}}=4m$\\
    \\
    \hspace{-0.5cm}\includegraphics[width=80mm]{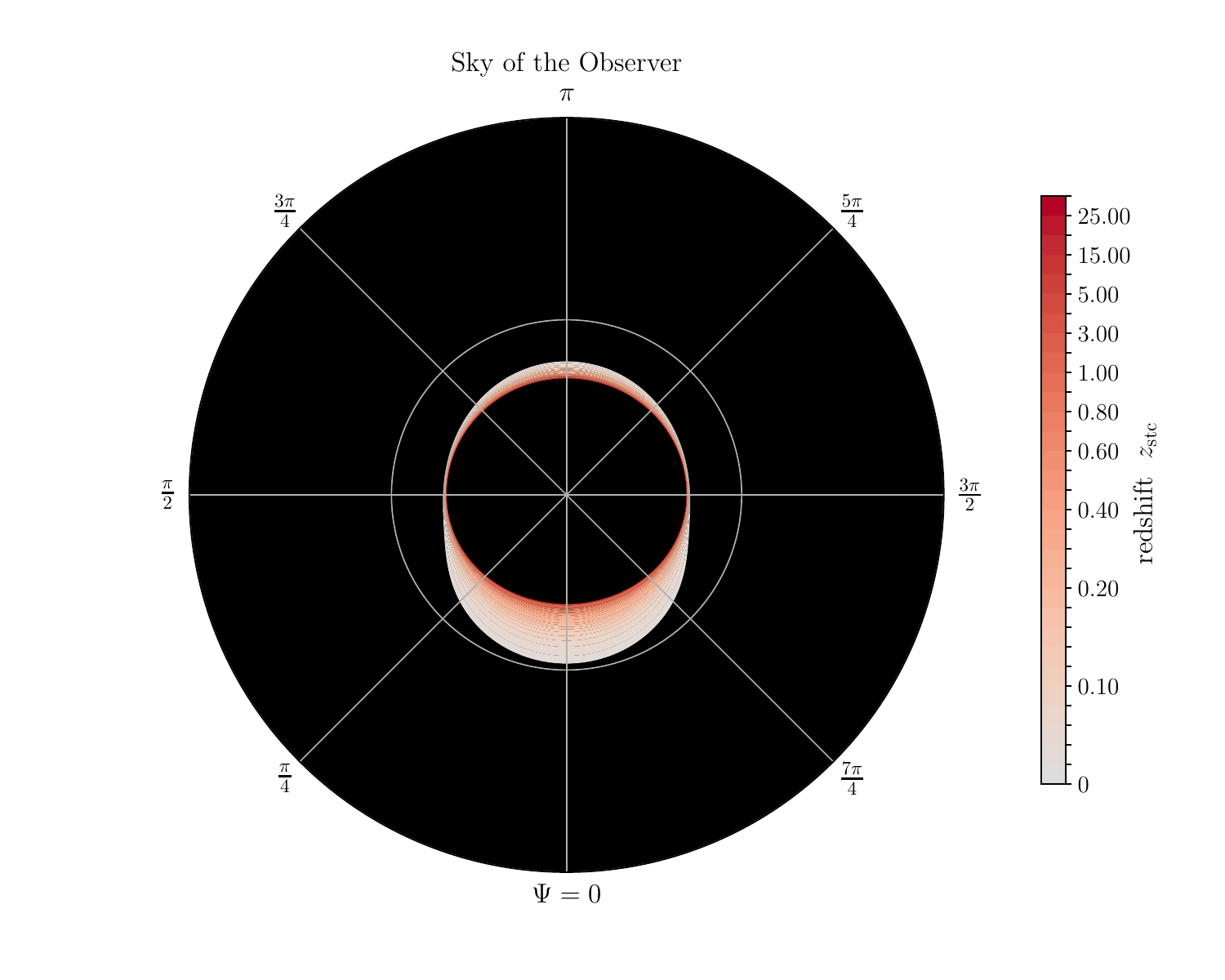} &   \hspace{0.5cm}\includegraphics[width=80mm]{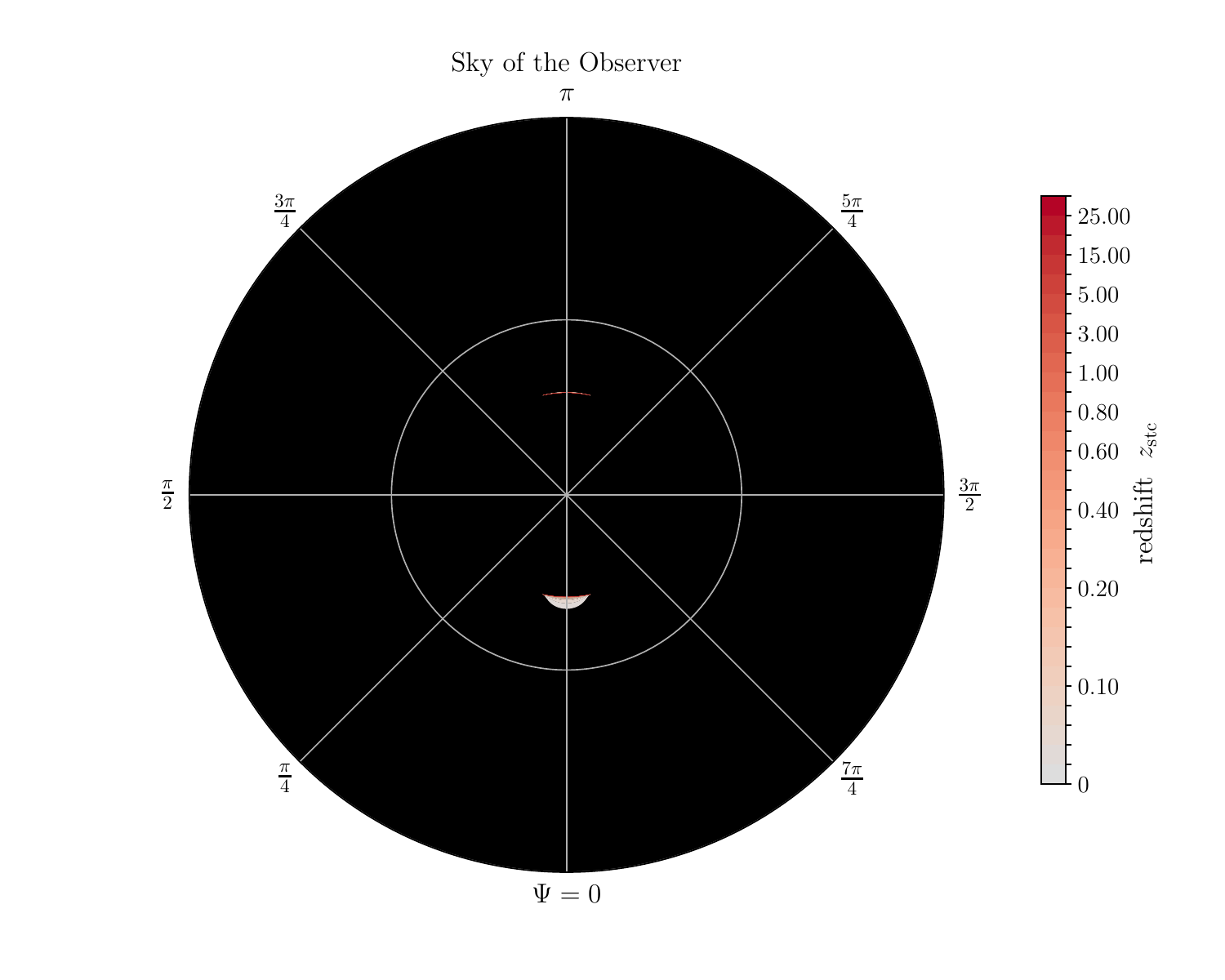} \\
  \end{tabular}
	\caption{Redshift maps for the first-order photon rings in the Schwarzschild spacetime for light rays travelling in vacuum (top left panel), through a homogeneous plasma (top right panel), and through an inhomogeneous plasma described by the distribution $E_{\text{pl}}(r,\vartheta)$ given by (\ref{eq:PlasmaEn}) with $\omega_{\text{p}}=m$ (middle left panel), $\omega_{\text{p}}=2m$ (middle right panel), $\omega_{\text{p}}=3m$ (bottom left panel), and $\omega_{\text{p}}=4m$ (bottom right panel). The observer is located at $r_{O}=40m$ and $\vartheta_{O}=\pi/4$ and the luminous disk is located in the equatorial plane between $r_{\text{in}}=2m$ and $r_{\text{out}}=20m$. For the light rays travelling through one of the plasmas the energy measured at the position of the observer is $E_{O}=\sqrt{53/50}E_{\text{C}}$.}
\end{figure*}

\begin{figure*}\label{fig:RSPhotonRing2E1}
  \begin{tabular}{cc}
    \hspace{-0.5cm}Vacuum & \hspace{0.5cm} Homogeneous Plasma ($\omega_{\text{p}}=0$)\\
\\
    \hspace{-0.5cm}\includegraphics[width=80mm]{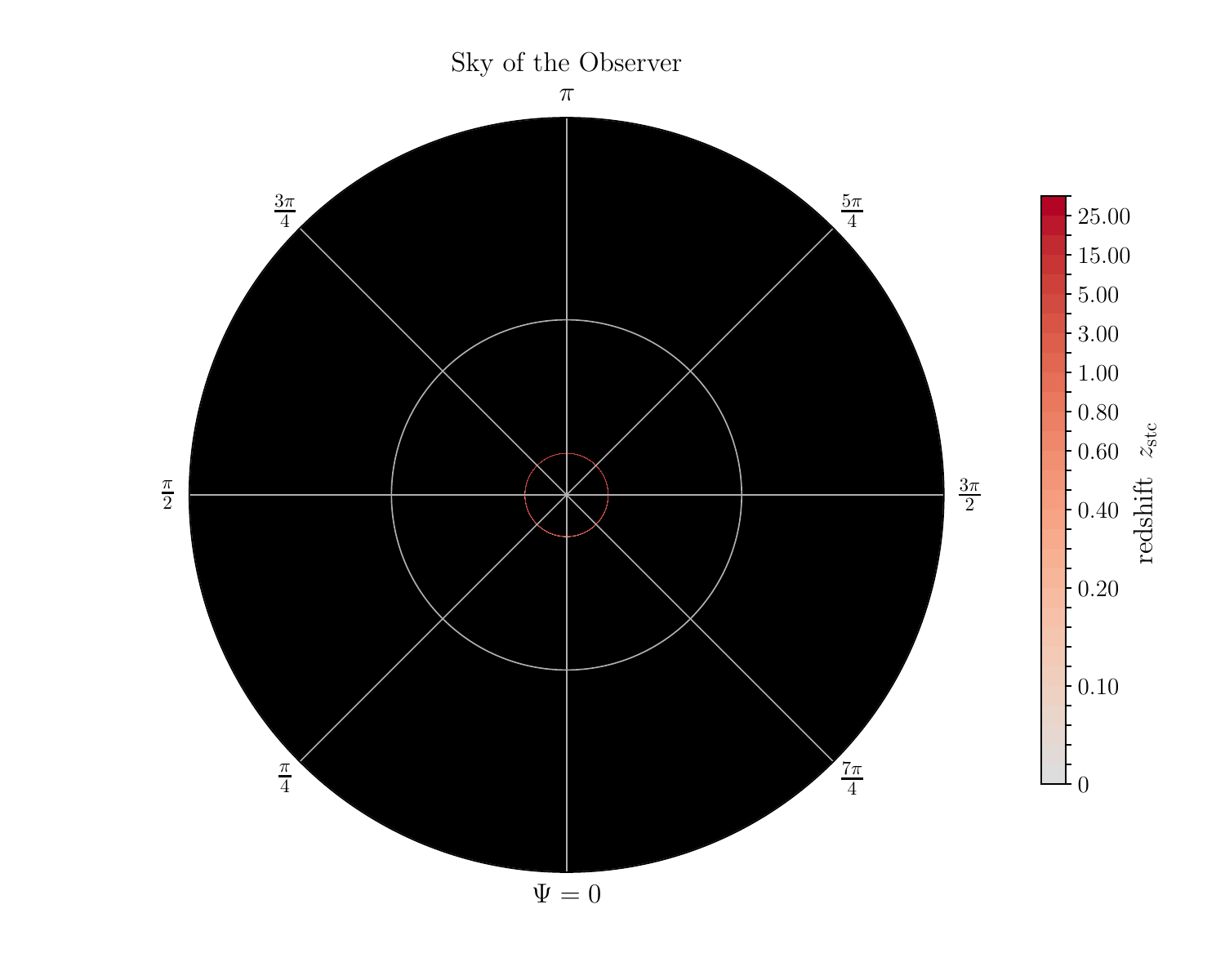} &   \hspace{0.5cm}\includegraphics[width=80mm]{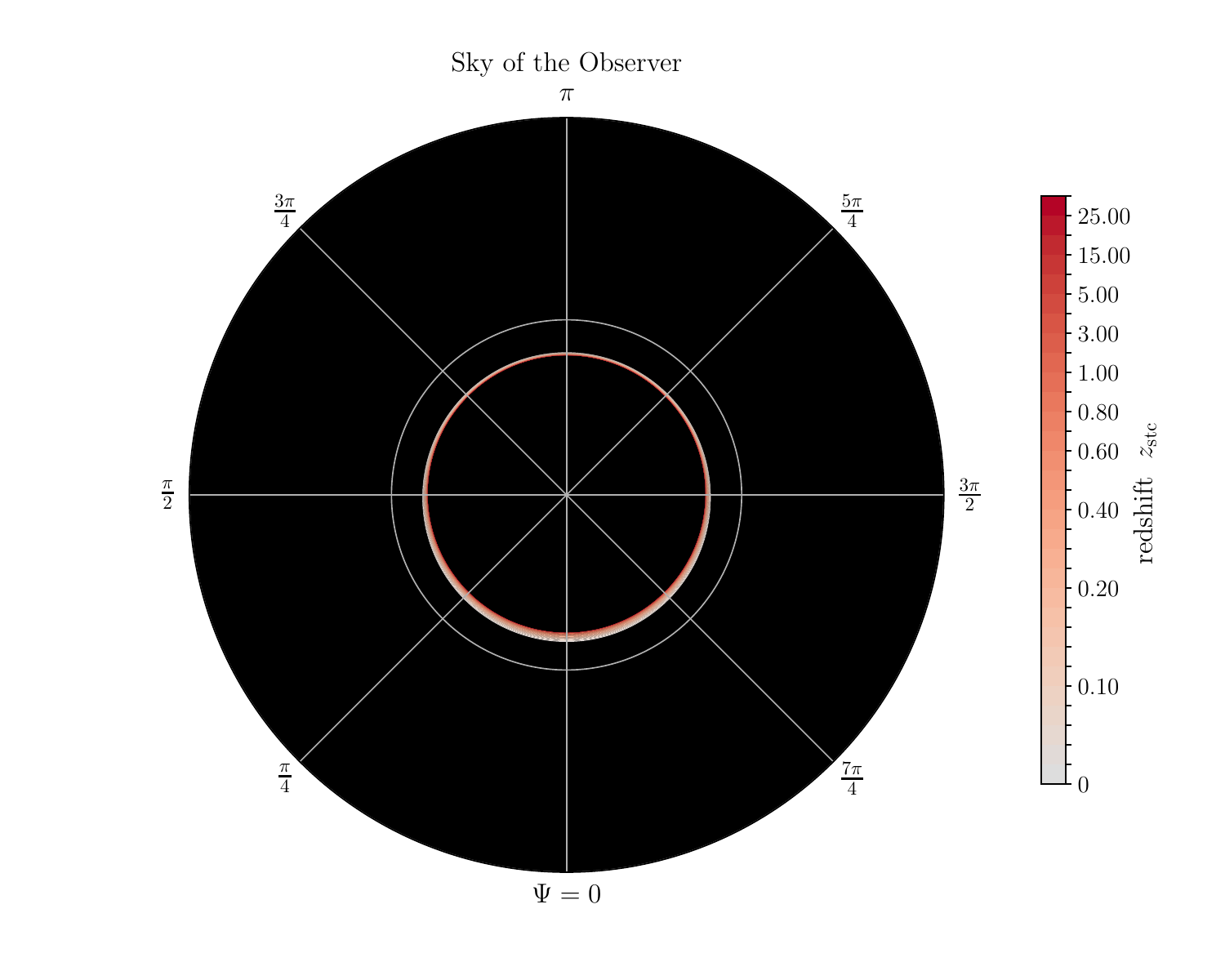} \\
\\
    \hspace{-0.5cm}Inhomogeneous Plasma with $\omega_{\text{p}}=m$ & \hspace{0.5cm} Inhomogeneous Plasma with $\omega_{\text{p}}=2m$\\
\\
    \hspace{-0.5cm}\includegraphics[width=80mm]{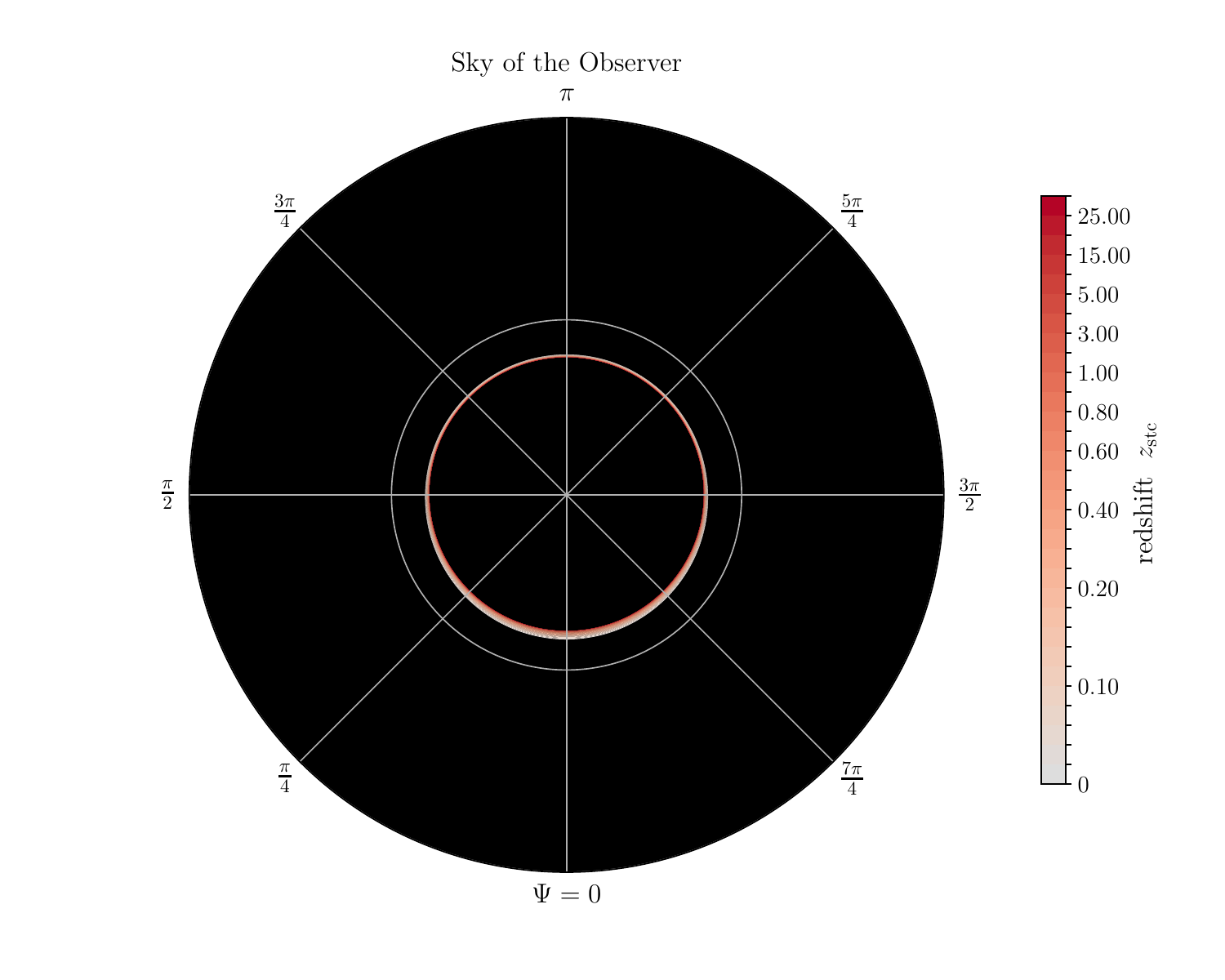} &   \hspace{0.5cm}\includegraphics[width=80mm]{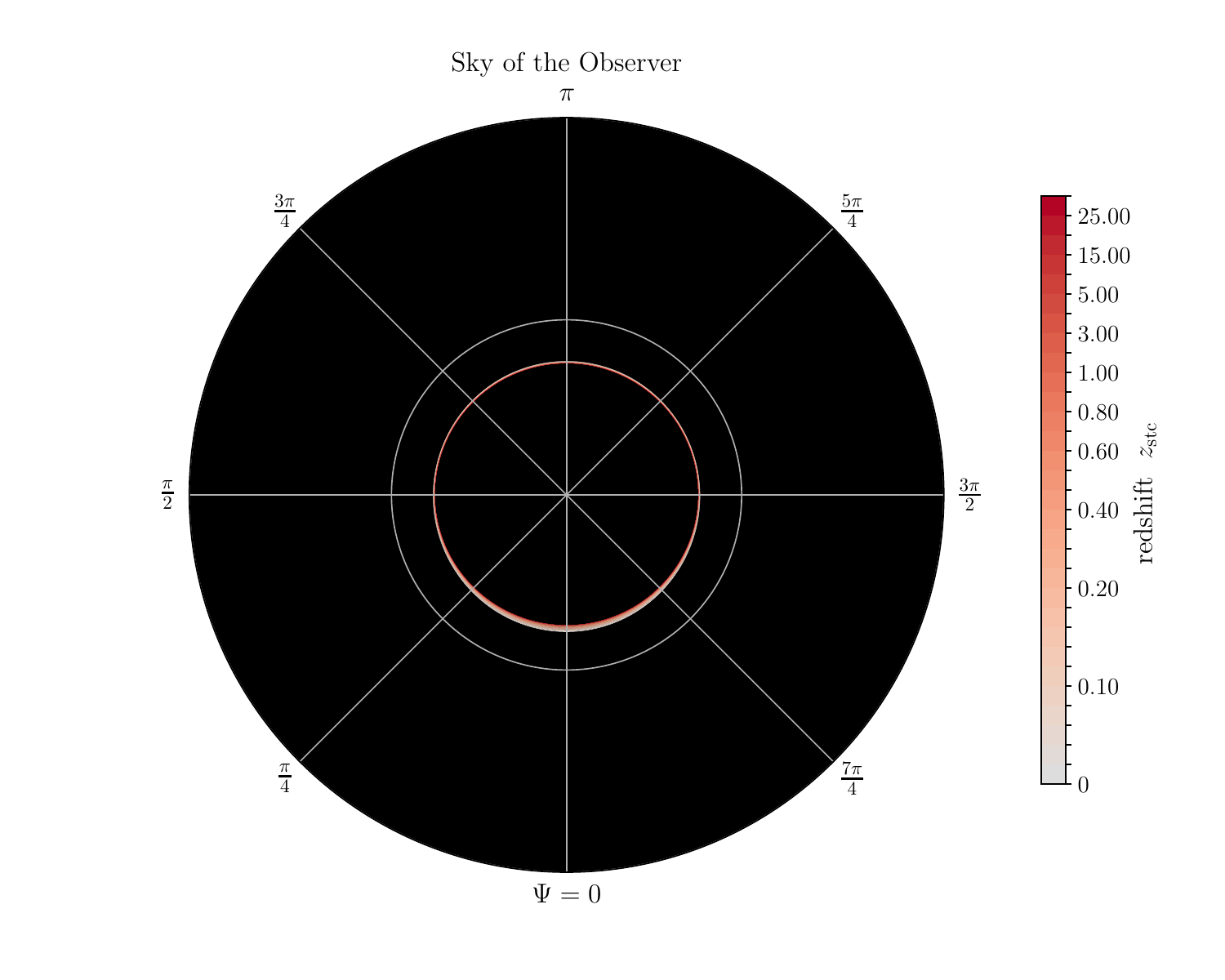} \\
    \hspace{-0.5cm}Inhomogeneous Plasma with $\omega_{\text{p}}=3m$ & \hspace{0.5cm} Inhomogeneous Plasma with $\omega_{\text{p}}=4m$\\
    \\
    \hspace{-0.5cm}\includegraphics[width=80mm]{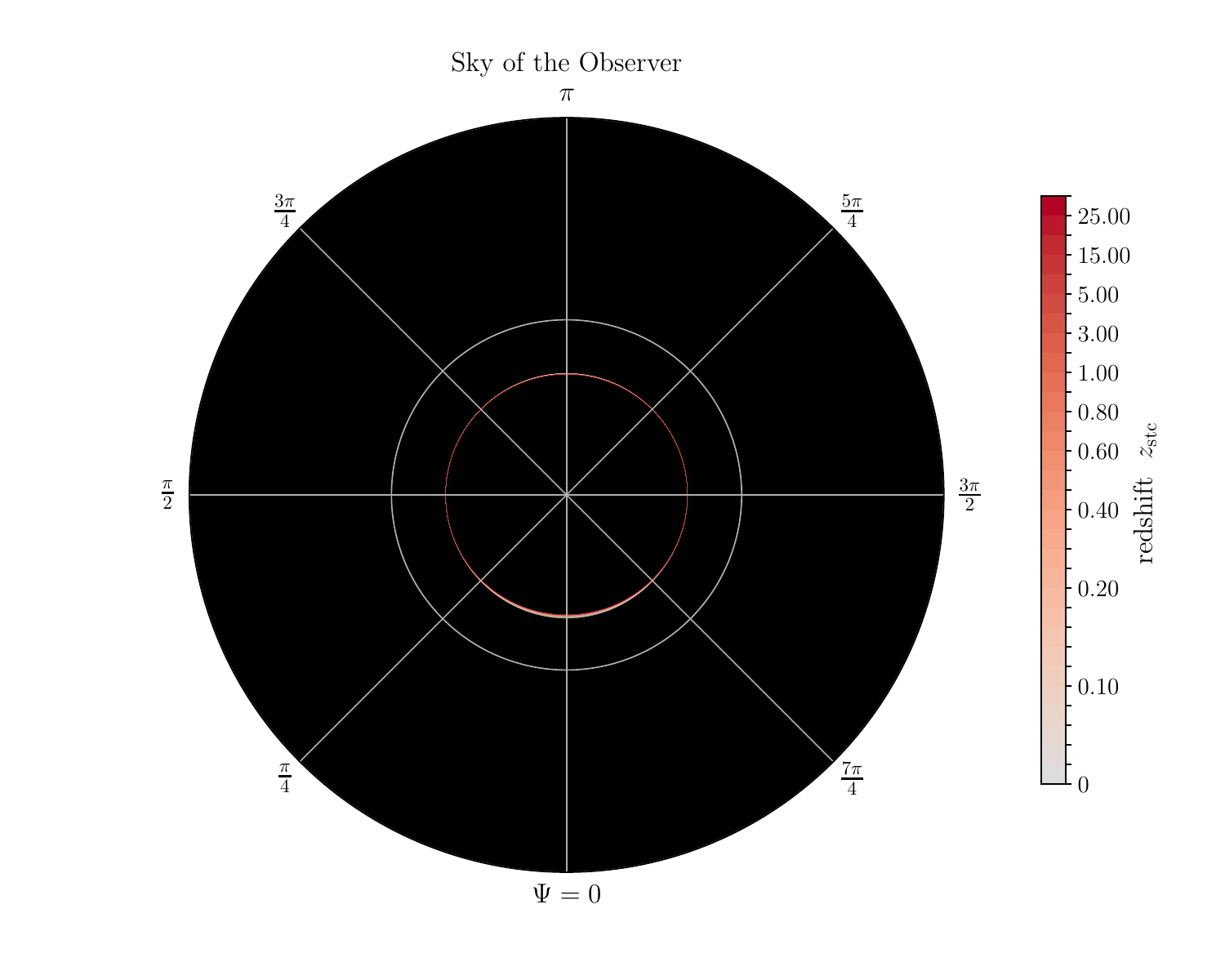} &   \hspace{0.5cm}\includegraphics[width=80mm]{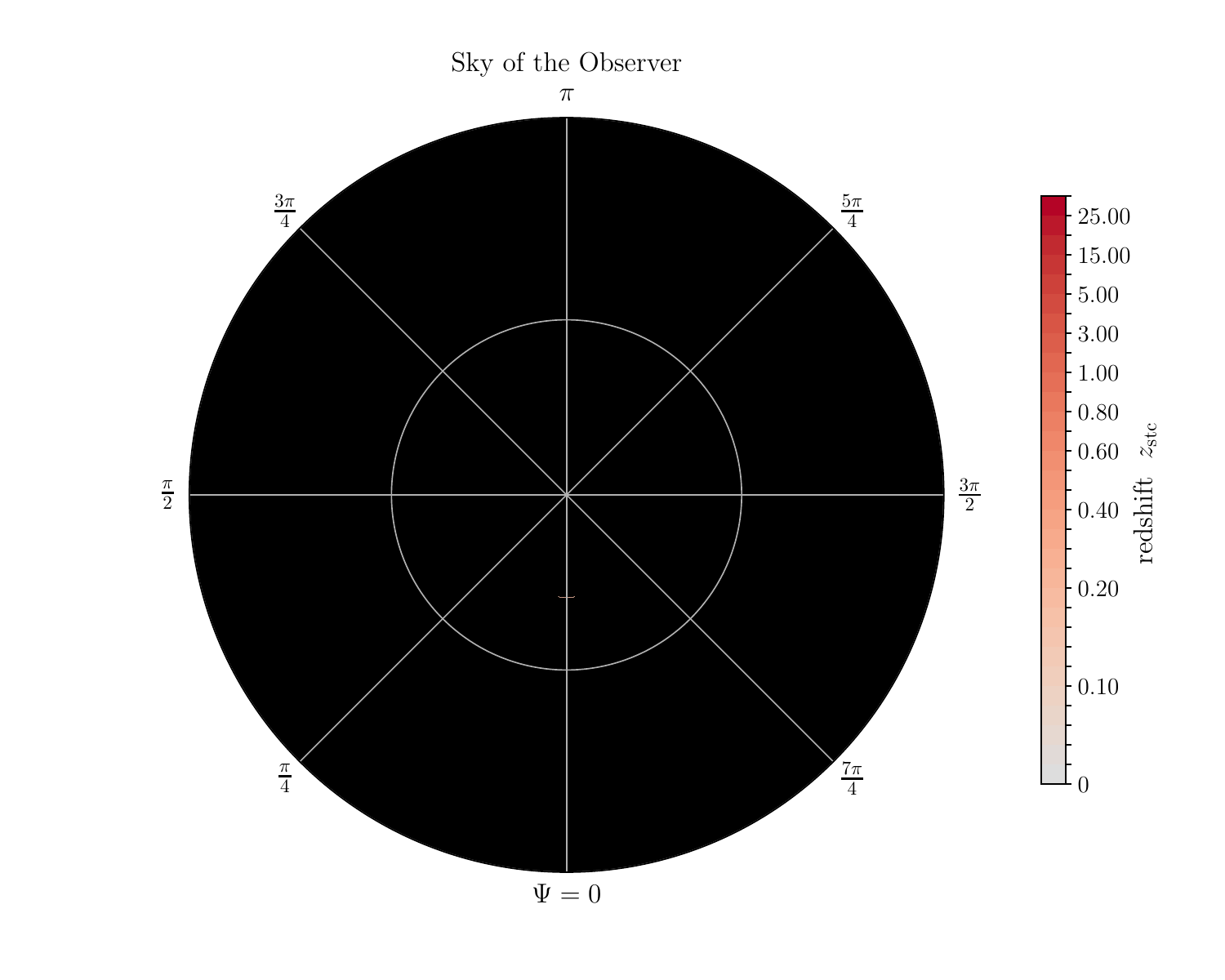} \\
  \end{tabular}
	\caption{Redshift maps for the second-order photon rings in the Schwarzschild spacetime for light rays travelling in vacuum (top left panel), through a homogeneous plasma (top right panel), and through an inhomogeneous plasma described by the distribution $E_{\text{pl}}(r,\vartheta)$ given by (\ref{eq:PlasmaEn}) with $\omega_{\text{p}}=m$ (middle left panel), $\omega_{\text{p}}=2m$ (middle right panel), $\omega_{\text{p}}=3m$ (bottom left panel), and $\omega_{\text{p}}=4m$ (bottom right panel). The observer is located at $r_{O}=40m$ and $\vartheta_{O}=\pi/4$ and the luminous disk is located in the equatorial plane between $r_{\text{in}}=2m$ and $r_{\text{out}}=20m$. For the light rays travelling through one of the plasmas the energy measured at the position of the observer is $E_{O}=\sqrt{53/50}E_{\text{C}}$.}
\end{figure*}

Besides the geometric structures and the lens equation of the direct image and the photon rings, the second observable which is of interest for observations is the redshift, or more precisely the redshift factor. It relates the energy of a light ray measured in the reference frame of the observer to the energy of the light ray in the reference frame of its source. In its most general form it reads
\begin{eqnarray}
z=\frac{E_{S}}{E_{O}}-1.
\end{eqnarray}
Please note that since in our case we assume that the sources in the luminous disk are static the results cannot directly be transferred to real observations, however, it can provide us with a first idea how the redshift structure of an observed image changes for light rays travelling through a plasma in comparison to light rays travelling in vacuum.

Now we first have to rewrite $E_{O}$ and $E_{S}$. For this purpose we use that we have $E_{O}=E/\sqrt{P(r_{O})}$ and $E_{S}=E/\sqrt{P\left(r_{S}(\Sigma,\Psi)\right)}$ and obtain
\begin{eqnarray}
z=\sqrt{\frac{P(r_{O})}{P\left(r_{S}\left(\Sigma,\Psi\right)\right)}}-1.
\end{eqnarray}
As we can easily see via $r_{S}$ the redshift factor depends indirectly on the latitude-longitude coordinates on the observer's celestial sphere. 

The computational evaluation of the redshift was carried out using the same set of Julia codes which was also used for calculating the geometric structure and the lens equation for the direct image and the first- and second-order photon rings. 

Figs.~8--10 show redshift maps for the direct image and the first- and second-order photon rings for light rays travelling in vacuum (top left panels), through a homogeneous plasma (top right panels; $\omega_{\text{p}}=0$), and through the inhomogeneous plasma described by the distribution given by Eq.~(\ref{eq:PlasmaEn}) with $\omega_{\text{p}}=m$ (middle left panels), $\omega_{\text{p}}=2m$ (middle right panels), $\omega_{\text{p}}=3m$ (bottom left panels), and $\omega_{\text{p}}=4m$ (bottom right panels). In the case of the redshift maps for light rays travelling through the homogeneous and the inhomogeneous plasmas the photon energy is $E_{O}=\sqrt{53/50}E_{\text{C}}$. Again the luminous disk extends from the horizon ($r_{\text{in}}=r_{\text{H}}=2m$) to $r_{\text{out}}=20m$, and the observer is located at $r_{O}=40m$ and $\vartheta_{O}=\pi/4$. Since the redshift maps were calculated for the same setups as the lens maps for the direct image, and the first- and second-order photon rings in Figs.~2--4 it is not surprising that they show the same geometric structures. Since these were already discussed above the following discussion will focus on the visible structures in the redshift maps.

We will again start our discussion with the maps for the direct images. They are shown in Fig.~8. In all six redshift maps the observable redshift generally increases with decreasing celestial latitude towards the boundary of the inner shadow. However, while for light rays travelling in vacuum high redshifts can only be found very close to the centre of the map and, without zooming in, in this region the redshift levels are only very difficult to distinguish for light rays travelling through the different plasma distributions this region becomes much better visible and the different redshift levels easier to distinguish. In the first five maps the levels of constant redshift form closed rings around the inner shadow. For light rays travelling through a homogeneous plasma and the inhomogeneous plasma with $\omega_{\text{p}}=m$ while we have small changes in the size of the direct images it is relatively difficult to identify changes in the redshift. However, this is different when we increase the plasma parameter to $\omega_{\text{p}}=2m$ and $\omega_{\text{p}}=3m$. Here, in particular around the meridian, the area covered with high redshifts becomes less dominant. However, the most drastic change occurs when we increase the plasma parameter to $\omega_{\text{p}}=4m$. In this case the areas with high redshifts have nearly completely disappeared and only around the meridian and the antimeridian areas with slightly higher redshifts remain. In addition, one can easily see that in the inner region of the direct image the levels of constant redshift do not form closed rings around the inner shadow anymore. Instead slightly north of the celestial equator we can see that with decreasing celestial latitude the redshift first increases, reaches a maximum at a celestial latitude which is a bit lower than $\Sigma=\pi/6$, and then it decreases towards the boundary of the inner shadow. Since in the Schwarzschild spacetime the redshift increases when the radius coordinate of the source approaches the event horizon this indicates that the associated light rays were emitted in parts of the luminous disk with a larger radius coordinate. In addition, we can now see two crescent-like structures around the meridian and the antimeridian.

When we now turn to the photon rings of first and second order we see relatively similar structures in the redshift maps, however, the boundary of the dark central region can now be found closer to the boundary of the shadow. Here, overall the main differences are that for $\omega_{\text{p}}=3m$ at the celestial equator the photon rings are very thin and thus with decreasing celestial latitude the redshift increases very rapidly while for $\omega_{\text{p}}=4m$ in these regions we will not be able to observe any redshifts since we cannot observe photons from the luminous disk anymore. In addition, in both cases the redshift increases more rapidly with decreasing celestial latitude around the antimeridian (in the case of $\omega_{\text{p}}=4m$ only for the first-order photon ring) than at the meridian. 
 
How can the observed structures in the redshift maps now help us to determine the properties of the plasma? While overall for the direct image the redshift maps are relatively similar the most likely chance to extract information about the plasma from the redshift maps is by combining the geometric changes of the direct image and the photon rings with latitudinal changes in the redshift. Here, in particular plasmas with high electron densities close to the horizon showed very characteristic features. On one hand for $\omega_{\text{p}}=3m$ and $\omega_{\text{p}}=4m$ the photon rings become thinner around the celestial equator or completely disappear and thus in the case of the former the latitudinal change in redshift will be relatively high. In addition, for $\omega_{\text{p}}=4m$ the redshift maps showed very characteristic features. When we now increase the energy of the photons in all cases the photon ring structures as well as the structures in the redshift maps slowly approach the structures of the direct image and the photon rings in vacuum. Here, how they change while approaching the vaccum structures depends on the properties of the plasma and thus while it will be a challenging task extracting these information has the potential to provide us with additional information about the properties of plasmas in accretion disks surrounding astrophysical black holes.

However, in real astrophysical scenarios when we want to measure the redshift there are two difficulties. On one hand in accretion disks we usually only have few characteristic atomic emission lines which allow us to directly infer the redshift and, in particular, these emission lines are located at very high frequencies, namely in the X-ray band, see, e.g., M\"{u}ller and Camenzind \cite{Mueller2004}. On the other hand, as already mentioned above, the particles in an accretion disk are orbiting the black hole and as a consequence we also have to consider the Doppler red- and blueshift due to the motion of the plasma particles. However, the general changes induced by the presence of a plasma will still be the same. Therefore, in real astrophysical scenarios it will be easier to determine the plasma properties via the changes in intensity induced by the redshift instead of the redshift itself. Here, calculating intensity maps is already a common practise, see, e.g., the works of Desire, C\'{a}rdenas-Avenda\~{n}o, and Chael \cite{Desire2025} or of Kobialko, Gal'tsov, and Molchanov \cite{Kobialko2025}, however, deriving separate redshift maps even for idealised models has the potential to help us better understand which effects may be introduced due to the propagation of the light rays through the plasma and which may be introduced by the emission process of the photons. 
\newline

\subsection{The Travel Time}
\begin{figure*}\label{fig:TTDI}
  \begin{tabular}{cc}
    \hspace{-0.5cm}Vacuum & \hspace{0.5cm} Homogeneous Plasma ($\omega_{\text{p}}=0$)\\
\\
    \hspace{-0.5cm}\includegraphics[width=80mm]{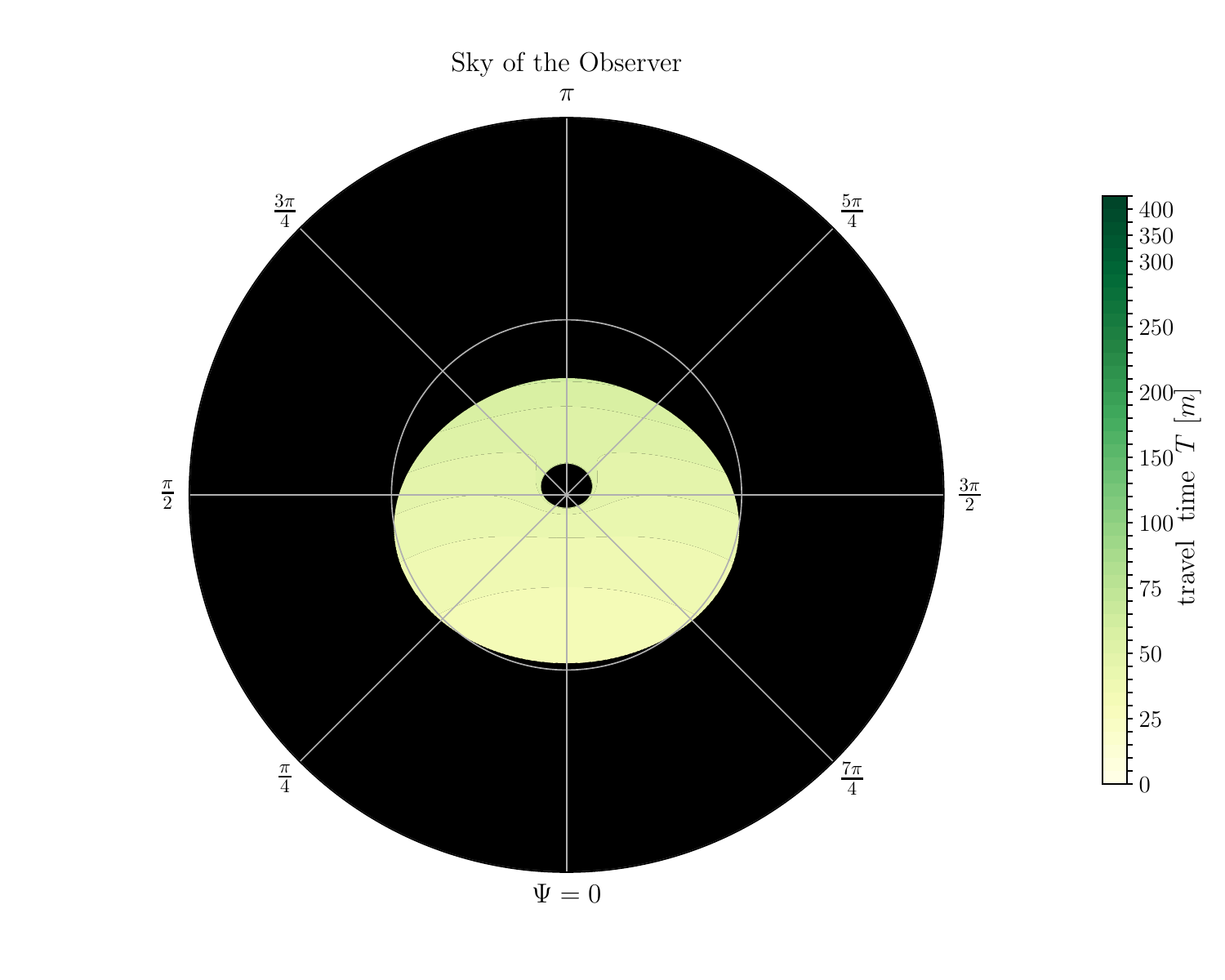} &   \hspace{0.5cm}\includegraphics[width=80mm]{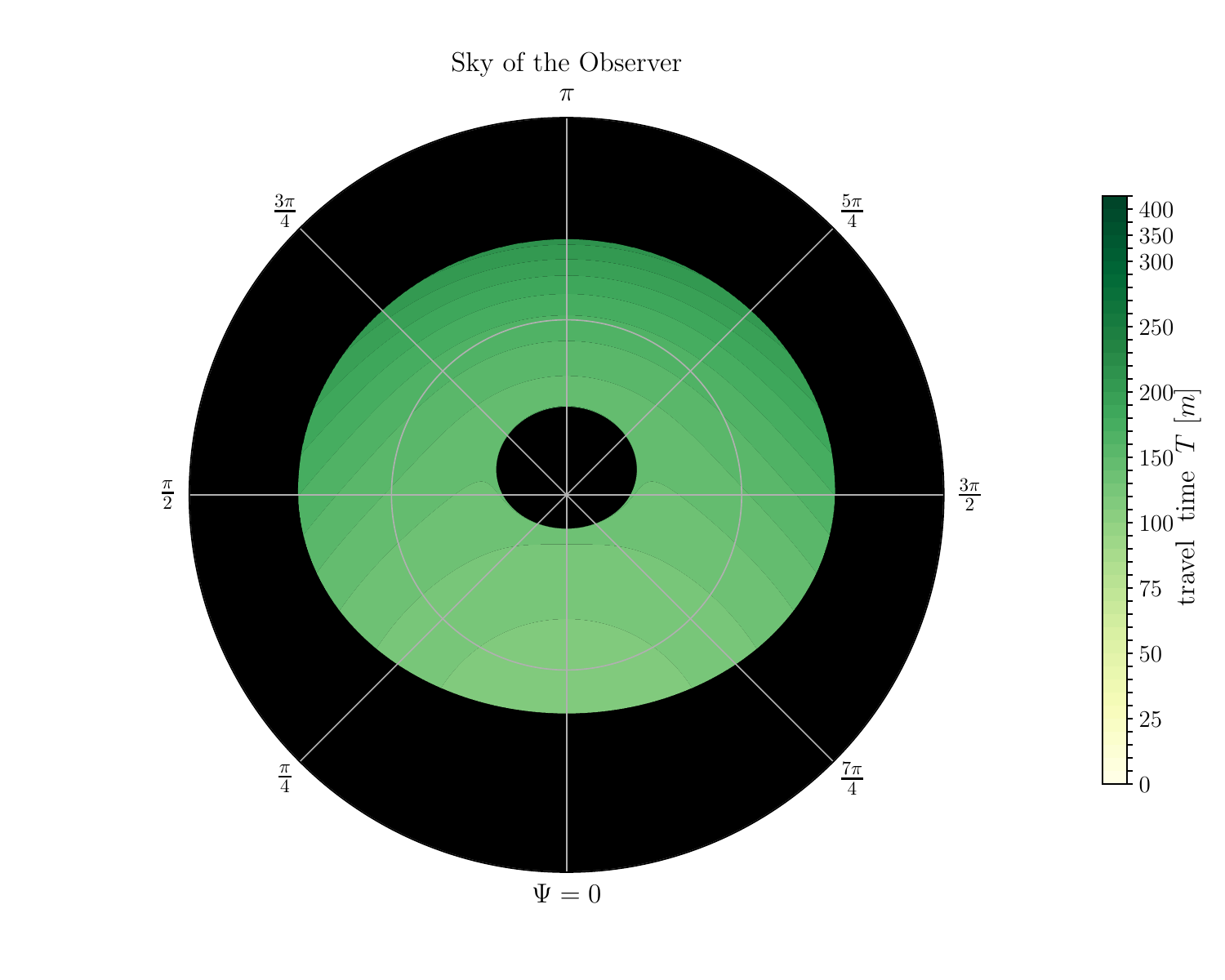} \\
\\
    \hspace{-0.5cm}Inhomogeneous Plasma with $\omega_{\text{p}}=m$ & \hspace{0.5cm} Inhomogeneous Plasma with $\omega_{\text{p}}=2m$\\
\\
    \hspace{-0.5cm}\includegraphics[width=80mm]{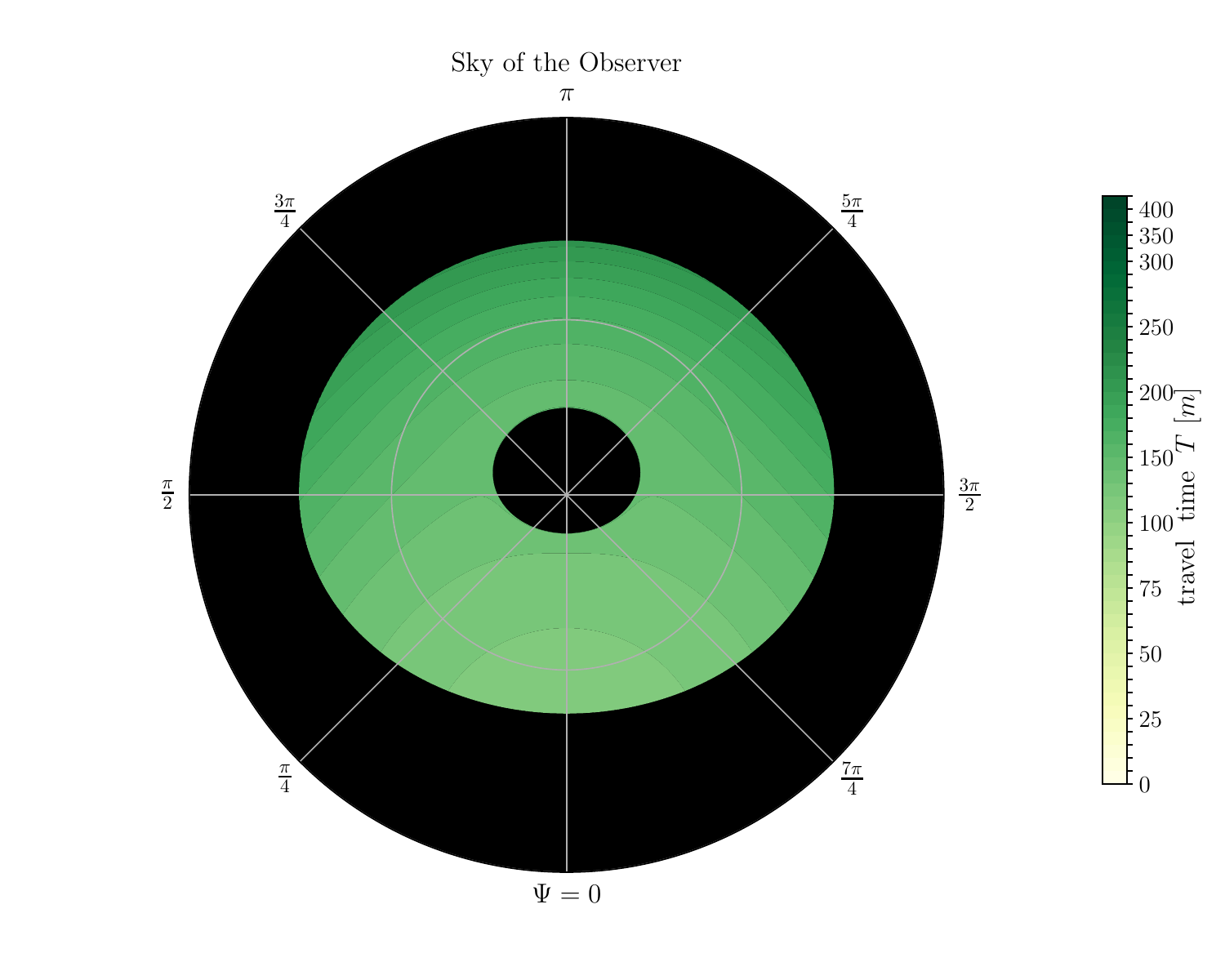} &   \hspace{0.5cm}\includegraphics[width=80mm]{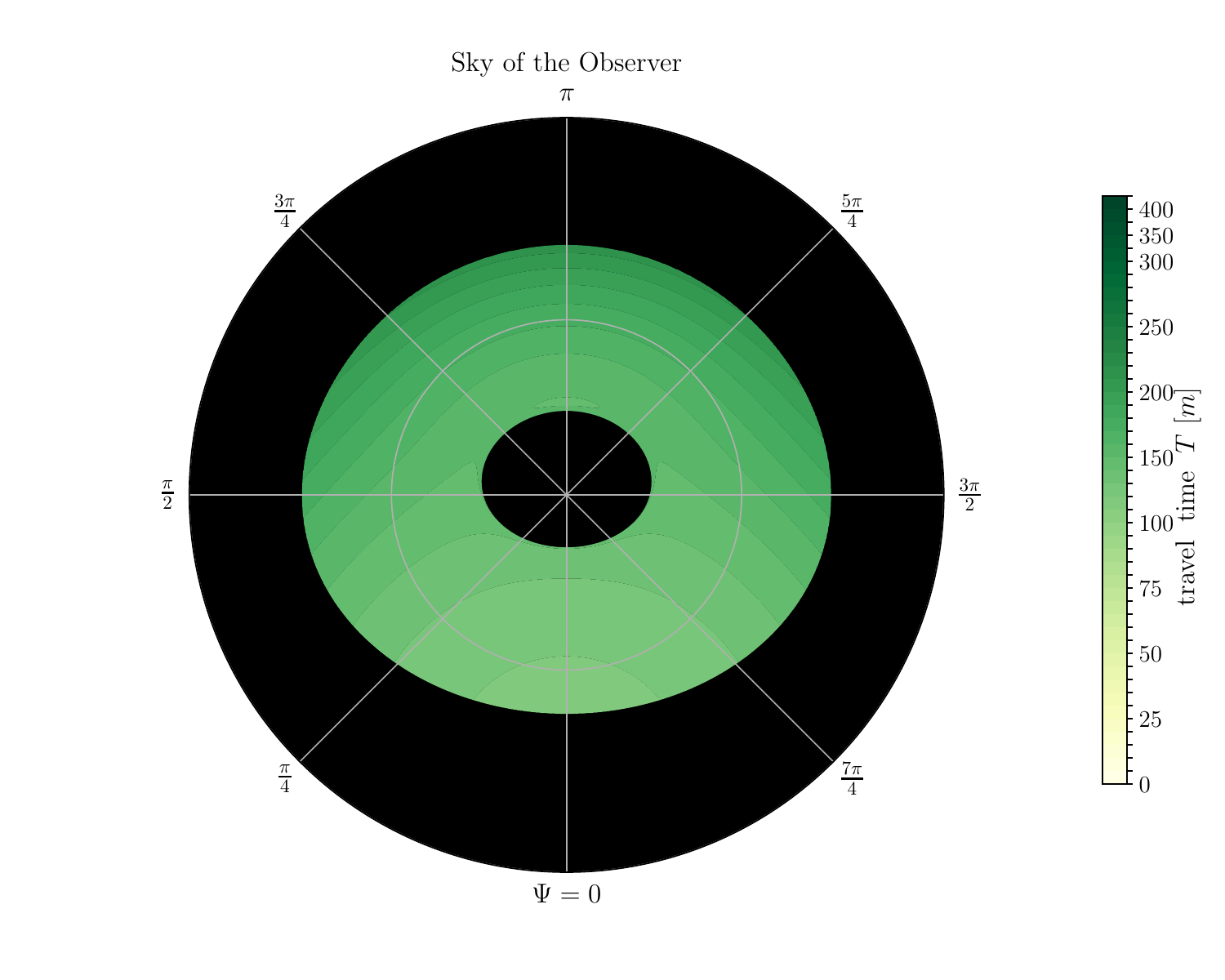} \\
    \hspace{-0.5cm}Inhomogeneous Plasma with $\omega_{\text{p}}=3m$ & \hspace{0.5cm} Inhomogeneous Plasma with $\omega_{\text{p}}=4m$\\
    \\
    \hspace{-0.5cm}\includegraphics[width=80mm]{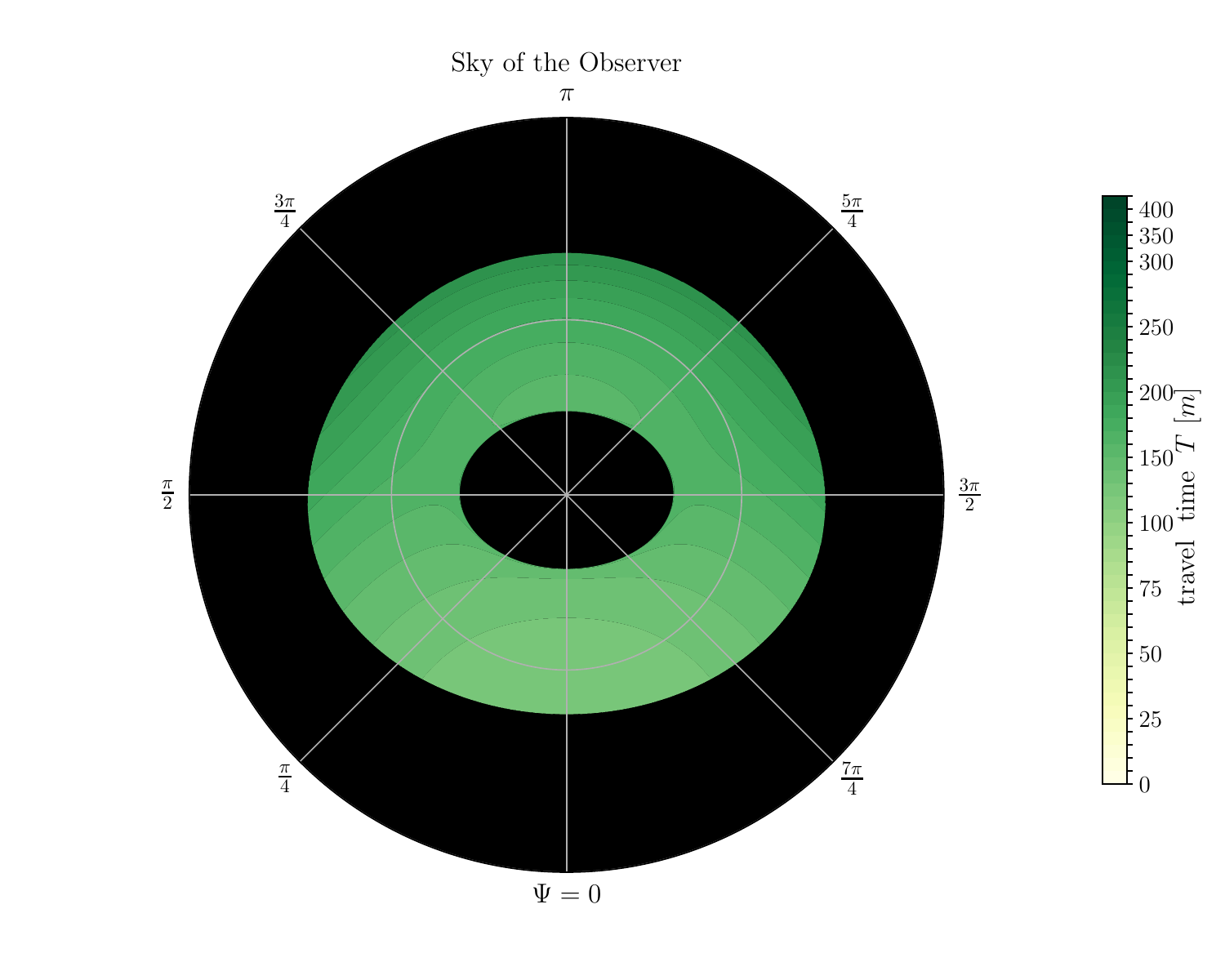} &   \hspace{0.5cm}\includegraphics[width=80mm]{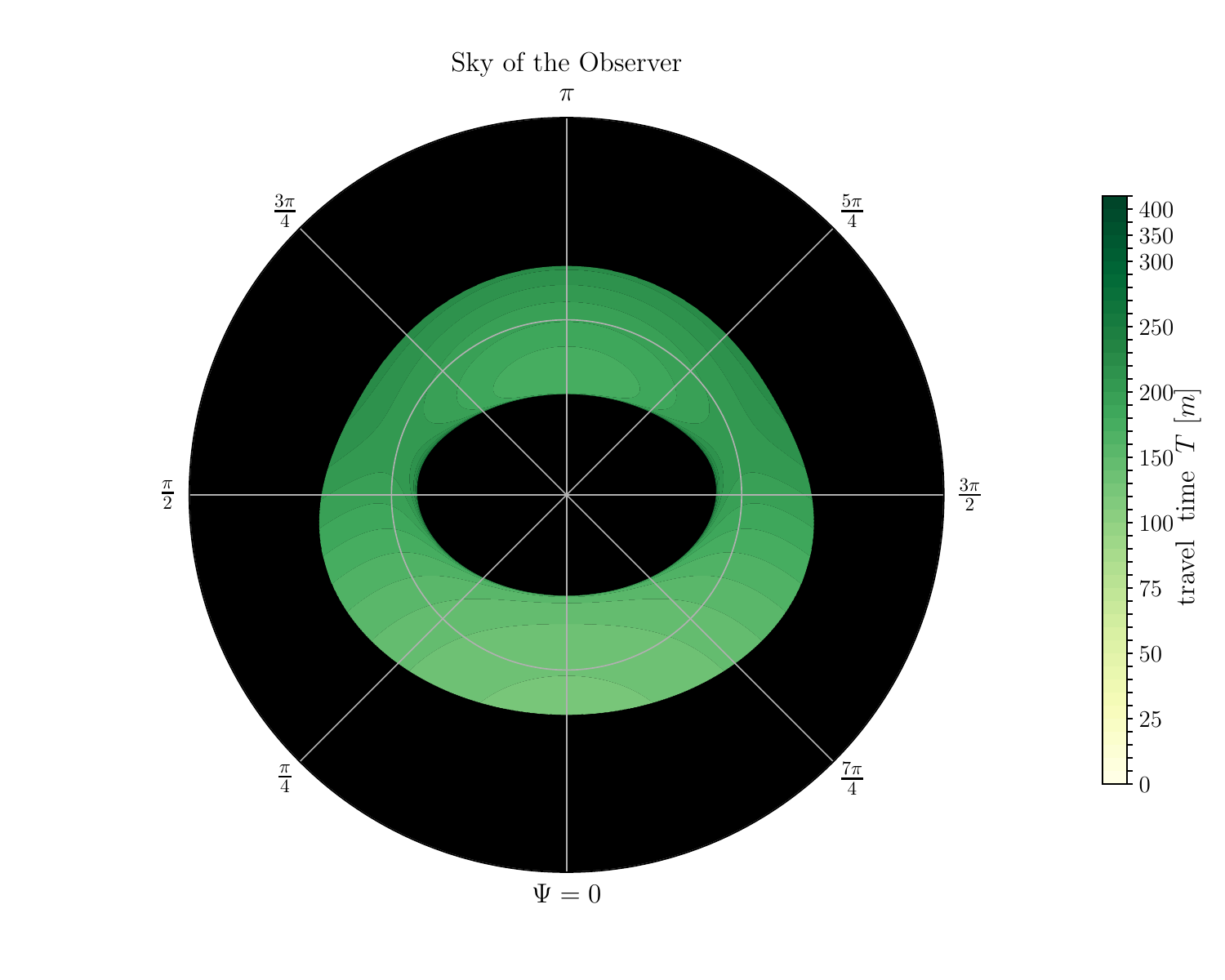} \\
  \end{tabular}
	\caption{Travel time maps for the direct images in the Schwarzschild spacetime for light rays travelling in vacuum (top left panel), through a homogeneous plasma (top right panel), and through an inhomogeneous plasma described by the distribution $E_{\text{pl}}(r,\vartheta)$ given by (\ref{eq:PlasmaEn}) with $\omega_{\text{p}}=m$ (middle left panel), $\omega_{\text{p}}=2m$ (middle right panel), $\omega_{\text{p}}=3m$ (bottom left panel), and $\omega_{\text{p}}=4m$ (bottom right panel). The observer is located at $r_{O}=40m$ and $\vartheta_{O}=\pi/4$ and the luminous disk is located in the equatorial plane between $r_{\text{in}}=2m$ and $r_{\text{out}}=20m$. For the light rays travelling through one of the plasmas the energy measured at the position of the observer is $E_{O}=\sqrt{53/50}E_{\text{C}}$.}
\end{figure*}

\begin{figure*}\label{fig:TTPhotonRing1E1}
  \begin{tabular}{cc}
    \hspace{-0.5cm}Vacuum & \hspace{0.5cm} Homogeneous Plasma ($\omega_{\text{p}}=0$)\\
\\
    \hspace{-0.5cm}\includegraphics[width=80mm]{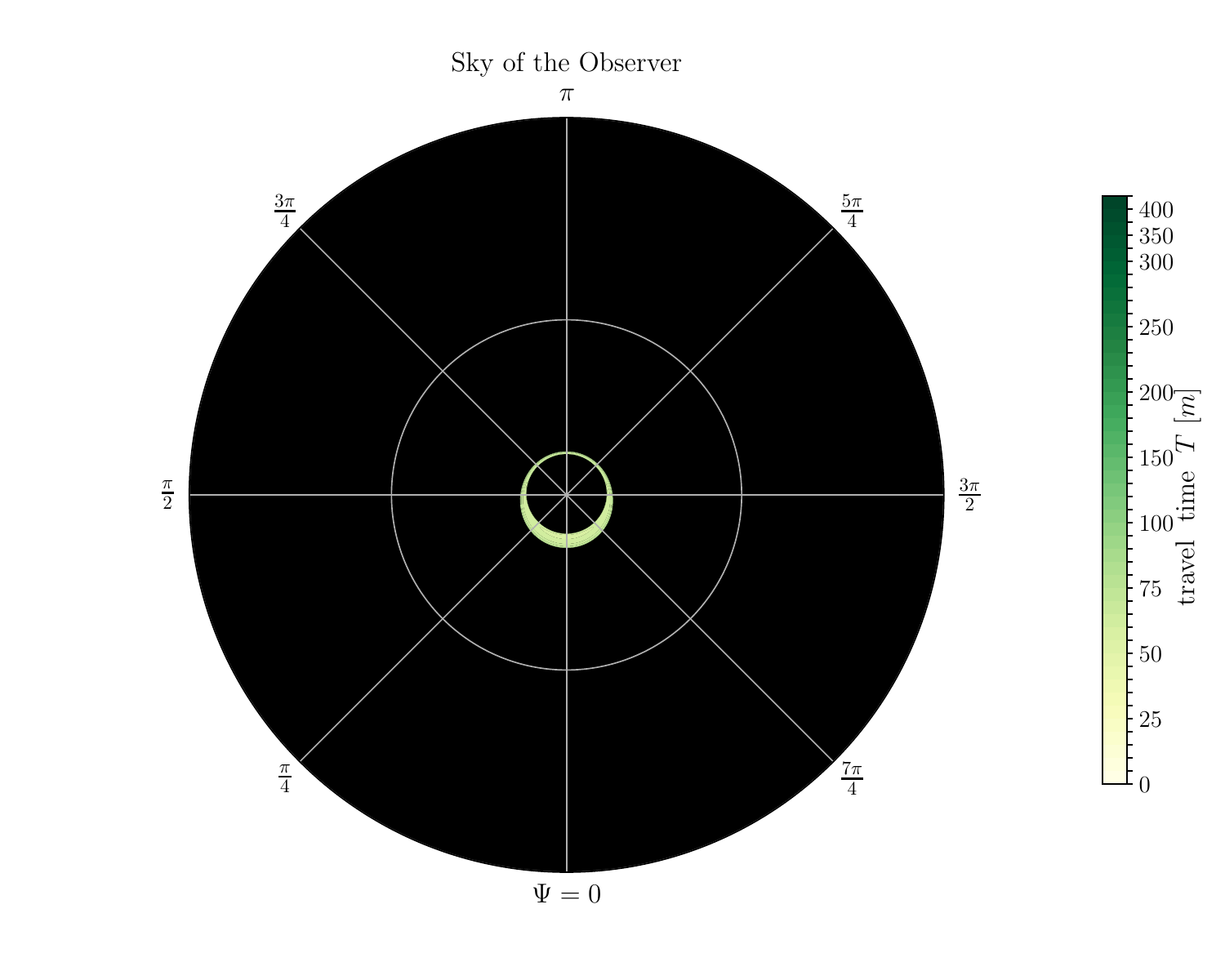} &   \hspace{0.5cm}\includegraphics[width=80mm]{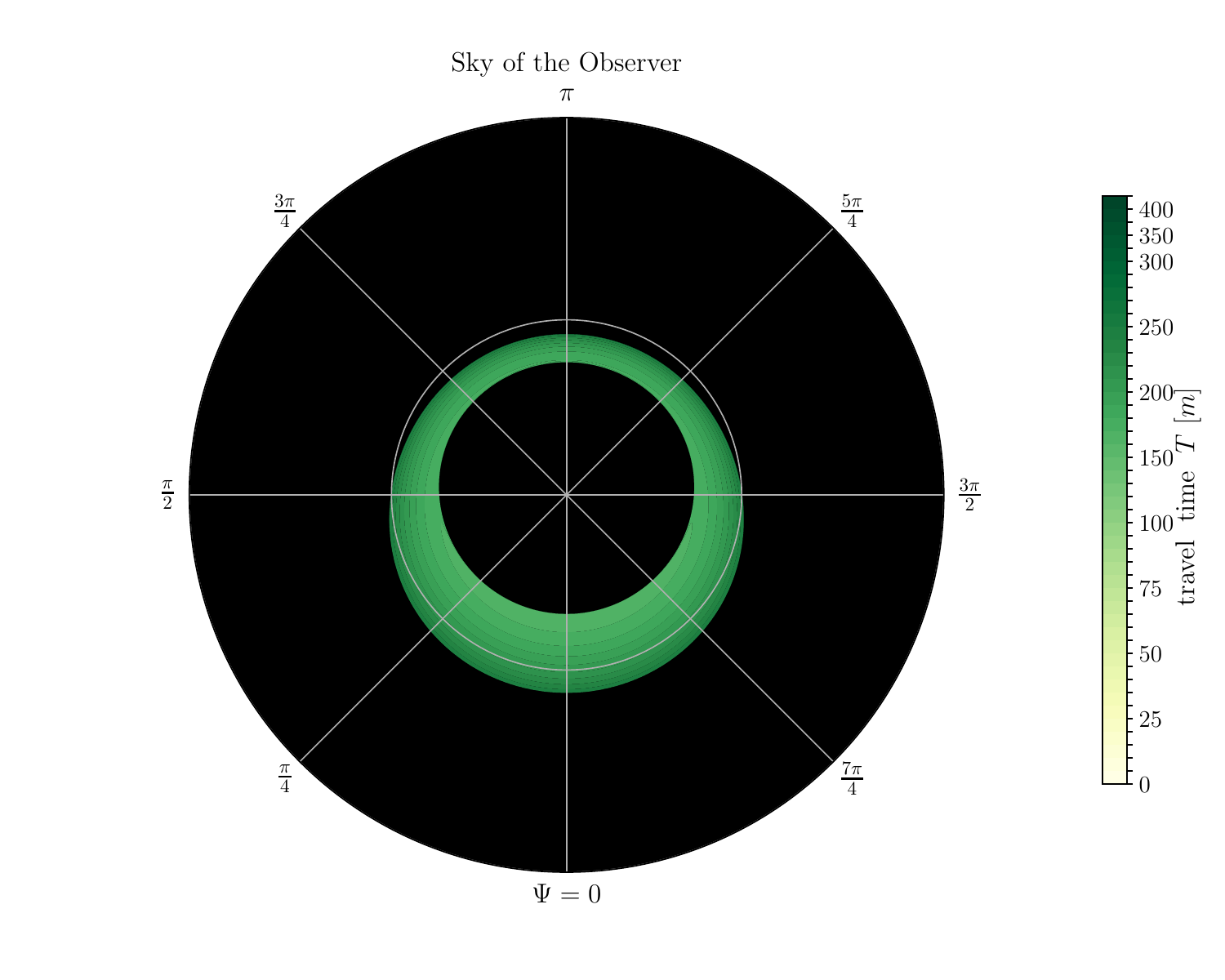} \\
\\
    \hspace{-0.5cm}Inhomogeneous Plasma with $\omega_{\text{p}}=m$ & \hspace{0.5cm} Inhomogeneous Plasma with $\omega_{\text{p}}=2m$\\
\\
    \hspace{-0.5cm}\includegraphics[width=80mm]{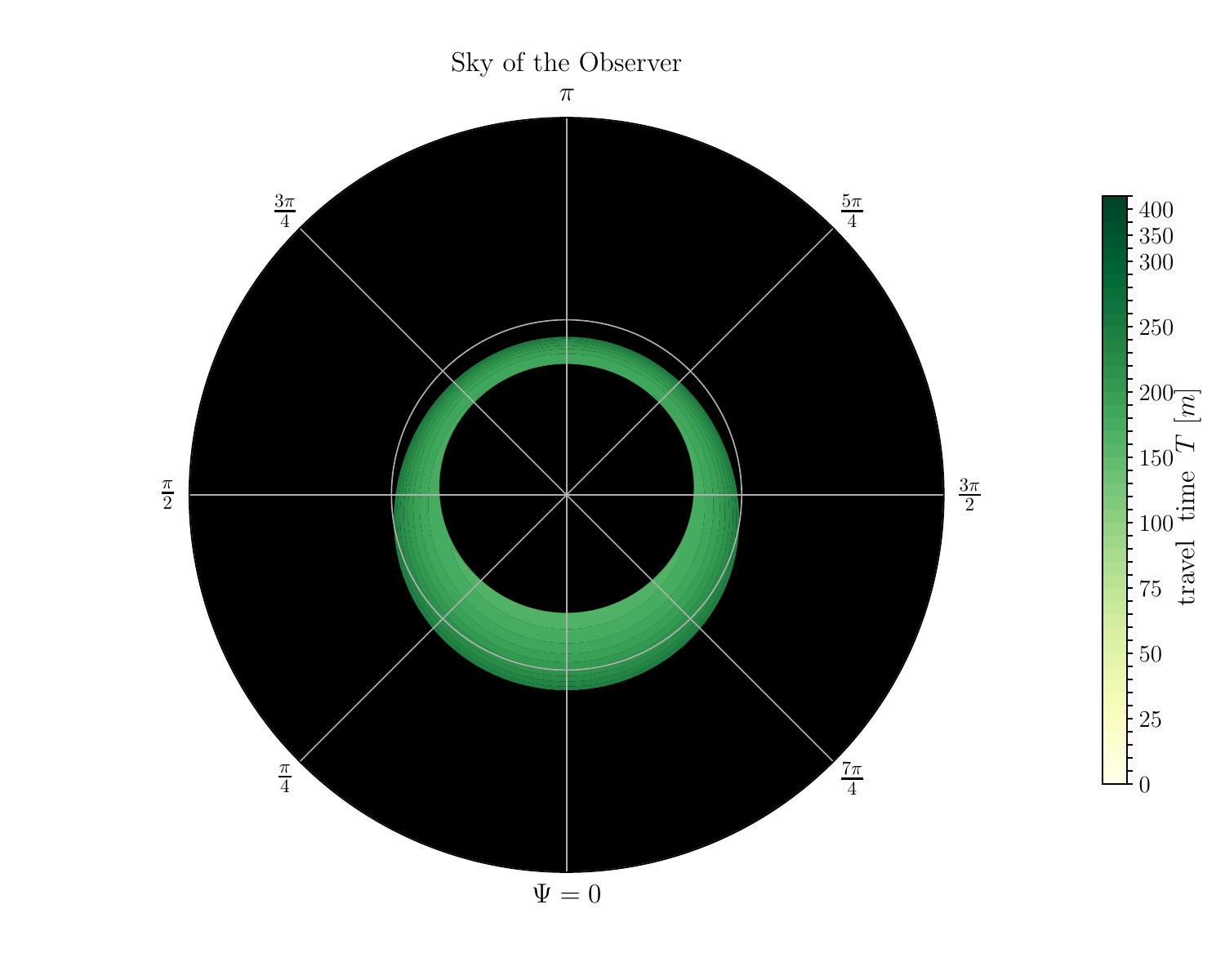} &   \hspace{0.5cm}\includegraphics[width=80mm]{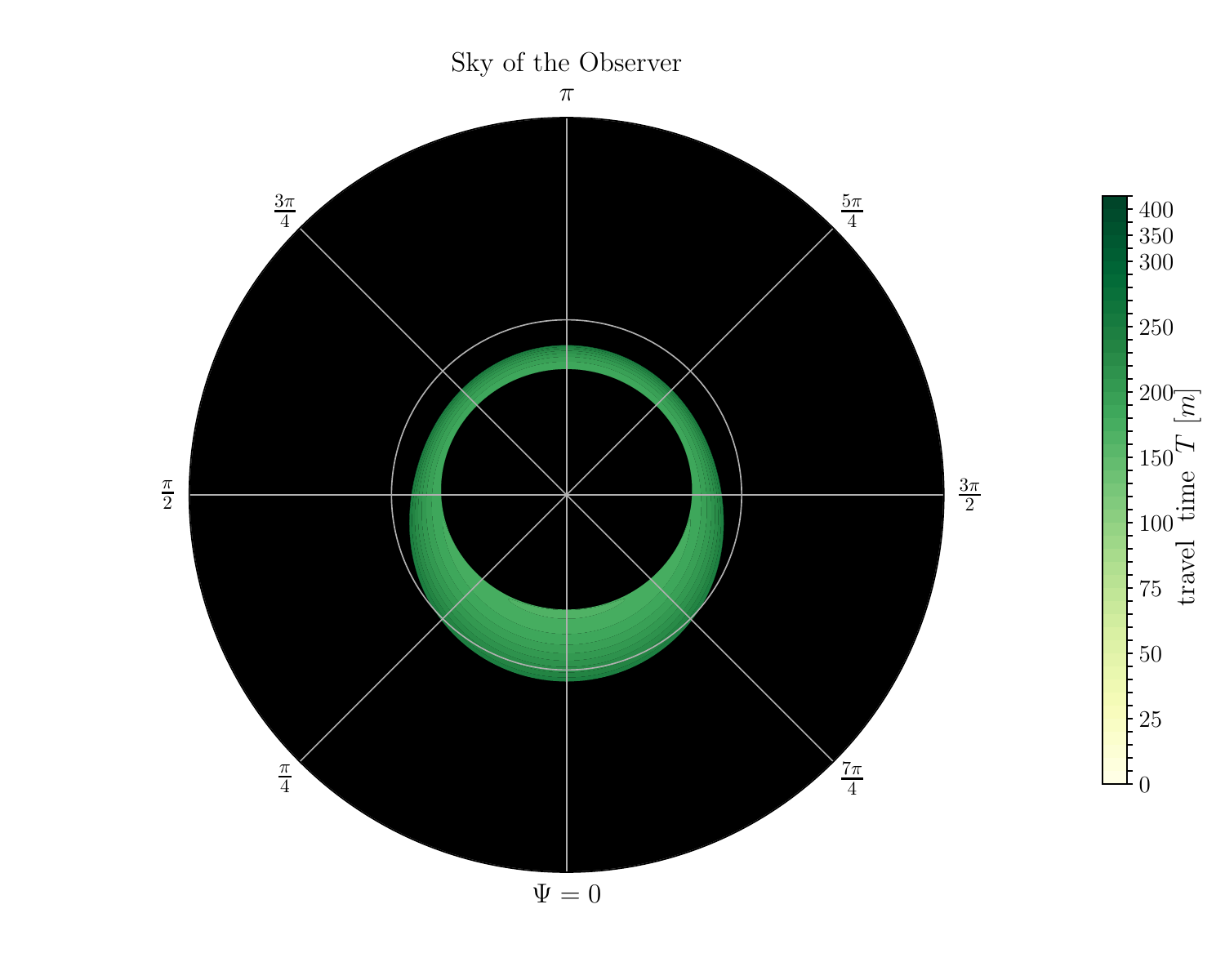} \\
    \hspace{-0.5cm}Inhomogeneous Plasma with $\omega_{\text{p}}=3m$ & \hspace{0.5cm} Inhomogeneous Plasma with $\omega_{\text{p}}=4m$\\
    \\
    \hspace{-0.5cm}\includegraphics[width=80mm]{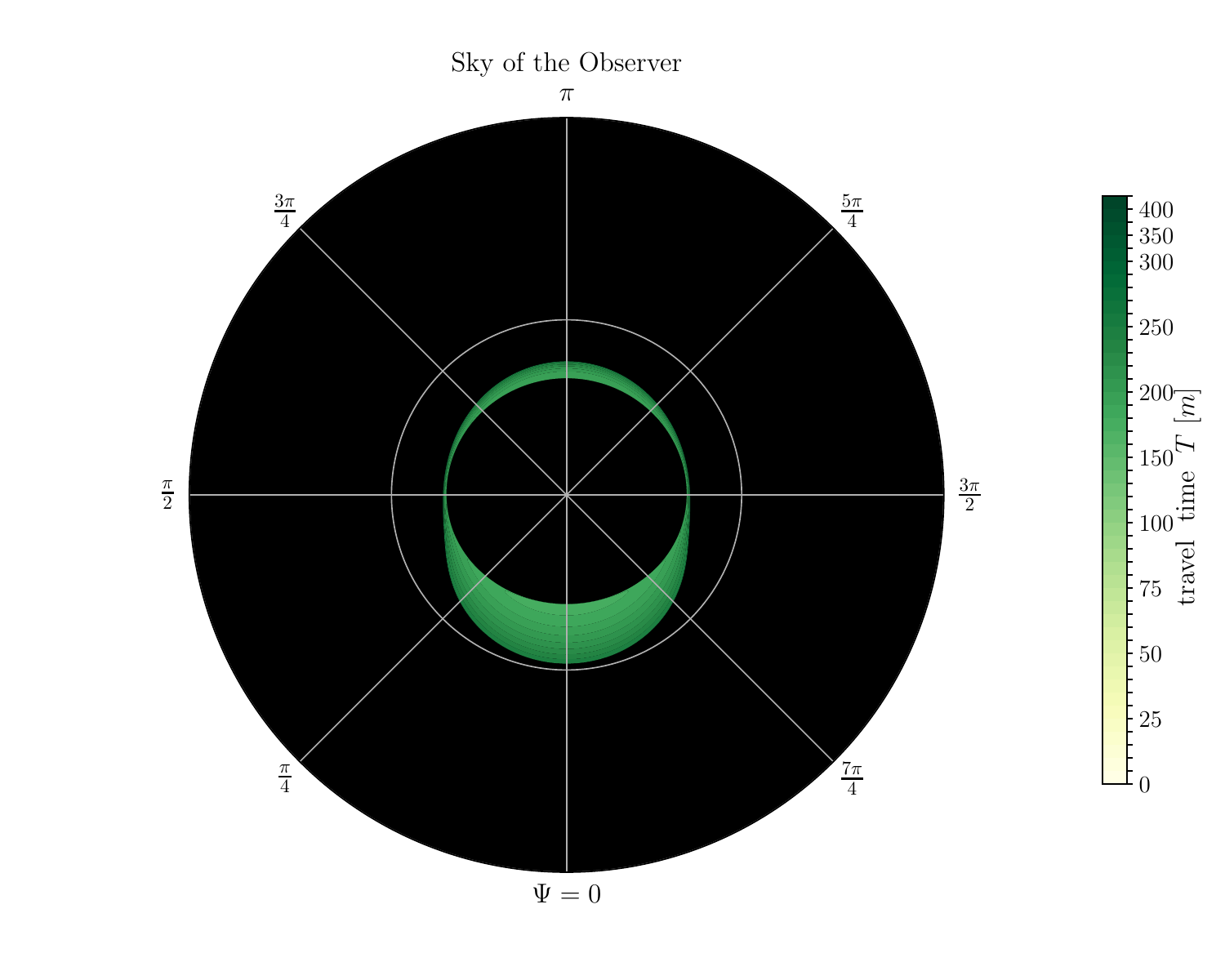} &   \hspace{0.5cm}\includegraphics[width=80mm]{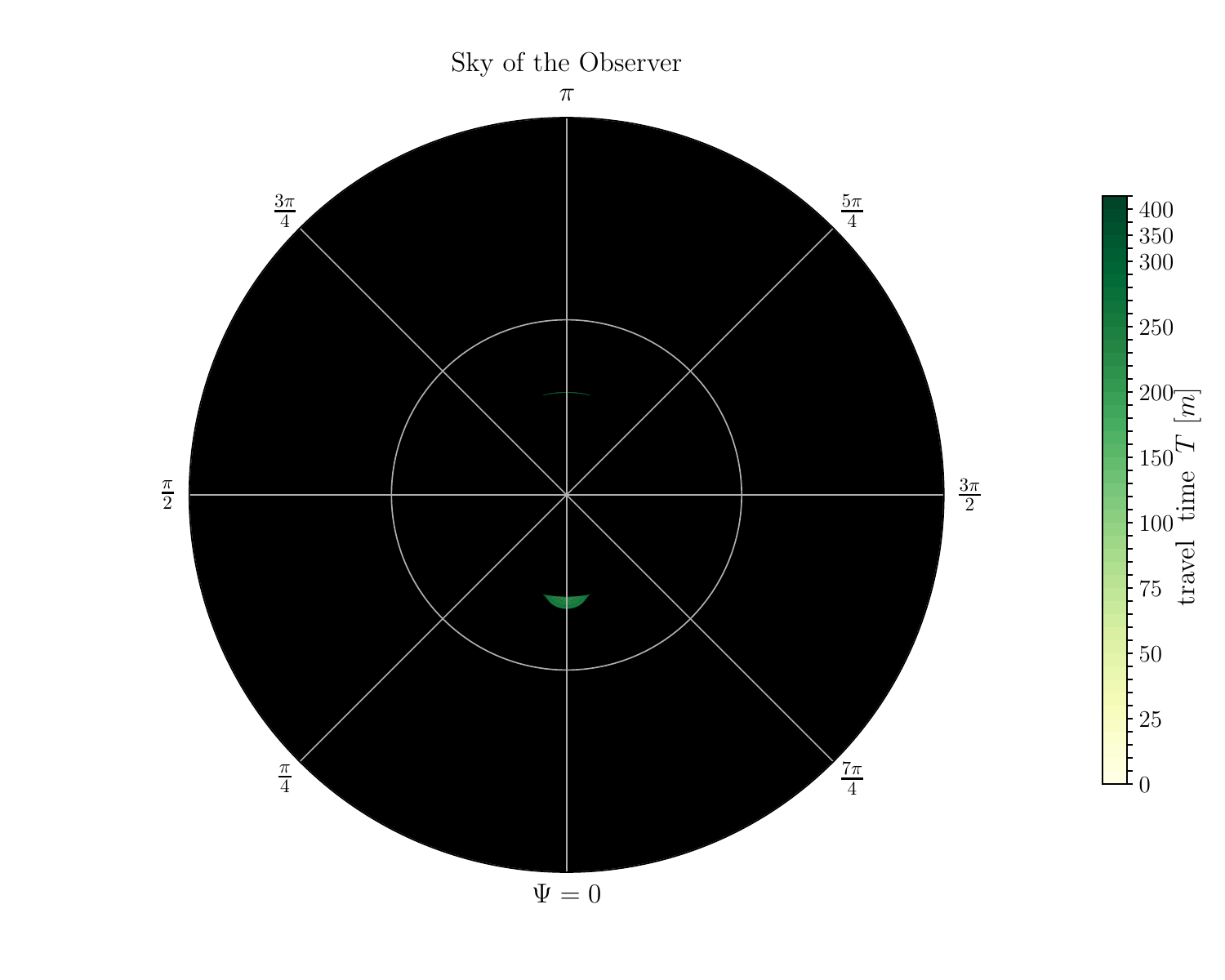} \\
  \end{tabular}
	\caption{Travel time maps for the first-order photon rings in the Schwarzschild spacetime for light rays travelling in vacuum (top left panel), through a homogeneous plasma (top right panel), and through an inhomogeneous plasma described by the distribution $E_{\text{pl}}(r,\vartheta)$ given by (\ref{eq:PlasmaEn}) with $\omega_{\text{p}}=m$ (middle left panel), $\omega_{\text{p}}=2m$ (middle right panel), $\omega_{\text{p}}=3m$ (bottom left panel), and $\omega_{\text{p}}=4m$ (bottom right panel). The observer is located at $r_{O}=40m$ and $\vartheta_{O}=\pi/4$ and the luminous disk is located in the equatorial plane between $r_{\text{in}}=2m$ and $r_{\text{out}}=20m$. For the light rays travelling through one of the plasmas the energy measured at the position of the observer is $E_{O}=\sqrt{53/50}E_{\text{C}}$.}
\end{figure*}

\begin{figure*}\label{fig:TTPhotonRing2E1}
  \begin{tabular}{cc}
    \hspace{-0.5cm}Vacuum & \hspace{0.5cm} Homogeneous Plasma ($\omega_{\text{p}}=0$)\\
\\
    \hspace{-0.5cm}\includegraphics[width=80mm]{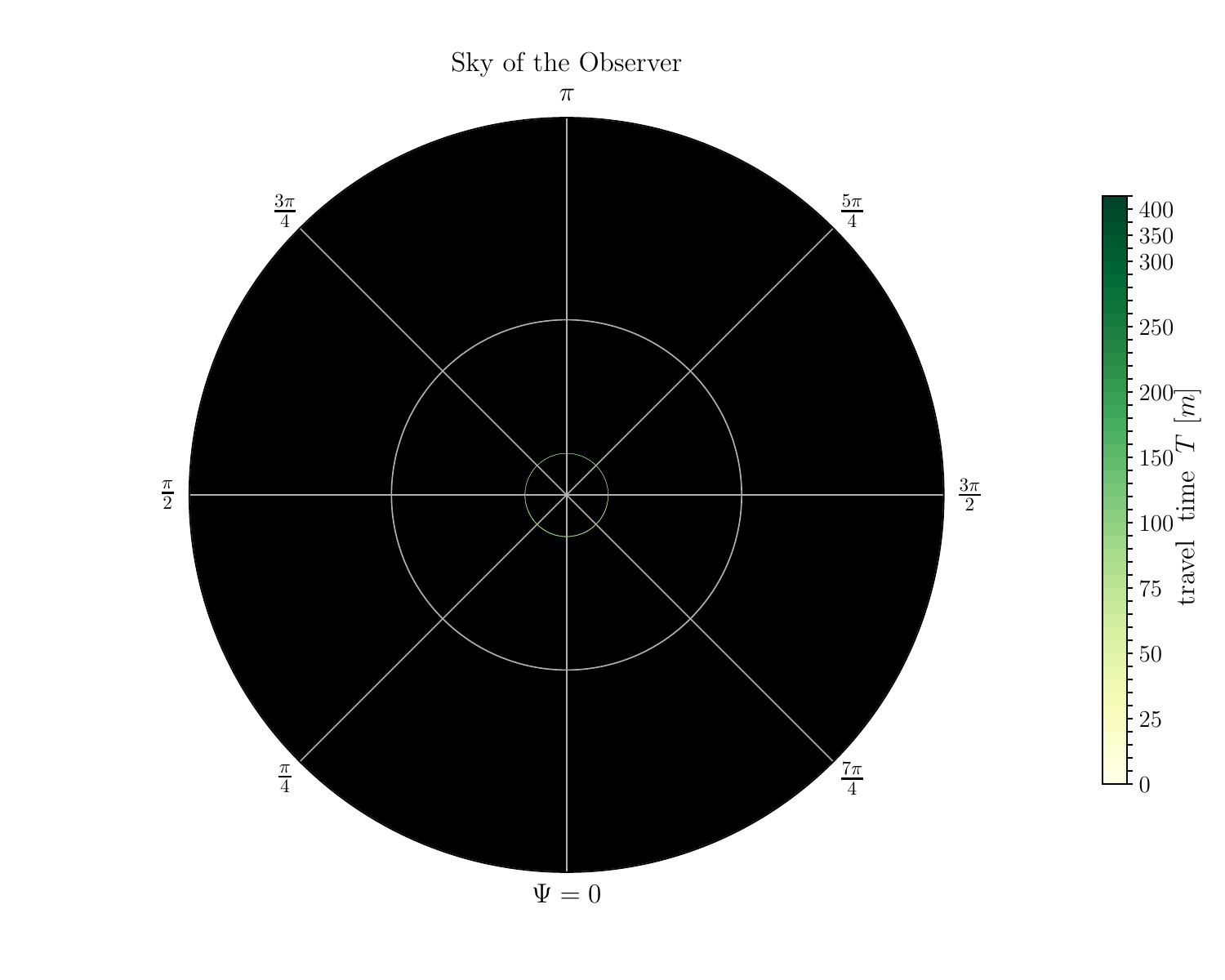} &   \hspace{0.5cm}\includegraphics[width=80mm]{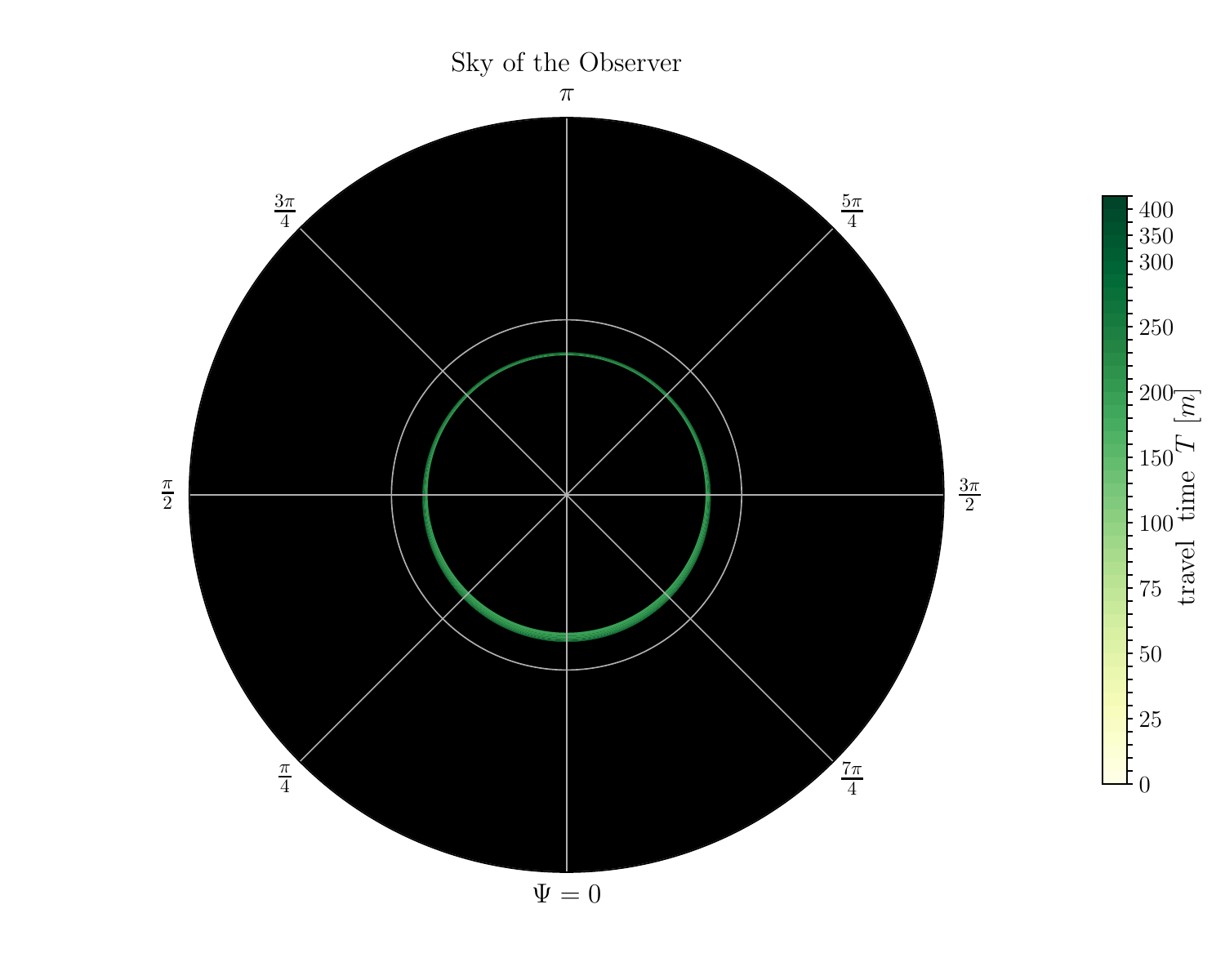} \\
\\
    \hspace{-0.5cm}Inhomogeneous Plasma with $\omega_{\text{p}}=m$ & \hspace{0.5cm} Inhomogeneous Plasma with $\omega_{\text{p}}=2m$\\
\\
    \hspace{-0.5cm}\includegraphics[width=80mm]{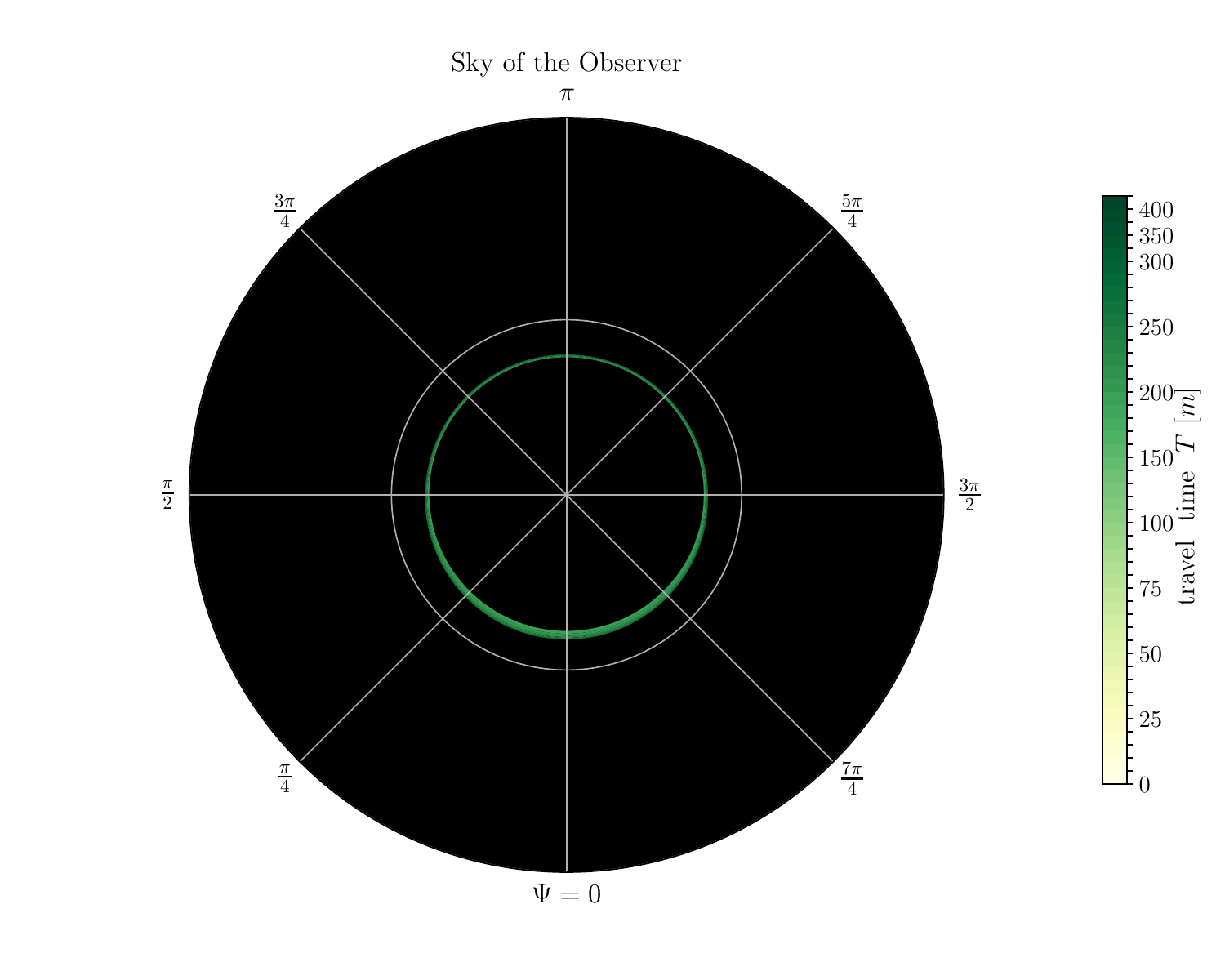} &   \hspace{0.5cm}\includegraphics[width=80mm]{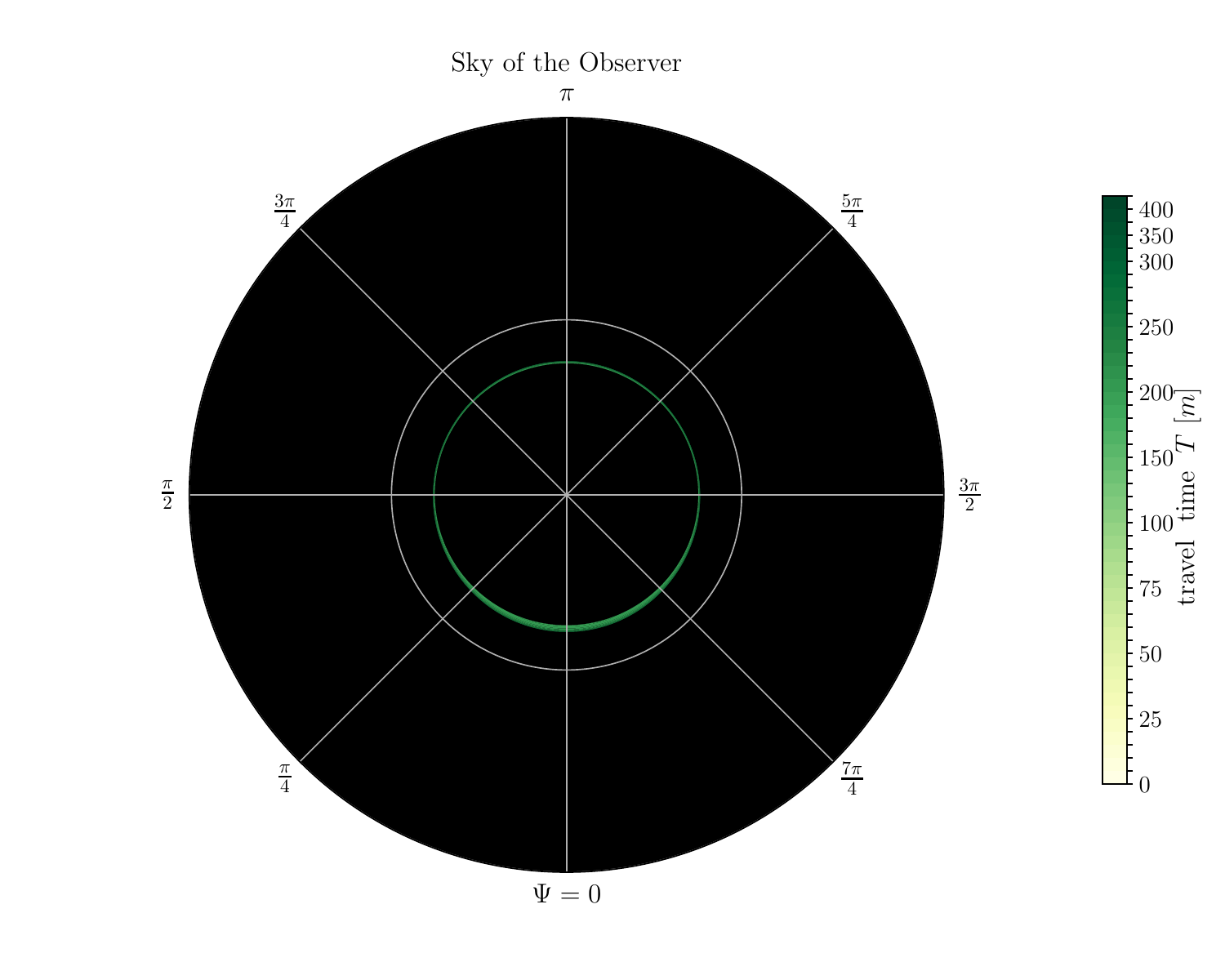} \\
    \hspace{-0.5cm}Inhomogeneous Plasma with $\omega_{\text{p}}=3m$ & \hspace{0.5cm} Inhomogeneous Plasma with $\omega_{\text{p}}=4m$\\
    \\
    \hspace{-0.5cm}\includegraphics[width=80mm]{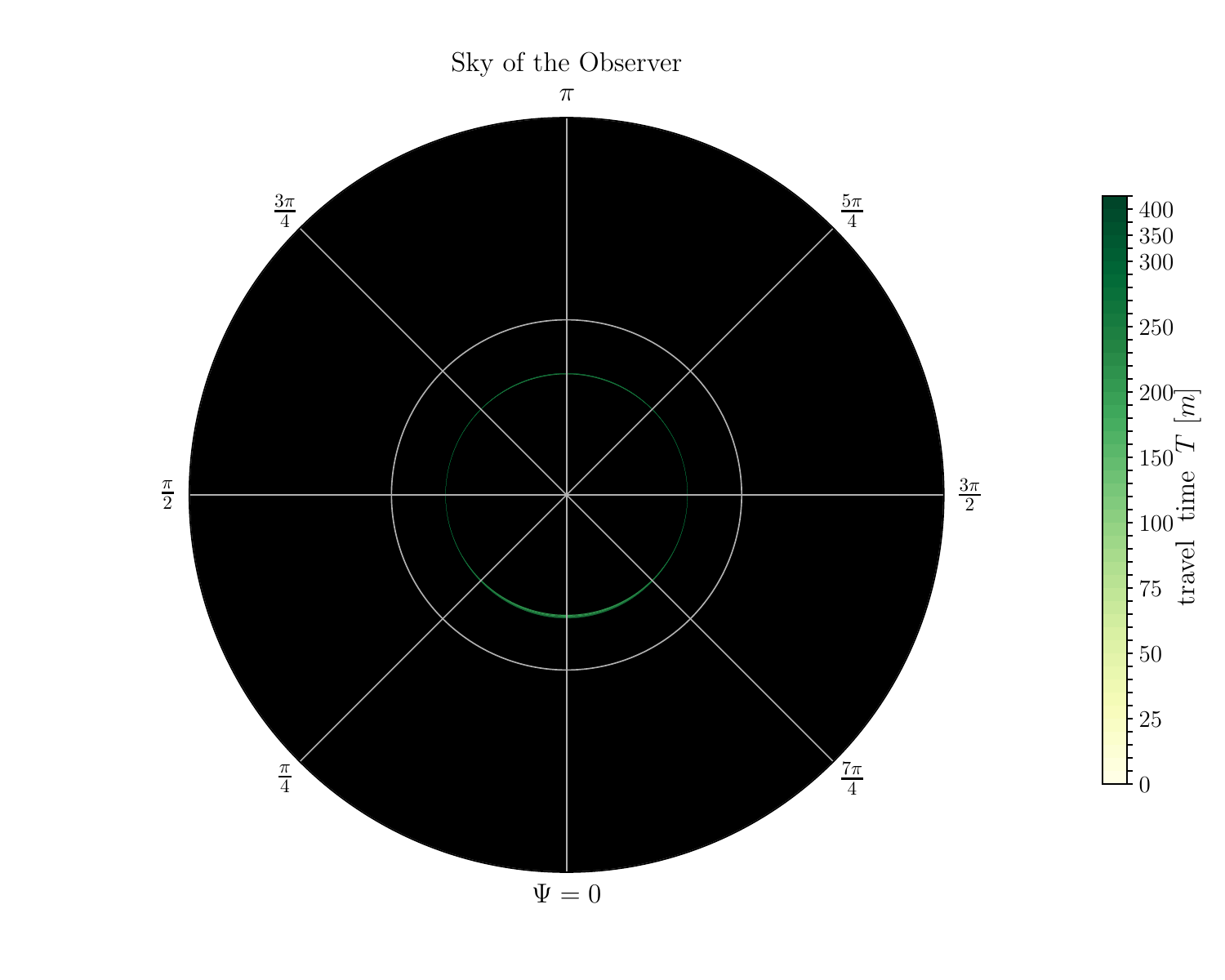} &   \hspace{0.5cm}\includegraphics[width=80mm]{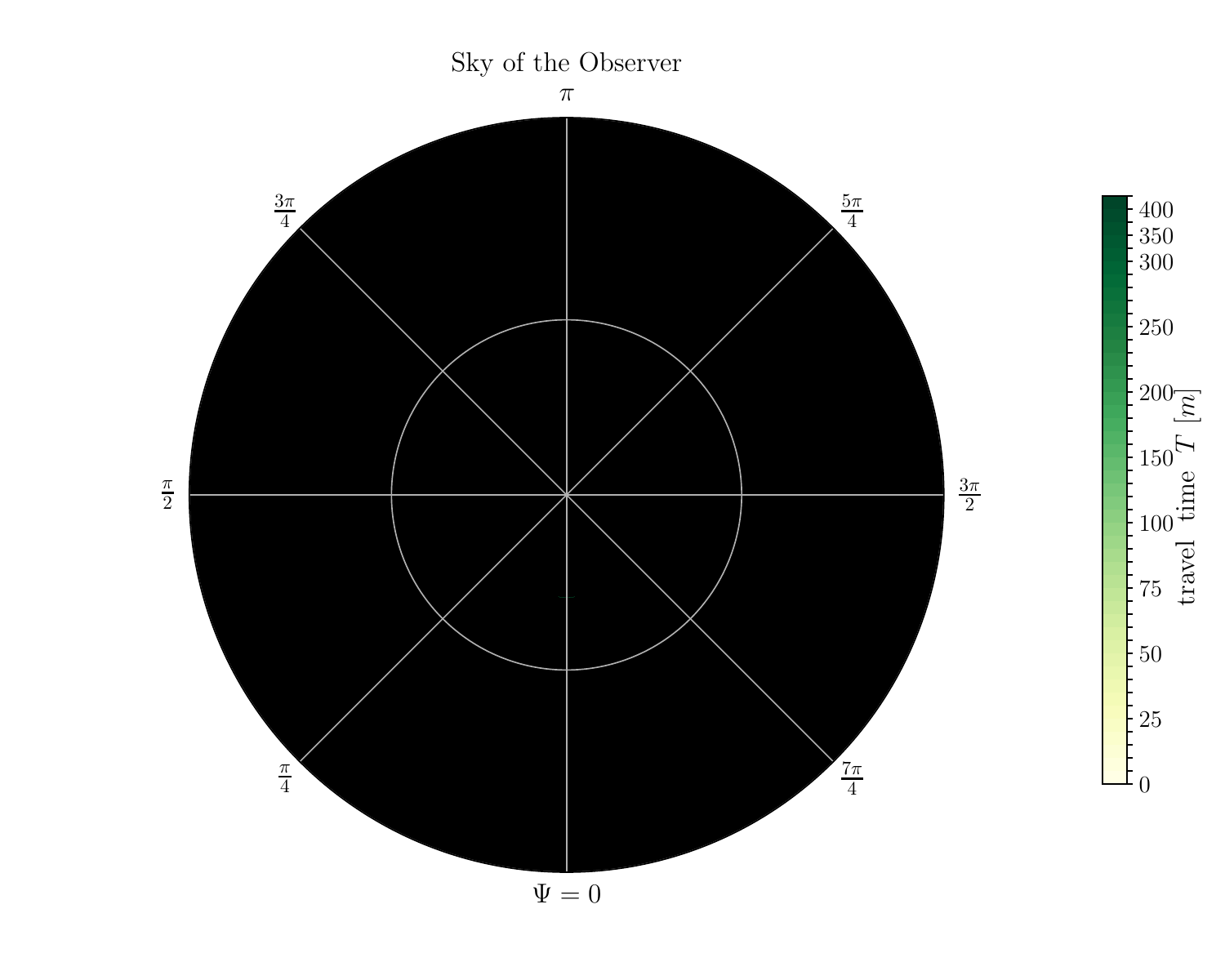} \\
  \end{tabular}
	\caption{Travel time maps for the second-order photon rings in the Schwarzschild spacetime for light rays travelling in vacuum (top left panel), through a homogeneous plasma (top right panel), and through an inhomogeneous plasma described by the distribution $E_{\text{pl}}(r,\vartheta)$ given by (\ref{eq:PlasmaEn}) with $\omega_{\text{p}}=m$ (middle left panel), $\omega_{\text{p}}=2m$ (middle right panel), $\omega_{\text{p}}=3m$ (bottom left panel), and $\omega_{\text{p}}=4m$ (bottom right panel). The observer is located at $r_{O}=40m$ and $\vartheta_{O}=\pi/4$ and the luminous disk is located in the equatorial plane between $r_{\text{in}}=2m$ and $r_{\text{out}}=20m$. For the light rays travelling through one of the plasmas the energy measured at the position of the observer is $E_{O}=\sqrt{53/50}E_{\text{C}}$.}
\end{figure*}

The last quantity of interest is the travel time. It measures in terms of the time coordinate $t$ the time a light ray needs to travel from the source to the observer. It is not directly observable, however, when we have characteristic structures in an accretion disk, such as, e.g., a hotspot, we can search for them in the images of the photon rings of the different orders. When these structures now have characteristic features in their emission profile we can use them to derive travel time differences and compare them with model results for different black hole spacetimes and plasma profiles or, in the ideal case, they may allow us to infer information about the plasma directly from the observations without any further model assumptions. In terms of the time coordinate at the position of the observer $t_{O}$ and the time coordinate at which the light ray was emitted by the source $t_{S}$ the travel time $T$ reads
\begin{eqnarray}
T=t_{O}-t_{S}.
\end{eqnarray}
Here, when an observer detects two images of the same structure in the photon rings and is able to identify the same characteristic emission signatures in these images at the proper times $\tau_{1}$ and $\tau_{2}$ in his reference frame, where we for simplicity assume that we have $\tau_{1}<\tau_{2}$, he can calculate the travel time difference in terms of the time coordinate from the determined proper times via
\begin{eqnarray}\label{eq:difftrav}
\Delta T=t_{2}-t_{1}=\frac{\tau_{2}-\tau_{1}}{\sqrt{P(r_{O})}}=\frac{\Delta\tau}{\sqrt{P(r_{O})}}.
\end{eqnarray}
In the following we will now derive the travel time for the direct image and the photon rings of first and second order for light rays travelling in vacuum as well as through the homogeneous and inhomogeneous plasmas specified in Section~\ref{Sec:SST}, and investigate how they change for different plasma distributions. Let us for this purpose first write down the travel time integral. In our case we can choose $t_{O}=0$ without loss of generality and thus the travel time integral reads
\begin{widetext}
\begin{eqnarray}
T(\Sigma,\Psi)=\int_{r_{O}...}^{...r_{S}\left(\Sigma,\Psi\right)}\frac{\sqrt{P(r_{O})}E_{O}r'^2\text{d}r'}{P(r')\sqrt{P(r_{O})E_{O}^2r'^4-E_{\text{C}}^2 r'^4P(r')-r'^2P(r')(r_{O}^2(E_{O}^2-E_{\text{C}}^2)\sin^2\Sigma+\omega_{\text{p}}^2E_{\text{c}}^2\sin^2\vartheta_{O}\cos^2\Sigma)}},
\end{eqnarray}
\end{widetext}
where the dots in the limits of the integral shall indicate that when the light ray passes through a turning point we have to split the integral into two separate integrals, and we then have to evaluate each integral separately. Here, for each integral the sign of the root in the denominator has to be chosen according to the direction of the $r$ motion along the corresponding section of the trajectory. 

For the explicit evaluation we first rewrite the travel time integral in terms of elementary functions as well as Legendre's elliptic integrals using the steps outlined in Sec.~\ref{Sec:EoMtint}. Again the computational evaluation of the travel time for the direct image as well as the photon rings of first and second order was carried out using the same set of Julia codes as for their geometric structures as well as the lens and redshift maps.

Figs.~11--13 show travel time maps for the direct image and the first- and second-order photon rings for light rays travelling in vacuum (top left panels), through a homogeneous plasma (top right panels; $\omega_{\text{p}}=0$), and through the inhomogeneous plasma described by the distribution given by Eq.~(\ref{eq:PlasmaEn}) with $\omega_{\text{p}}=m$ (middle left panels), $\omega_{\text{p}}=2m$ (middle right panels), $\omega_{\text{p}}=3m$ (bottom left panels), and $\omega_{\text{p}}=4m$ (bottom right panels). Since the maps are again structurally the same as the lens maps in Figs.~2--4 we will again only discuss the structures in the travel time maps and not the structures of the direct images and the photon rings of first and second order. 

We again start our discussion with the maps for the direct images. We can easily see that all six travel time maps have roughly the same structure. The travel times tend to be shorter on the southern hemisphere while they are longer on the northern hemisphere. In addition, in particular around the meridian and the antimeridian, the travel time increases with decreasing celestial latitude on the southern hemisphere while it decreases with decreasing celestial latitude on the northern hemisphere. Furthermore, when we compare the travel time maps for light rays in vacuum with the travel time maps for light rays travelling through one of the plasmas, we see that for light rays travelling through one of the plasmas the travel time is significantly longer than for light rays travelling in vacuum. In addition, the travel time increases with increasing plasma parameter $\omega_{\text{p}}$. Furthermore, for $\omega_{\text{p}}=2m$ the travel time map shows a separate area with slightly lower travel times at the antimeridian close to the boundary of the inner shadow. When we increase the plasma parameter to $\omega_{\text{p}}=3m$ and then to $\omega_{\text{p}}=4m$ this area grows. In addition, the travel time map for $\omega_{\text{p}}=4m$ shows a similar pattern as the redshift map in the bottom right panel of Fig.~8. On the northern hemisphere close to the celestial equator we can see a region where the travel time first decreases with decreasing celestial latitude, reaching a minimum around $\Sigma=\pi/6$, and then it increases again towards the boundary of the inner shadow. Here, the areas with low redshifts in the bottom right panel of Fig.~8 are at the same positions as the areas with the high travel times in the bottom right panel of Fig.~11. Thus these light rays were emitted in parts of the luminous disk located at larger radius coordinates. 

When we now turn to the travel time maps for the photon rings we can easily see that the travel times generally get longer. However, we also see that they generally decrease towards the boundary of the inner shadow. In addition, in the travel time maps for the first-order photon rings we cans see that we have a crescent-like area of lower travel times on the southern hemisphere close to the boundary of the inner shadow and when we move further in the travel time increases again. The reason for this feature is quite simple. When we compare the travel time maps with the redshift maps we can see that light rays associated with the longer travel times were emitted in parts of the accretion disk for which they experience a lower redshift when they travel to the observer. Thus these light rays are emitted in the outer part of the luminous disk. The light rays associated with lower travel times are emitted at radius coordinates closer to the photon sphere and thus experience a higher redshift.  

Overall when we compare the travel time maps for the photon rings, similar to the redshift maps, the main differences are that for $\omega_{\text{p}}=3m$ at the celestial equator the photon rings are very thin and thus with decreasing celestial latitude the travel time changes very rapidly while for $\omega_{\text{p}}=4m$ in these regions we will not detect any photons. In addition, in both cases the travel time decreases more rapidly with decreasing celestial latitude around the antimeridian (in the case of $\omega_{\text{p}}=4m$ only for the first-order photon ring) than at the meridian. 

For the travel time the question how it can help us to determine the plasma properties is a bit easier to answer than for the redshift. The travel time is not directly accessible, however, travel time differences are. While in vacuum we can only calculate the travel time differences between two different images of the same source for light rays travelling through a plasma we now have three different option. On one hand we can hope that we find the signature of the same source, e.g., in the direct image and the first-order photon ring or the first-order photon ring and the second-order photon ring. When the source has a characteristic emission signature we can calculate the travel time differences in the coordinate system of the observer and then translate it to the travel time. For a static observer this can easily be done using (\ref{eq:difftrav}). The second option is that we calculate travel time differences for signals of the same source, e.g., in the first-order photon ring, but for photons with different energies. Here, the basic procedure will be the same as for the travel time differences between different images of the same source. The third option is that we calculate travel time differences at different energies for signatures of the same source in, e.g., the direct image and the photon ring of first order. While one or even a few travel time differences may not be unique this will very likely change with the number of travel time differences at the same and different energies we are able to calculate. In addition, when we combine the travel time differences with the geometric structures of the direct image and the photon rings it may allow us to place more tight constraints on the plasma properties than the observation of the geometric structures of the direct images and the photon rings alone. However, in all three cases it is still very likely that for interpreting the obatined results one still has to compare them with the output of models for different black holes and accretion disks.

\section{Summary}\label{Sec:SC}
Astrophysical black hole candidates are usually surrounded by an accretion disk which at least partially consists of a plasma. The plasma does not only emit electromagnetic radiation but it also affects the paths of photons travelling through it. While it is a common approach to assume a specific model to determine the properties of the plasma, see, e.g., Refs.~\cite{EHTCollaboration2019b} and \cite{EHTCollaboration2021}, it is desirable to find methods which allow to derive the plasma properties from astronomical observations without assuming a specific model. For this purpose it is particularly important to identify effects which are characteristic enough so that they allow to extract the properties of a plasma without prior knowledge about the plasma. 

The work presented in this paper now represents a step in this direction. Using the example of a luminous disk composed of static light sources surrounding a Schwarzschild black hole in the equatorial plane we investigated how the presence of a plasma affects the direct image and, in particular, the photon rings. 

For this purpose in the first part of this paper we derived the equations of motion for light rays travelling in vacuum, through a homogeneous plasma, and through a generalised version of the inhomogeneous plasma distribution originally introduced by Perlick \cite{Perlick2023}. Then we solved the equations of motion analytically using elementary and Jacobi's elliptic functions as well as Legendre's elliptic integrals. 

In the second part of the paper we then used the solutions to calculate the geometric structures of the direct image of the luminous disk and the first- and second-order photon rings. For this purpose we fixed a static observer in the domain of outer communication and wrote down a lens equation. In addition, we also calculated the redshift and the travel time. 

Here, the geometric structures of the direct images as well as the photon rings, and the lens maps showed the most promising effects for extracting information about the properties of a plasma. For light rays travelling in vacuum and for light rays travelling through a homogeneous plasma in the lens maps the images of the sources from the different quadrants of the luminous disk were clearly separated by straight lines at the celestial equator, the meridian, and the antimeridian. For the inhomogeneous plasma at the celestial equator these lines were on one hand curved and on the other hand shifted to the southern hemisphere. In addition, the strength of the curvature and the amount by which the borders between the image regions were shifted to the southern hemisphere increased with the plasma parameter $\omega_{\text{p}}$. Thus when we observe these features around an astrophysical supermassive black hole candidate, in particular for the photon rings, they will allow us to draw conclusions about the distribution of the plasma as well as the electron density and therefore offer one way to infer information about the properties of the plasma surrounding the black hole. 

In addition, in the case of the homogeneous plasma for light rays with frequencies close to the plasma frequency (in our case expressed as energies) we found two more very characteristic features. On one hand for an observer at $\vartheta_{O}=\pi/4$ the direct image, which is asymmetric for light rays travelling in vacuum, became significantly more circular and on the other hand the size of the inner shadow increased. Since with increasing frequency the structure of the direct image for light rays travelling through a homogeneous plasma approaches the structure of the direct image for light rays travelling in vacuum, when we conduct multifrequency observations and observe a change from a nearly circular direct image at low frequencies to an image which is asymmetric with respect to the celestial equator at high frequencies this likely indicates that a supermassive black hole candidate is surrounded by a homogeneous plasma. 

In addition, we also found one very characteristic effect for the inhomogeneous plasma. In the case of the inhomogeneous plasma when $\omega_{\text{p}}$ is particularly large (in our case we had $\omega_{\text{p}}=4m$) parts of the photon rings can start to disappear. Thus observing this effect and in particular how the photon rings disappear across different frequencies also has the potential provide us with valuable information about the structure of the plasma. 

In addition to these relatively clear signatures we also found that the width of the photon rings and the size and the shape of the (inner) shadow will carry characteristic information about the plasma properties. However, in this case they will be much harder to extract since on one hand currently we are not yet able to observe these features with the resolution which is required to extract these information, and on the other hand for extracting these information very systematic data analyses are required. 

Also the redshift maps were affected by the presence of the plasma, however, the introduced changes are much more difficult to utilise for extracting information about the properties of a plasma. On one hand because around black holes we only have very few characteristic emission lines in the X-ray band and on the other hand because in our case we only considered static sources. However, accretion disks around astrophysical black hole candidates are moving and thus we will have an additional Doppler red- or blueshift which will make the extraction of information much more challenging. However, the redshift also affects observed intensities and thus the effects introduced by the plasma may leave an imprint in them and a comparison with redshift maps calculated for different black hole and plasma models may help to identify them.  

The last quantity we calculated was the travel time. Here, besides some structures we will not be able to resolve, the travel time maps showed that the presence of a plasma leads to a significant increase in the travel time. While the travel time itself is not directly accessible to observations, when we have characteristic singatures in the direct image of an accretion disk and also in the photon rings we can observe them at different frequencies and construct travel time differences between them. In the presence of a plasma these travel time differences will depend on the plasma properties and thus the more travel time differences we are able to calculate the higher the chance that we can use them to extract information about these properties.

What will now be the next steps? First, it will be important to construct more realistic analytical plasma models. Here, it will likely not be possible to formulate a plasma model which can be used to describe the whole plasma profile around a black hole. Thus we will have to combine different plasma models to create a realistic plasma profile. Here, the plasma distributions described by these models will affect the light rays in different ways. In some of them the effects which were found in this study may be amplified while other effects may be weakened or they may not occur at all. A first starting point would be, e.g., to combine a homogeneous plasma model for the plasma profile in the close vicinity of a black hole with an appropriate power-law profile to describe the outer region of the plasma profile. As already mentioned in the introduction both have already investigated separately but it would be an interesting question to investigate how the effects of both profiles combine, and in particular how they affect the geometric structure of the photon rings.

\section*{Acknowledgments}
I would like to thank Che-Yu Chen and Xian Chen for their comments and our discussions. I acknowledge funding from the National Key Research and Development Program of China (Grant No. 2021YFC2203002).

\appendix
\section{Roots for the $r$ and $\vartheta$ Motion}
In this appendix we briefly summarise the basic equations for calculating the roots of the $r$ and $\vartheta$ motion in the Schwarzschild spacetime which were not already provided in Sec.~\ref{Sec:SST}.
\subsection{Roots of the $r$ Motion}\label{Sec:rRoots}
For calculating the roots of the $r$ motion we start with writing down the conditional equation. In the presence of a plasma with the distribution specified in (\ref{eq:PlasmaEn}) it reads
\begin{eqnarray}
r^4+\frac{2mE_{\text{C}}^2}{E^2-E_{\text{C}}^2}r^3-\frac{K}{E^2-E_{\text{C}}^2}r^2+\frac{2mK}{E^2-E_{\text{C}}^2}r=0.
\end{eqnarray}
We can easily see that in the vacuum case with $E_{\text{C}}=0$ the second term vanishes and otherwise the equation structurally remains the same. Thus we can treat both cases at the same time. Now we can easily see that one root is always located at $r=0$ and thus effectively we only have to calculate the roots of the third order polynomial 
\begin{eqnarray}
r^3+\frac{2mE_{\text{C}}^2}{E^2-E_{\text{C}}^2}r^2-\frac{K}{E^2-E_{\text{C}}^2}r+\frac{2mK}{E^2-E_{\text{C}}^2}=0.
\end{eqnarray}
For calculating them we use Cardano's method. In the first step we substitute 
\begin{eqnarray}
r=z-\frac{2mE_{\text{C}}^2}{3\left(E^2-E_{\text{C}}^2\right)}.
\end{eqnarray}
In the next step we introduce two new constants of motion $p$ and $q$. They read
\begin{eqnarray}
&p=-\frac{4m^2E_{\text{C}}^4+3\left(E^2-E_{\text{C}}^2\right)K}{9\left(E^2-E_{\text{C}}^2\right)^2}
\end{eqnarray}
and
\begin{eqnarray}
&q=\frac{m\left(E_{\text{C}}^2\left(8m^2E_{\text{C}}^4+9\left(E^2-E_{\text{C}}^2\right)K\right)+27\left(E^2-E_{\text{C}}^2\right)^2 K\right)}{27\left(E^2-E_{\text{C}}^2\right)^3}.
\end{eqnarray}
In the next step we introduce a third constant of motion $\Delta=p^3+q^2$, which reads
\begin{widetext}
\begin{eqnarray}
\Delta=\frac{27m^2\left(E^2-E_{\text{C}}^2\right)^2 K^2-\left(K-18m^2E_{\text{C}}^2\right)\left(E^2-E_{\text{C}}^2\right)K^2+m^2E_{\text{C}}^4 K\left(16m^2E_{\text{C}}^2-K\right)}{27\left(E^2-E_{\text{C}}^2\right)^4}.
\end{eqnarray}
\end{widetext}
Using these constants of motion we now define two new quantities $r_{+}$ and $r_{-}$. They read
\begin{eqnarray}
r_{+}=\sqrt[3]{-q+\sqrt{\Delta}}
\end{eqnarray}
and
\begin{eqnarray}
r_{-}=-\sqrt[3]{q+\sqrt{\Delta}}.
\end{eqnarray}
Note that in our case we choose the third root such that when we have $\Delta<0$ $r_{+}$ and $r_{-}$ are complex conjugates. Now we use both quantities and get for the remaining three roots
\begin{eqnarray}
r_{1}=r_{+}+r_{-}-\frac{2mE_{\text{C}}^2}{3\left(E^2-E_{\text{C}}^2\right)},
\end{eqnarray}
\begin{eqnarray}
\hspace{-0.3cm}r_{2}=\frac{-1+i\sqrt{3}}{2}r_{+}-\frac{1+i\sqrt{3}}{2}r_{-}-\frac{2mE_{\text{C}}^2}{3\left(E^2-E_{\text{C}}^2\right)},
\end{eqnarray}
and
\begin{eqnarray}
\hspace{-0.3cm}r_{3}=-\frac{1+i\sqrt{3}}{2}r_{+}+\frac{-1+i\sqrt{3}}{2}r_{-}-\frac{2mE_{\text{C}}^2}{3\left(E^2-E_{\text{C}}^2\right)}.
\end{eqnarray}
Using our definition for $r_{+}$ and $r_{-}$ we can easily see that $r_{1}$ is always real and negative. On the other hand we see that the roots $r_{2}$ and $r_{3}$ are real and positive for $\Delta\leq 0$ and complex conjugates for $\Delta>0$. In the former case we now sort and relabel the roots such that we have $r_{4}\leq r_{3}=0\leq r_{2}\leq r_{1}$. In the latter case on the other hand we sort and label the roots such that for the real roots we have $r_{2}<r_{1}=0$. In addition, we choose the imaginary part of the complex conjugate roots such that we have $r_{3}=\bar{r}_{4}=R_{3}+iR_{4}$ with $0<R_{4}$.

\subsection{Roots of the $\vartheta$ Motion for Light Rays in the Inhomogeneous Plasma}\label{Sec:thetaRoots}
Calculating the roots of the $\vartheta$ motion is straight forward. As first step we take (\ref{eq:EoMx}) and write down the conditional equation for the roots. For light rays propagating through the inhomogeneous plasma described by the distribution (\ref{eq:PlasmaEn}) it is a biquadratic polynomial and reads
\begin{eqnarray}\label{eq:thetacond}
x^4+\frac{K-2\omega_{\text{p}}^2 E_{\text{C}}^2}{\omega_{\text{p}}^2 E_{\text{C}}^2}x^2-\frac{K-\omega_{\text{p}}^2 E_{\text{C}}^2-L_{z}^2}{\omega_{\text{p}}^2 E_{\text{C}}^2}=0.
\end{eqnarray} 
Now the roots of this polynomial are easy to derive. In this paper we are only interested in light rays that can come from a luminous disk in the equatorial plane. Therefore, as stated in Sec.~\ref{Sec:EoMtheta} vortical geodesics are effectively not relevant for our discussion. In addition, the photon rings are not well-defined for light rays in the equatorial plane and thus we also exclude this case from our discussion. As a consequence only light rays with $0<K-\omega_{\text{p}}^2 E_{\text{C}}^2-L_{z}^2$ are relevant for our discussion. We can easily read from (\ref{eq:thetacond}) that in this case we have only two real roots. The other two roots are complex conjugates and purely imaginary. 

Now we solve (\ref{eq:thetacond}) and obtain the roots. In the case of the real roots we sort and label them such that we have $x_{2}<0<x_{1}$. They read
\begin{eqnarray}\label{eq:xminIP}
&x_{1}=\cos\vartheta_{\text{min}}=\sqrt{\frac{2\omega_{\text{p}}^2 E_{\text{C}}^2-K+\sqrt{K^2-4\omega_{\text{p}}^2 E_{\text{C}}^2 L_{z}^2}}{2\omega_{\text{p}}^2 E_{\text{C}}^2}}
\end{eqnarray}
and
\begin{eqnarray}\label{eq:xmaxIP}
&\hspace{-0.5cm} x_{2}=\cos\vartheta_{\text{max}}=-\sqrt{\frac{2\omega_{\text{p}}^2 E_{\text{C}}^2-K+\sqrt{K^2-4\omega_{\text{p}}^2 E_{\text{C}}^2 L_{z}^2}}{2\omega_{\text{p}}^2 E_{\text{C}}^2}}.
\end{eqnarray}
In the case of the complex conjugate roots on the other hand we sort and label them such that we have $x_{3}=\bar{x}_{4}=X_{3}+iX_{4}$, where the real and imaginary parts $X_{3}$ and $X_{4}$ are given by
\begin{eqnarray}
X_{3}=0
\end{eqnarray}
and 
\begin{eqnarray}
X_{4}=\sqrt{\frac{K-2\omega_{\text{p}}^2 E_{\text{C}}^2+\sqrt{K^2-4\omega_{\text{p}}^2 E_{\text{C}}^2 L_{z}^2}}{2\omega_{\text{p}}^2 E_{\text{C}}^2}}.
\end{eqnarray}
\section{Elementary and Elliptic Integrals}
When we integrated the equations of motion in Section~\ref{Sec:EoM} we needed to evaluate several elementary and elliptic integrals. In this appendix we provide a brief overview and summarise how to evaluate them.  

\subsection{Elementary Integrals}\label{Sec:EI}
In total we need to evaluate four different elementary integrals. After applying the coordinate transformation (\ref{eq:suby}) we can rewrite all of them in one of the following forms 
\begin{eqnarray}\label{eq:EI1}
I_{1}=\int_{y_{i}}^{y}\frac{\text{d}y'}{\left(y'-y_{a}\right)\sqrt{y'-y_{1}}},
\end{eqnarray}
\begin{eqnarray}\label{eq:EI2}
I_{2}=\int_{y_{i}}^{y}\frac{\text{d}y'}{\left(y'-y_{a}\right)^2\sqrt{y'-y_{1}}},
\end{eqnarray}
where in the first integral the parameter $y_{a}$ can be $-K/12$, $y_{\text{H}}$, or $y_{\text{ph}}$, while in the second integral it can only be $-K/12$. Here $y_{\text{H}}$ and $y_{\text{ph}}$ are related to $r_{\text{H}}=2m$ and $r_{\text{ph}}$ by (\ref{eq:suby}). 

We will start with evaluating the first integral $I_{1}$. When we have $y_{a}=-K/12$ we have $y_{a}<y$. In this case we first substitute $z=y-y_{a}$. In the next step we integrate, and then we rewrite the obtained function in terms of the radius coordinate $r$. The result reads
\begin{widetext} 
\begin{eqnarray}\label{eq:EII11}
I_{1_{1}}=4\sqrt{\frac{r_{3}-r_{4}}{2mK}}\left(\text{arcoth}\left(\sqrt{\frac{r_{i}-r_{4}}{r_{i}}}\right)-\text{arcoth}\left(\sqrt{\frac{r-r_{4}}{r}}\right)\right).
\end{eqnarray}
\end{widetext}
In the next step we evaluate $I_{1}$ for $y_{a}$ given by $y_{\text{H}}$ or $y_{\text{ph}}$. In this case we have $y<y_{a}$. We substitute $z=y-y_{1}$ and integrate. In the next step we again rewrite the obtained functions in terms of the radius coordinate $r$. The results read
\begin{widetext} 
\begin{eqnarray}\label{eq:EII12}
I_{1_{2}}=4\sqrt{\frac{r_{4}}{K \left(r_{4}-2m\right)}}\left(\text{artanh}\left(\sqrt{\frac{2m\left(r_{i}-r_{4}\right)}{\left(2m-r_{4}\right)r_{i}}}\right)-\text{artanh}\left(\sqrt{\frac{2m\left(r-r_{4}\right)}{\left(2m-r_{4}\right)r}}\right)\right)
\end{eqnarray}
and
\begin{eqnarray}\label{eq:EII13}
I_{1_{3}}=4\sqrt{\frac{r_{\text{ph}}r_{4}}{2mK \left(r_{4}-r_{\text{ph}}\right)}}\left(\text{artanh}\left(\sqrt{\frac{r_{\text{ph}}\left(r_{i}-r_{4}\right)}{\left(r_{\text{ph}}-r_{4}\right)r_{i}}}\right)-\text{artanh}\left(\sqrt{\frac{r_{\text{ph}}\left(r-r_{4}\right)}{\left(r_{\text{ph}}-r_{4}\right)r}}\right)\right).
\end{eqnarray}
\end{widetext} 
In the case of the second integral $I_{2}$ we only have $y_{a}=-K/12$ and thus $y_{a}<y$. We again substitute $z=y-y_{a}$ and integrate. After the integration we again rewrite the result in terms of the radius coordinate $r$ and get
\begin{widetext}
\begin{eqnarray}\label{eq:EII2}
I_{2}=\frac{4}{mK}\sqrt{-\frac{r_{4}}{2mK}}\left(r_{4}\left(\text{arcoth}\left(\sqrt{\frac{r_{i}-r_{4}}{r_{i}}}\right)-\text{arcoth}\left(\sqrt{\frac{r-r_{4}}{r}}\right)\right)+\sqrt{r_{i}\left(r_{i}-r_{4}\right)}-\sqrt{r\left(r-r_{4}\right)}\right).
\end{eqnarray}
\end{widetext}
\subsection{Elliptic Integrals}\label{Sec:ELF}
When we solved the equations of motion in Sec.~\ref{Sec:EoM} we encountered several elliptic integrals. In this paper we will rewrite them in terms of elementary functions and Legendre's elliptic integrals of the first, second, and third kind. However, before we use them to solve the equations of motion we need to define them. This is the purpose of the first part of this appendix. In the second part we will then rewrite three different nonstandard elliptic integrals in terms of elementary functions and Legendre's elliptic integrals of the first, second, and third kind.

We will start with defining Legendre's incomplete elliptic integrals of the first, second, and third kind. They read
\begin{eqnarray}\label{eq:FL}
F_{L}\left(\chi,k\right)=\int_{0}^{\chi}\frac{\text{d}\chi'}{\sqrt{1-k\sin^2\chi'}},
\end{eqnarray}
\begin{eqnarray}\label{eq:EL}
E_{L}\left(\chi,k\right)=\int_{0}^{\chi}\sqrt{1-k\sin^2\chi'}\text{d}\chi',
\end{eqnarray}
and 
\begin{eqnarray}\label{eq:PiL}
\Pi_{L}\left(\chi,k,n\right)=\int_{0}^{\chi}\frac{\text{d}\chi'}{\left(1-n\sin^2\chi'\right)\sqrt{1-k\sin^2\chi'}}.
\end{eqnarray}
Here, $\chi$ is the amplitude of the elliptic integrals, $k$ is the square of the elliptic modulus, and $n$ is an arbitrary real parameter. Now when the amplitude takes the value $\chi=\pi/2$ the three elliptic integrals are referred to as complete elliptic integrals and one often omits the amplitude in the argument. In addition, in the case of Legendre's elliptic integral it is also a convention to write it as $F_{L}\left(\pi/2,k\right)=K_{L}\left(k\right)$. In addition, one can easily see that when we replace $\chi \rightarrow -\chi$ we get
\begin{eqnarray}
F_{L}\left(-\chi,k\right)=-F_{L}\left(\chi,k\right),
\end{eqnarray}
\begin{eqnarray}
E_{L}\left(-\chi,k\right)=-E_{L}\left(\chi,k\right),
\end{eqnarray}
and 
\begin{eqnarray}
\Pi_{L}\left(-\chi,k,n\right)=-\Pi_{L}\left(\chi,k,n\right).
\end{eqnarray}
Furthermore, we can also easily see that the integrand of Legendre's elliptic integral of the third kind has a singularity at $\chi=\text{arcsin}\left(1/\sqrt{n}\right)$. Thus when the amplitude approaches this characteristic value Legendre's elliptic integral of the third kind diverges. In our case this divergence always occurs for the time coordinate when the radius coordinate $r$ approaches infinity or the radius coordinate of the horizon, and in addition for the inhomogeneous plasma distribution specified in (\ref{eq:PlasmaEn}) for the $\varphi$ coordinate when the $\vartheta$ coordinate approaches $\vartheta=0$ or $\vartheta=\pi$. 

In the case that we have to integrate over the singularity, and also in the case that one integrates from a position close to the singularity and wants to perform very high-resolution calculations, one can now use equations (17.7.7) and (17.7.8) in the formula collection of Abramowitz and Stegun \cite{MilneThomson1972} to rewrite Legendre's elliptic integral of the third kind as 
\begin{widetext}
\begin{eqnarray}\label{eq:EITK}
\Pi_{L}\left(\chi,k,n\right)=F_{L}\left(\chi,k\right)-\Pi_{L}\left(\chi,k,\frac{k}{n}\right)+\frac{1}{2\tilde{p}}\ln\left(\frac{\cos\chi\sqrt{1-k\sin^2\chi}+\tilde{p}\sin\chi}{\left|\cos\chi\sqrt{1-k\sin^2\chi}-\tilde{p}\sin\chi\right|}\right),
\end{eqnarray}
\end{widetext}
where $\tilde{p}$ is defined by
\begin{eqnarray}
\tilde{p}=\sqrt{\frac{\left(n-1\right)\left(n-k\right)}{n}}.
\end{eqnarray}
While some of the integrals we encountered in Sec.~\ref{Sec:EoM} can directly be rewritten in terms of Legendre's elliptic integrals of the first and third kind, when we integrated the time coordinate $t$ we also encountered three nonstandard elliptic integrals which do not immediately take one of the standard forms given by (\ref{eq:FL})--(\ref{eq:PiL}). They read 
\begin{eqnarray}\label{eq:IL1}
&\hspace{-0.3cm}I_{L_{1}}\left(\chi_{i},\chi,k,n\right)=\int_{\chi_{i}}^{\chi}\frac{\text{d}\chi'}{\left(1+n\cos\chi'\right)\sqrt{1-k\sin^2\chi'}},
\end{eqnarray}
\begin{eqnarray}\label{eq:IL2}
&\hspace{-0.3cm}I_{L_{2}}\left(\chi_{i},\chi,k,n\right)=\int_{\chi_{i}}^{\chi}\frac{\text{d}\chi'}{\left(1+n\cos\chi'\right)^2\sqrt{1-k\sin^2\chi'}},
\end{eqnarray}
and
\begin{eqnarray}\label{eq:IL3}
&\hspace{-0.3cm}I_{L_{3}}\left(\chi_{i},\chi,k,n\right)=\int_{\chi_{i}}^{\chi}\frac{\text{d}\chi'}{\left(1-n\sin^2\chi'\right)^2\sqrt{1-k\sin^2\chi'}}.
\end{eqnarray}
In the following we will now briefly demonstrate how to rewrite these three integrals in terms of elementary functions and Legendre's elliptic integrals of the first, second, and third kind. We will start with the integrals $I_{L_{1}}\left(\chi_{i},\chi,k,n\right)$ and $I_{L_{2}}\left(\chi_{i},\chi,k,n\right)$. In this case the two amplitudes $\chi_{i}$ and $\chi$ are related to $r_{i}$ and $r(\lambda)$ via (\ref{eq:chi1}). Note that in the following we will omit the Mino parameter for the corresponding amplitude for brevity. The square of the elliptic modulus $k$ on the other hand is given by (\ref{eq:k1}). The parameter $n$ on the other hand can be given by 
\begin{eqnarray}\label{eq:n1}
n_{1}=\frac{\bar{R}+R}{\bar{R}-R}
\end{eqnarray}
or
\begin{eqnarray}\label{eq:n2}
n_{2}=\frac{2m\bar{R}+\left(2m-r_{2}\right)R}{2m\bar{R}-\left(2m-r_{2}\right)R},
\end{eqnarray}
where $R$ and $\bar{R}$ are given by (\ref{eq:R}) and (\ref{eq:Rbar}), respectively. In the case of the first integral $I_{L_{1}}\left(\chi_{i},\chi,k,n\right)$ we first expand by $1-n\cos\chi'$ in the integrand and get
\begin{widetext}
\begin{eqnarray}
&I_{L_{1}}\left(\chi_{i},\chi,k,n\right)=\frac{1}{1-n^2}\left(\Pi_{L}\left(\chi,k,\frac{n^2}{n^2-1}\right)-\Pi_{L}\left(\chi_{i},k,\frac{n^2}{n^2-1}\right)-n\int_{\chi_{i}}^{\chi}\frac{\cos\chi'\text{d}\chi'}{\left(1-\frac{n^2}{n^2-1}\sin^2\chi'\right)\sqrt{1-k\sin^2\chi'}}\right),
\end{eqnarray}
\end{widetext}
where we already rewrote the first term in terms of Legendre's elliptic integral of the third kind. The second term on the other hand is an elementary integral. Its evaluation requires a case by case analysis and we will not reproduce it here. After evaluating the elementary integral we insert the obtained result and get 
\begin{widetext}
\begin{eqnarray}\label{eq:EI1EV}
I_{L_{1}}\left(\chi_{i},\chi,k,n\right)=\frac{\Pi_{L}\left(\chi,k,\frac{n^2}{n^2-1}\right)-\Pi_{L}\left(\chi_{i},k,\frac{n^2}{n^2-1}\right)}{1-n^2}+\frac{n\left(\tilde{I}_{L}(\chi,k,n)-\tilde{I}_{L}(\chi_{i},k,n)\right)}{2\sqrt{\left(n^2-1\right)\left(n^2\left(1-k\right)+k\right)}},
\end{eqnarray}
\end{widetext}
where the function $\tilde{I}_{L}\left(\chi',k,n\right)$ is given below. Now we turn to the second integral $I_{L_{2}}\left(\chi_{i},\chi,k,n\right)$. Again we follow the same procedure. We first expand with $\left(1-n\cos\chi'\right)^2$ in the integrand and simplify the terms. After a few steps we can easily see that we can rewrite $I_{L_{2}}\left(\chi_{i},\chi,k,n\right)$ in the following form
\begin{widetext}
\begin{eqnarray}\label{eq:IL2M}
&I_{L_{2}}\left(\chi_{i},\chi,k,n\right)=\frac{2}{\left(n^2-1\right)^2}\left(\int_{\chi_{i}}^{\chi}\frac{\text{d}\chi'}{\left(1-\frac{n^2}{n^2-1}\sin^2\chi'\right)^2\sqrt{1-k\sin^2\chi'}}-n\int_{\chi_{i}}^{\chi}\frac{\cos\chi'\text{d}\chi'}{\left(1-\frac{n^2}{n^2-1}\sin^2\chi'\right)^2\sqrt{1-k\sin^2\chi'}}\right)\\
&+\frac{\Pi_{L}\left(\chi,k,\frac{n^2}{n^2-1}\right)-\Pi_{L}\left(\chi_{i},k,\frac{n^2}{n^2-1}\right)}{n^2-1}.\nonumber
\end{eqnarray}
\end{widetext}
Here, the first integral is another nonstandard elliptic integral and we can easily see that it has the form of (\ref{eq:IL3}). We will rewrite it in terms of elementary functions and Legendre's elliptic integrals of the first, second, and third kind below. The second term is an elementary integral. Again evaluating it requires a case by case analysis, which will not be reproduced here. The third term again consists of elliptic integrals, which we already rewrote in terms of Legendre's elliptic integral of the third kind. After evaluating the elementary integral we use the obtained result and (\ref{eq:EI3EV}) below to rewrite (\ref{eq:IL2}) as  
\begin{widetext}
\begin{eqnarray}\label{eq:EI2EV}
&\hspace{-0.7cm}I_{L_{2}}\left(\chi_{i},\chi,k,n\right)=\frac{n^3}{\left(n^2-1\right)\left(n^2\left(1-k\right)+k\right)}\left(\frac{\sin\chi\sqrt{1-k\sin^2\chi}}{1+n\cos\chi}-\frac{\sin\chi_{i}\sqrt{1-k\sin^2\chi_{i}}}{1+n\cos\chi_{i}}\right)+\frac{n\left(n^2\left(1-2k\right)+2k\right)\left(\tilde{I}_{L}\left(\chi_{i},k,n\right)-\tilde{I}_{L}\left(\chi,k,n\right)\right)}{2\left(\left(n^2-1\right)\left(n^2\left(1-k\right)+k\right)\right)^{\frac{3}{2}}}\\
&+\frac{F_{L}\left(\chi,k\right)-F_{L}\left(\chi_{i},k\right)}{n^2-1}+\frac{n^2\left(E_{L}\left(\chi_{i},k\right)-E_{L}\left(\chi,k\right)\right)}{\left(n^2-1\right)\left(n^2\left(1-k\right)+k\right)}+\frac{\left(n^2\left(1-2k\right)+2k\right)\left(\Pi_{L}\left(\chi,k,\frac{n^2}{n^2-1}\right)-\Pi_{L}\left(\chi_{i},k,\frac{n^2}{n^2-1}\right)\right)}{\left(n^2-1\right)^2\left(n^2\left(1-k\right)+k\right)},\nonumber
\end{eqnarray}
\end{widetext}
where here and also in (\ref{eq:EI1EV}) above the function $\tilde{I}_{L}\left(\chi',k,n\right)$ reads
\newpage
\begin{eqnarray}
&\hspace{-0.9cm}\tilde{I}_{L}\left(\chi',k,n\right)=\ln\left(\frac{\sin\chi'\sqrt{\frac{n^2\left(1-k\right)+k}{n^2-1}}+\sqrt{1-k\sin^2\chi'}}{\left|\sin\chi'\sqrt{\frac{n^2\left(1-k\right)+k}{n^2-1}}-\sqrt{1-k\sin^2\chi'}\right|}\right).
\end{eqnarray}
Note that here we cannot avoid to integrate over the singularity of the integrand of Legendre's elliptic integral of the third kind. Therefore, we now use (\ref{eq:EITK}) to rewrite Legendre's elliptic integral of the third kind such that the associated terms remain finite. 

The last remaining nonstandard elliptic integral we have to rewrite is given by (\ref{eq:IL3}). It occurred when we rewrote (\ref{eq:IL2}) as (\ref{eq:EI2EV}) and when we rewrote the time coordinate $t$ in terms of elementary functions and Legendre's elliptic integrals of the first, second, and third kind for the $r$ motion with a turning point at $r_{\text{min}}$ in the domain of outer communication. In the first case $\chi_{i}$ and $\chi$ are again related to $r_{i}$ and $r(\lambda)$ by (\ref{eq:chi1}) and the square of the elliptic modulus $k$ is given by (\ref{eq:k1}). In the second case $\chi_{i}$ and $\chi$ are related to $r_{i}$ and $r(\lambda)$ by (\ref{eq:chi2}) and the square of the elliptic modulus $k$ is given by (\ref{eq:k2}). Again we can rewrite the nonstandard elliptic integral in terms of elementary functions and Legendre's elliptic integrals of the first, second, and third kind. The result reads
\begin{widetext}
\begin{eqnarray}\label{eq:EI3EV}
&I_{L_{3}}\left(\chi_{i},\chi,k,n\right)=\frac{n^2}{4\left(n-k\right)\left(n-1\right)}\left(\frac{\sin\left(2\chi\right)\sqrt{1-k\sin^2\chi}}{1-n\sin^2\chi}-\frac{\sin\left(2\chi_{i}\right)\sqrt{1-k\sin^2\chi_{i}}}{1-n\sin^2\chi_{i}}\right)+\frac{F_{L}\left(\chi,k\right)-F_{L}\left(\chi_{i},k\right)}{2\left(n-1\right)}\\
&+\frac{n\left(E_{L}\left(\chi_{i},k\right)-E_{L}\left(\chi,k\right)\right)}{2\left(n-k\right)\left(n-1\right)}+\frac{\left(n\left(n-2\right)-\left(2n-3\right)k\right)\left(\Pi_{L}\left(\chi,k,n\right)-\Pi_{L}\left(\chi_{i},k,n\right)\right)}{2\left(n-k\right)\left(n-1\right)}.\nonumber
\end{eqnarray}
\end{widetext}
Note that here for light rays that can pass through the turning point $r_{\text{min}}=r_{1}$ in the domain of outer communication we can use (\ref{eq:EI3EV}) as is. However, when we use (\ref{eq:EI3EV}) to rewrite (\ref{eq:IL2}) as (\ref{eq:EI2EV}) we have to replace $n\rightarrow n^2/(n^2-1)$.
\section{Jacobi's Elliptic Functions and Their Application to Solving Differential Equations}\label{eq:JEF}
In this paper we use Jacobi's elliptic functions to solve the equations of motion for $r$ and $\vartheta$ when we do not have multiple roots. Since not all readers may be familiar with Jacobi's elliptic functions in this appendix we will briefly define them and show how we can use them to solve differential equations of the type
\begin{eqnarray}\label{eq:DiffJac}
\left(\frac{\text{d}z}{\text{d}\lambda}\right)^2=a_{4}z^4+a_{3}z^3+a_{2}z^2+a_{1}z+a_{0},
\end{eqnarray}
where the five coefficients $a_{0}$--$a_{4}$ are all real and $\lambda$ is describing the evolution of $z$ and is for now not related to the Mino parameter. Note that in the case that we have $a_{4}=a_{3}=0$ or multiple roots we do not need Jacobi's elliptic functions to solve this equation. In this case it can be solved using elementary functions.

Let us now first define Jacobi's elliptic functions. We start with writing down Legendre's elliptic integral of the first kind
\begin{eqnarray}\label{eq:IntLeg}
\tilde{\lambda}=F_{L}\left(\chi,k\right)=\int_{0}^{\chi}\frac{\text{d}\chi'}{\sqrt{1-k\sin^2\chi'}},
\end{eqnarray}
where again $\chi$ is the amplitude and $k$ is the square of the elliptic modulus. This equation relates $\tilde{\lambda}$ on the left-hand side with the ampitude $\chi$ on the right-hand side and the common termonilogy is that one says that $\chi$ is the amplitude of $\tilde{\lambda}$, in short $\chi=\text{am}\tilde{\lambda}$. Now we can use this relation to define Jacobi's elliptic functions. For this purpose we start with the sine and cosine functions from trigonometry. We first define Jacobi's elliptic sn function, also called the \emph{sinus amplitudinis}. It relates the amplitude $\chi$ to $\tilde{\lambda}$ via
\begin{eqnarray}
\sin\chi=\sin\text{am}\tilde{\lambda}=\text{sn}\left(\tilde{\lambda},k\right).
\end{eqnarray}
Analogously we can define Jacobi's elliptic cn function, also called \emph{cosinus amplitudinis}, via
\begin{eqnarray}
\cos\chi=\cos\text{am}\tilde{\lambda}=\text{cn}\left(\tilde{\lambda},k\right).
\end{eqnarray}
The last of the elliptic functions is Jacobi's elliptic dn function, also called \emph{delta amplitude}. It does not have a trigonometric analogue and is defined by
\begin{eqnarray}
\sqrt{1-k\sin^2\chi}=\text{dn}\left(\tilde{\lambda},k\right).
\end{eqnarray}
The three elliptic functions form a set of basic elliptic functions. However, in addition to these elliptic functions one often defines several associated elliptic functions. In this paper we only need one of them, Jacobi's elliptic sd function. It is defined by
\begin{eqnarray}
\text{sd}\left(\tilde{\lambda},k\right)=\frac{\text{sn}\left(\tilde{\lambda},k\right)}{\text{dn}\left(\tilde{\lambda},k\right)}.
\end{eqnarray}
All of the elliptic functions now have the property that they solve the differential equation
\begin{eqnarray}\label{eq:LegDiff}
\left(\frac{\text{d}\chi}{\text{d}\tilde{\lambda}}\right)^2=1-k\sin^2\chi.
\end{eqnarray}
Now it is easy to see that when we want to solve differential equations with the same form as (\ref{eq:DiffJac}) we have to transform (\ref{eq:DiffJac}) to the form given above. Here, it is interesting to note that when the right-hand side of (\ref{eq:DiffJac}) does not have multiple roots, we can always find a coordinate transformation, in our case, e.g., $z=f(\sin\chi)$, $z=f(\cos\chi)$, or $z=f(\sin\chi,\cos\chi)$ (note that in the latter case we will see that we can rewrite the solution in terms of Jacobi's elliptic sd function), that transforms (\ref{eq:DiffJac}) to 
\begin{eqnarray}\label{eq:LegDiffTrans}
\left(\frac{\text{d}\chi}{\text{d}\lambda}\right)^2=a_{4}C_{L}\left(1-k\sin^2\chi\right),
\end{eqnarray}
where $C_{L}$ is a constant whose exact form depends on the coordinate transformation we used to put (\ref{eq:DiffJac}) into the form of (\ref{eq:LegDiffTrans}). As we can see this form is already very similar to (\ref{eq:LegDiff}). Now we separate variables and integrate. We obtain
\begin{eqnarray}\label{eq:LFE}
\lambda-\lambda_{i}=\frac{i_{\chi_{i}}}{\sqrt{a_{4}C_{L}}}\int_{\chi_{i}}^{\chi}\frac{\text{d}\chi'}{\sqrt{1-k\sin^2\chi'}},
\end{eqnarray}
where $i_{\chi_{i}}=\text{sgn}\left(\left.\text{d}\chi/\text{d}\lambda\right|_{\chi=\chi_{i}}\right)$. In the next step we move all terms containing the initial conditions to the left-hand side and get
\begin{widetext}
\begin{eqnarray}
\tilde{\lambda}\left(\lambda\right)=i_{\chi_{i}}\sqrt{a_{4}C_{L}}\left(\lambda-\lambda_{i}\right)+F_{L}\left(\chi_{i},k\right)=\int_{0}^{\chi}\frac{\text{d}\chi'}{\sqrt{1-k\sin^2\chi'}},
\end{eqnarray}
\end{widetext}
where on the left-hand side we defined $\tilde{\lambda}(\lambda)$ such that the equation takes the same form as equation (\ref{eq:IntLeg}). Now we can easily see that in this case we have $\chi=\text{am}\tilde{\lambda}(\lambda)$ and thus, in our case, the solutions to (\ref{eq:LegDiff}) are given by $\text{sn}\left(\tilde{\lambda}(\lambda),k\right)$, $\text{cn}\left(\tilde{\lambda}(\lambda),k\right)$, or $\text{sd}\left(\tilde{\lambda}(\lambda),k\right)$. As a result in our case the solutions to (\ref{eq:DiffJac}) are then given by $z(\lambda)=f\left(\text{sn}\left(\tilde{\lambda}(\lambda),k\right)\right)$, $z(\lambda)=f\left(\text{cn}\left(\tilde{\lambda}(\lambda),k\right)\right)$, or $z(\lambda)=f\left(\text{sd}\left(\tilde{\lambda}(\lambda),k\right)\right)$.

\bibliography{Photon_Rings_Plasma_Schwarzschild.bib}

\end{document}